\documentclass[prd,superscriptaddress,showpacs]{revtex4}
\usepackage{amsmath}
\usepackage{graphicx}
\usepackage{dsfont}
\usepackage{subfigure}
\def \D {\tilde{\nabla}}
\def \curl {\mbox{curl}\,}
\def \ep {\varepsilon}
\def\l{\label}

\def\Th{\Theta}
\def\dd{\mathcal D}

\def\sig{\sigma}
\def\om{\omega}
\def\udot{\dot{u}}
\def\nab{\nabla}
\def\3nab{\tilde{\nabla}}

\def\lgl{\langle}
\def\rgl{\rangle}

\def\nn{\nonumber}
\def\c{\mbox{curl}}

\def\hsp5{\hspace{5mm}}

\def\case#1/#2{\textstyle\frac{#1}{#2}}

\def\be {\begin{equation}}
\def\ee {\end{equation}}
\def\ber {\begin{eqnarray}}
\def\eer {\end{eqnarray}}
\def\bea {\begin{eqnarray}}
\def\eea {\end{eqnarray}}

\def\bc {\begin{center}}
\def\ec {\end{center}}
\def\case#1/#2{\frac{#1}{#2}}
\def\rf#1{(\ref{#1})}

\def\cqg{{\it Class. Quantum Grav.}\ }

\def\etal\;{{\it et al.}}

\begin{document}

\title{A detailed analysis of  structure growth in  $f(R)$ theories of  gravity}

 \author{Kishore N. Ananda}
\affiliation{ Department of Mathematics and Applied\ Mathematics,
University of Cape Town, South Africa.}
\author{Sante Carloni}
 \affiliation{ Department of Mathematics and Applied\ Mathematics,
University of Cape Town, South Africa.}
 \author{Peter K S Dunsby}
\affiliation{ Department of Mathematics and Applied\ Mathematics,
University of Cape Town, South Africa.}
\affiliation{\ South African
Astronomical Observatory, Observatory Cape Town, South Africa.}

\begin{abstract}
We investigate the connection between dark energy and fourth order gravity by analyzing the behavior of scalar perturbations around a Friedmann-Robertson-Walker  background. The evolution equations for scalar perturbation are derived using the covariant and gauge invariant  approach and applied to two widely studied $f(R)$ gravity models. The structure of the general {\it fourth order} perturbation equations and the analysis of scalar perturbations lead to the discovery of a characteristic signature of fourth order gravity in the matter power spectrum, the details of which have not seen before in other works in this area. This could provide a crucial test for fourth order gravity on cosmological scales.
\end{abstract}

\date{\today}
 \maketitle

\section{Introduction}
In spite of all the efforts made so far, the problem of the nature of Dark Energy (DE) is still far from a completely satisfactory resolution. Among the many theoretical frameworks proposed, the idea of a geometrical origin of Dark Energy has recently received a great deal of attention. The main reason for this popularity can be found in the fact that these type of theories of gravity, which are suggested by the low energy limit of very fundamental schemes \cite{stringhe,birrell}, lead to cosmologies which admit naturally a Dark Energy era \cite{SalvSolo,revnostra,Odintsov, Carroll,star2007} (and possibly even an inflationary one \cite{kerner1,star80,Cognola1}) without the introduction of any additional cosmological fields.

Much of the investigation performed up to now on the idea of Geometric Dark Energy has been focused on the so-called fourth order theories of gravity. In these theories the Hilbert-Einstein action is modified with terms that are at most of order four in the metric tensor. The features of fourth order gravity have been analyzed with different methods \cite{OurDynSys, BarrowDyn,Barrow Hervik,OdintRec,CognolaDyn} and it has been shown that their cosmologies can give rise to a phase of accelerated expansion which is considered the footprint of Dark Energy.

Although these results are very encouraging there are still some important open problems to be addressed. One of them is the analysis of  the evolution of the linear perturbations and their comparison with observations. Over the past year this problem has been studied by a number of authors,  by (1) considering different ways of parameterizing the non-Einstein modifications of gravity or (2) by simplifying the underlying fourth-order perturbation equations using a quasi-static approximation or a combination of (1) and (2) \cite{Li:2008ai,HuSawicki,Bertschinger:2008zb,otherperts}.

In a number of recent papers \cite{SantePert, K1} we derived the evolution equations for scalar and tensor perturbations of a subclass of fourth order theories of gravity characterized by an action which is a general analytic function of the Ricci scalar.
In our work we study the dynamics of linear scalar perturbations using the covariant and gauge invariant approach developed for General Relativity (GR) in \cite{EllisCovariant,EB,EBH,BDE,DBE,BED,DBBE}. This approach has the  advantage of using perturbation variables with a clear geometrical and physical interpretation. Furthermore, we use a specific recasting of the field equations that will make the development of the cosmological perturbation theory even more transparent, allowing one to integrate the perturbation equations exactly for a given $f(R)$ model without making any additional approximations.

The preliminary results obtained in \cite{SantePert} showed some interesting features. First of all the evolution of scalar perturbations is determined by a fourth order differential equation rather than a second order one. This implies that the evolution of the density fluctuations contains, in general, four modes rather that two and can give rise to a more complex evolution than the one of General Relativity (GR).  Secondly, the perturbations are found to depend on the scale for any equation of state for standard matter (while in GR the evolution of the dust perturbations are not scale dependent). This means that, for example, in this framework the evolution of super-horizon and sub-horizon  perturbations is different. Third, and more surprisingly, we found  that growth of large density fluctuations can occur also in backgrounds in which the expansion rate is increasing in time. This is in striking contrast with what one finds in GR and what one would naively expect, but at the same time suggests new ways to tackle the DE problem.

The  features mentioned above imply that the evolution of perturbations in this framework can be completely different from the one we are familiar with. Yet this does not necessarily mean that they are incompatible with observations. Rather, they are a sign of the fact that in dealing with these models one has to resist the temptation of using assumptions which work well in GR. In this paper, following in this spirit, we will analyze further what was found in \cite{SantePert} with the aim of achieving  a clearer understanding of the physics of the matter dominated era in fourth order gravity. In order to do this, we will rewrite the perturbation equations in a more physically meaningful way and will develop a series of tools which will make the analysis of the evolution of  density perturbations easier to understand and to compare with GR.

The paper is organized as follows. In section II we will give some  basic equations and we will present briefly the covariant gauge invariant formalism we use to develop the perturbation theory. In section III, we give the background and the perturbation equations. In section IV we rewrite these equation in an interesting form allowing us to discuss their general structure. In section V  we propose some useful tools to understand the behavior of the perturbations and compare it with what one obtains in General Relativity.  In section VI we apply these tools to some simple specific examples. Section VII is dedicated to the conclusions.

Unless otherwise specified, natural units ($\hbar=c=k_{B}=8\pi G=1$)
will be used throughout this paper, Latin indices run from 0 to 3.
The symbol $\nabla$ represents the usual covariant derivative and
$\partial$ corresponds to partial differentiation. We use the
$-,+,+,+$ signature and the Riemann tensor is defined by
\begin{equation}
R^{a}{}_{bcd}=W^a{}_{bd,c}-W^a{}_{bc,d}+ W^e{}_{bd}W^a{}_{ce}-
W^f{}_{bc}W^a{}_{df}\;,
\end{equation}
where the $W^a{}_{bd}$ are the Christoffel symbols (i.e. symmetric in
the lower indices), defined by
\begin{equation}
W^a_{bd}=\frac{1}{2}g^{ae}
\left(g_{be,d}+g_{ed,b}-g_{bd,e}\right)\;.
\end{equation}
The Ricci tensor is obtained by contracting the {\em first} and the
{\em third} indices
\begin{equation}\label{Ricci}
R_{ab}=g^{cd}R_{acbd}\;.
\end{equation}
Finally the Hilbert--Einstein action in the presence of matter is
given by
\begin{equation}
{\cal A}=\int d x^{4} \sqrt{-g}\left[\frac{1}{2}R+ L_{m}\right]\;.
\end{equation}

\section{General equations for fourth order gravity.}
In four dimensional homogeneous and isotropic spacetimes i.e. Friedmann Lema\^{\i}tre Robertson Walker (FLRW) universes,  the most general action for fourth order gravity can be written as an analytic function of the Ricci scalar only:
\begin{equation}\label{lagr f(R)}
\mathcal{A}=\int d^4 x \sqrt{-g}\left[ f(R)+{\cal L}_{m}\right]\;,
\end{equation}
where $\mathcal{L}_m$ represents the matter contribution.
Varying the action with respect to the metric gives the generalization of the Einstein equations:
\begin{equation}\label{eq:einstScTn}
f'G_{ab}=f'\left(R_{ab}-\frac{1}{2}\,g_{ab} R\right)=T
_{ab}^{m}+\frac{1}{2}g_{ab} \left(f-R f'\right) +\nab_b\nab_a f'-
g_{ab}\nab_c\nab^c f'\;,
\end{equation}
where $f=f(R)$, $f'= \displaystyle{\frac{d f(R)}{dr}}$, and
$\displaystyle{T^{M}_{ab}=\frac{2}{\sqrt{-g}}\frac{\delta
(\sqrt{-g}\mathcal{L}_{m})}{\delta g_{ab}}}$ represents the
stress energy tensor of standard matter. These equations reduce to
the standard Einstein field equations when $f(R)=R$. It is crucial
for our purposes to be able to write \rf{eq:einstScTn} in the form
\begin{equation}
\label{eq:einstScTneff}
 G_{ab}=\tilde{T}_{ab}^{m}+T^{R}_{ab}=T^{tot}_{ab}\,,
 \end{equation}
where $\displaystyle{\tilde{T}_{ab}^{m}=\frac{ T_{ab}^{m}}{f'}}$ and
\begin{eqnarray}\label{eq:TenergymomentuEff}
T_{ab}^{R}=\frac{1}{f'}\left[\frac{1}{2}g_{ab} \left(f-R f'\right)
+\nab_b\nab_a f- g_{ab}\nab_c\nab^cf\right], \label{eq:semt}
\end{eqnarray}
represent two effective ``fluids": the  {\em curvature ``fluid"}\footnote{Some authors have recently adopted the name {\it Effective Dark Matter} for the curvature fluid \cite{Li:2008ai}.}
(associated with $T^{R}_{ab}$) and  the {\em effective matter
``fluid"} (associated with $\tilde{T}_{ab}^{m}$) \cite{SalvSolo, revnostra,SantePert}. This step is
important because it allows us to treat fourth order gravity as
standard Einstein gravity plus two ``effective"
fluids.  The details of the conservation properties of these effective fluids have been given in \cite{SantePert}. In particular, it has been shown that, no matter how the  effective fluids behave, standard matter still follows the usual conservation equations $T_{ab}^{m\ ;b}=0$.

The form \rf{eq:einstScTneff} of the field equations  allows us to use directly the covariant gauge invariant approach  \cite{EllisCovariant,EB,EBH,DBE,BED,BDE,DBBE}  in the same way presented  in \cite{SantePert}.
As usual the first step is to choose suitable frame, i.e., a 4-velocity $u_{a}$ of an
observer in spacetime. Following \cite{SantePert}, we will choose this frame to be the one comoving with  standard matter, which is also called matter energy frame and will be indicated in the following by $u^{m}_a$. We will also
assume that in $u^{m}_a$ standard matter is a barotropic perfect fluid with equation of state $p=w \rho$. Since the real observers
are attached to galaxies and the galaxies follow the standard matter geodesics, this frame choice appears to be best motivated  from a physical point of view.

Once the frame has been chosen the derivation of the kinematical
quantities can be obtained in a standard way \cite{EllisCovariant}. In particular the derivative along the matter fluid flow lines is defined by
$\dot{X}=u_a\nabla^aX$ and the projected covariant derivative operator orthogonal to $u^a$ is given by $\3nab_a=h^b{}_a\nabla_b$. With these definitions we can define the key kinematic quantities of the cosmological model: the
expansion $\Theta$, the shear $\sigma_{ab}$, the vorticity
$\omega_{ab}$ and the acceleration  $a_a = \dot{u}_a$. The general propagation equations for these kinematic variables in any spacetime correspond to the so called {\em 1+3 covariant equations} \cite{EllisCovariant} which are given in Appendix \ref{CovID}.

The definition of a frame $u_a$ also allows us  to obtain an irreducible
decomposition of the stress energy momentum tensor. In a general
frame and for a general tensor $T_{{a}{{b}}}$ one obtains:
\begin{equation}\label{Tdecomp}
T_{{a}{{b}}}=\mu u_a
u_{{b}}+ph_{{a}{{b}}}+2q_{(a}u_{{{b}})}+\pi_{{{a}}{{b}}}\,,
\end{equation}
where $\mu$ and $p$ are the energy density and isotropic pressure,
$q_{{{a}}}$ is the energy flux ($q_{{{a}}}=q_{\langle{{a}}\rangle}$)  and $\pi_{{{a}}{{b}}}$
is the anisotropic pressure ($\pi_{{{a}}{{b}}}=\pi_{\langle{{a}}{{b}}\rangle}$).

In this way, relative to $u_a^m$, $T^{tot}_{ab}$ can be decomposed as
\begin{eqnarray}\label{mutot}
\mu^{\rm tot}\,&=&T^{\rm tot}_{ab}u^{a}u^{b}\,=\,\tilde{\mu}^{\,
m}+\mu^{\,R}\,,\qquad
p^{\rm tot}\,=\frac{1}{3}T^{\rm tot}_{ab}h^{ab}\,=\,\tilde{p}^{\, m}+p^{\,R}\;,\\
q^{\rm tot}_{a}\,&=&-T^{\rm
tot}_{bc}h_{a}^{b}u^{c}\,=\,\tilde{q}^{\,
m}_{a}+q^{\,R}_{a}\,,\qquad \pi^{\rm tot}_{ab}\,=\,T^{\rm
tot}_{cd}h_{<a}^{c}h_{b>}^{d}\,=\,\tilde{\pi}^{\,
m}_{ab}+\pi^{\,R}_{ab}\,,
\end{eqnarray}
with
\begin{eqnarray}
\tilde{\mu}^{\,m}\,&=&\,\frac{\mu^{\,m}}{f'}\,,\qquad
\tilde{p}^{\,m}\,=\,\frac{p^{\,m}}{f'}\,,\qquad
\tilde{q}^{\,m}_{a}\,=\,\frac{q^{\,m}_{a}}{f'}\,,\qquad
\tilde{\pi}^{\,m}_{ab}=\,\frac{\pi^{\,m}_{ab}}{f'}\, .
\end{eqnarray}
Since we assume that  standard matter is a perfect fluid in $u_a^m$,
$q^{\,m}_{a}$ and $\pi^{\,m}_{ab}$ are zero, so that the last two
quantities above  also vanish.

The effective thermodynamical quantities for the curvature ``fluid"
are
\begin{eqnarray}
&&\mu^{R}\,=\,\frac{1}{f'}\left[\frac{1}{2}(R f'-f)-\Theta
f''\dot{R}+f''\tilde{\nabla}^2{R}+f''\,a_b\D^b{R}\right]\;,\\
&&p^{R}\,=\,\frac{1}{f'}\left[\frac{1}{2}(f-R
f')+f''\ddot{R}+f'''\dot{R}^2+\frac{2}{3}\Theta
f''\dot{R}-\frac{2}{3}f''\tilde{\nabla}^2{R} +\right. \nonumber\\
&& \qquad\left.  -\frac{2}{3}f'''\D^{a}{R}\D_{a}{R}-\frac{1}{3}f''
\,a_b\D^b{R}\right]\;,\\
&&q^{R}_a\,=\,-\frac{1}{f'}\left[f'''\dot{R}\D_{a}R+f''\D_{a}\dot{R}-\frac{1}{3}\Theta f''
\D_{a}R\right]\;,\\
&&\pi^{R}_{ab}\,=\,\frac{1}{f'}\left[f''\D_{\lgl
a}\D_{b\rgl}R+f'''\D_{\lgl a}{R}\D_{b\rgl}{R}-\sigma_{a
b}\dot{R}\right]\,.\label{piR}
\end{eqnarray}
The twice contracted Bianchi Identities lead to evolution equations for
$\mu^{\,m}$, $\mu^{R}$, $q^{R}_a$ and are given in Appendix  \ref{CovID}.

\section{Linearized Scalar Perturbations Equations}
Using the quantities defined above, and the equations given in Appendix  \ref{CovID}, we are able to write both the evolution equations for the background and  ones for scalar perturbations. As in \cite{SantePert}  we will consider a background that is homogeneous and isotropic, i.e., a FLRW model. In this background the cosmological equations for a generic $f(R)$ read:
\begin{eqnarray}\label{f}
&&\Theta^2\,=\,3\tilde{\mu}^{m} + 3\mu^{R}-\frac{3\tilde{R}}{2}\;,\\
&&\dot{\Theta}+{\textstyle\frac{1}{3}}\Theta^2
+{\textstyle\frac{1}{2}}(\tilde{\mu}^{m} + 3\tilde{p}^{m})
        +{\textstyle\frac{1}{2}}({\mu}^{R} + 3{p}^{R})=0\;,\\
&& \dot{\mu}^m\,+ \,\Theta\,(\mu^m+{p^m})=0\;,
\end{eqnarray}
where $\tilde{R}=6K/S^2$ is the 3-Ricci scalar, $K=0,\pm1$ and $S$ is the scale factor.
The structure of these equations shows clearly that  the effect of the introduction of higher order gravity on the background has a twofold nature. On one side, higher order gravity behaves like a  an additional fluid in the model. On the other, it influences the way in which standard matter interacts gravitationally.

Following \cite{SantePert}, we characterize scalar perturbations using the variables
\begin{equation}\label{ScaVar}
\Delta^{m}_{{a}}=\frac{S^2}{\mu^{m}}\3nab^2\mu^{m}\,,\qquad
Z=S^2\3nab^2\Theta\,,\qquad C=S^{3}\3nab^2\tilde{R}\,,\qquad{\cal R}=S^2\3nab^2 R\,,\qquad\Re=S^2\D_{{a}} \dot{R}\;,
\end{equation}
the first three variables, which are borrowed from GR, represent the scalar fluctuations in the matter energy density, in the expansion rate (which is associated with $\dot{\mu}$) and in the spatial curvature. The last two represent the fluctuation of the Ricci scalar and its momentum.

It is a relatively easy task to derive the propagation equations  for these variables in a FLRW background. Their form is  Appendix \ref{App2}.  If we focus on the evolution of scalar part of these variables, which is associated with the spherically symmetric collapse terms, these equations become
\begin{eqnarray}
&&\dot{\Delta}_m =w\Theta \Delta_m-(1+w)Z\,,\label{eqDelta}\\
&&\nn\dot{Z} = \left(\frac{\dot{R}
   f''}{f'}-\frac{2 \Theta }{3}\right)Z+
   \left[\frac{ (w -1) (3 w +2)}{2 (w +1)} \frac{\mu}{ f'} + \frac{2 w \Theta ^2
   +3 w (\mu^{R}+3  p^{R}) }{6 (w +1) }\right]
   \Delta_m+\frac{\Theta f''}{f'}\Re+\\&&+
   \left[\frac{1}{2}-\frac{1}{2} \frac{f}{f'}\frac{ f''}{f'}- \frac{f''}{f'} \frac{\mu}{
   f'} + \dot{R} \Theta  \left(\frac{f''}{f'}\right)^{2}+ \dot{R} \Theta \frac{ f^{(3)}}{ f'}\right]\mathcal{R}
   -\frac{w}{w +1} \3nab^{2}{\Delta}_m-\frac{ f''}{f'}\3nab^{2}\mathcal{R}\,,\\
&&\dot{{\cal R}}=\Re-\frac{w }{w +1}\dot{R}\;{\Delta}_m\,,\\
&&\nn\dot{\Re}=- \left(\Theta + 2\dot{R} \frac{
   f^{(3)}}{f''}\right)\Re- \dot{R} Z -
   \left[\frac{ (3 w -1)}{3} \frac{\mu}{f''} + \frac{w}{3(w +1)} \ddot{R} \right]{\Delta}_m+\\&&-\left[\frac{1}{3}\frac{f'}{f''}+\frac{f^{(4)}}{f'} \dot{R}^2+\Theta \dot{R} \frac{f^{(3)}}{f''}+\ddot{R} \frac{f^{(3)}}{f''}-\frac{R}{3}\right]\mathcal{R}+\3nab^{2}\mathcal{R}\,,\label{eqRho}
\end{eqnarray}
\begin{eqnarray}
&& \dot{C}=\nonumber k^2 \left[\frac{18  f'' \mathcal{R}}{S^2
\Theta f'}-\frac{18\Delta_{m}}{S^2  \Theta } \right] +K\left[\frac{3}{S^2\Theta}C
+\Delta_{m}\left(\frac{2 (w-1) \Theta }{w +1}+\frac{ 6\mu^{R}}{\Theta}\right)-\frac{6  f''}{\Theta f''}\3nab^{2}\mathcal{R}
+\frac{6 f''}{f'}\Re+\right.\\&&
\nonumber\left.+\frac{ 6 \dot{R} \Theta  f'
   f^{(3)}- f'' \left(3 f-2 \left(\Theta ^2-3 \mu^{R}\right) f'+6 \dot{R} \Theta f''\right)}{\Theta  (f')^2 }\mathcal{R}\right] +\3nab^{2}\left[\frac{4 w
    S^2 \Theta }{3 (w +1)}\Delta_{m}+\frac{2  S^2
   f''}{f'}\Re-\frac{2 S^2 \left(\Theta
    f''-3 \dot{R} f^{(3)}\right)}{3
    f'}\mathcal{R}\right]\,,\label{eqC}
\end{eqnarray}
together with the constraint
\begin{equation}\label{Gauss1}
  \frac{C}{S^2}+ \left(\frac{4  }{3}\Theta +\frac{2 \dot{R}
   f''}{f'}\right) Z-2\frac{
    \mu }{f'}{\Delta_{m}}_m+ \left[2 \dot{R}
   \Theta  \frac{f^{(3)}}{ f'}-\frac{f''}{ (f')^2} \left(f-2 \mu +2
   \dot{R} \Theta  f''\right)\right]\mathcal{R}+\frac{2 \Theta
   f''}{f'}\Re-\frac{2 f''}{f'}\3nab^{2}\mathcal{R}=0\,.
\end{equation}
Note that this system is made up of four first order differential equation, which means that the evolution of every single perturbation variable is determined by a fourth order differential equation. This has a profound influence in the dynamics of the perturbations and makes them potentially very different from what one obtains in standard GR.

Traditionally the analysis of the perturbation equations is simplified by using a harmonic decomposition.  In the 1+3 formalism this  can be done by developing the scalar quantities defined above using the eigenfunctions of the Laplace-Beltrami operator  \cite{BDE}:
\begin{eqnarray}\label{eq:harmonic}
  \3nab^{2}Q = -\frac{k^{2}}{S^{2}}Q\;,
\end{eqnarray}
where $k=2\pi S/\lambda$ is the wavenumber and $\dot{Q}=0$. Developing \rf{ScaVar} in terms of $Q$, (\ref{eqDelta}-\ref{Gauss1}) reduce to
\begin{eqnarray}
\dot{\Delta}_{m}^{(k)} &=&w\Theta \Delta_{m}^{(k)}-(1+w)Z^{(k)}\,,\label{eqDeltaHarm}\\
\dot{Z}^{(k)} &=& \left(\frac{\dot{R}
   f''}{f'}-\frac{2 \Theta }{3}\right)Z^{(k)}+
    \left[\frac{ (w -1) (3 w +2)}{2 (w +1)} \frac{\mu}{ f'} + \frac{2 w \Theta ^2
   +3 w (\mu^{R}+3  p^{R}) }{6 (w +1) }\right]   \Delta_{m}^{(k)}+\frac{\Theta f''}{f'}\Re^{(k)}+\nonumber \\&&+
   \left[\frac{1}{2}-\frac{ f''}{f'} \frac{k^2}{S^2}-\frac{1}{2} \frac{f}{f'}\frac{ f''}{f'}- \frac{f''}{f'} \frac{\mu}{
   f'} + \dot{R} \Theta  \left(\frac{f''}{f'}\right)^{2}+ \dot{R} \Theta \frac{ f^{(3)}}{ f'}\right]\mathcal{R}^{(k)}\,,\\
\dot{{\cal R}}^{(k)}&=&\Re^{(k)}-\frac{w }{w +1}\dot{R}\;{\Delta}_{m}^{(k)}\,,\label{eqZHarm}\\
\dot{\Re}^{(k)}&=&- \left(\Theta + 2\dot{R} \frac{f^{(3)}}{f''}\right)\Re^{(k)}- \dot{R} Z^{(k)} -
  \left[\frac{ (3 w -1)}{3} \frac{\mu}{f''} + \frac{w}{3(w +1)} \ddot{R} \right]{\Delta}_{m}^{(k)}+\nonumber\\ &&+\left[\frac{k^{2}}{S^2}-\left(\frac{1}{3}\frac{f'}{f''}+\frac{f^{(4)}}{f'} \dot{R}^2+\Theta \dot{R} \frac{f^{(3)}}{f''}+\ddot{R} \frac{f^{(3)}}{f''}-\frac{R}{3}\right)\right]\mathcal{R}^{(k)}\,,\label{eqRho2}\\
\dot{C}^{(k)}&=&\nn k^2 \left[\frac{18  f'' \mathcal{R}}{S^2
\Theta f'}-\frac{18\Delta_{m}}{S^2  \Theta } -6\frac{f''}{\Theta f'}\Re^{(k)}\right]  +K\left[\frac{3}{S^2\Theta}C
+\Delta\left(\frac{2 (w-1) \Theta }{w +1}-\frac{ 6\mu^{R}}{\Theta}\right)-\frac{6  f''}{\Theta f''}\3nab^{2}\mathcal{R}
+\frac{6 f''}{f'}\Re +\right.\\&&
\nonumber\left.+\frac{ 6 \dot{R} \Theta  f'
   f^{(3)}-6 k^2 f'' f'+ f'' \left(3 f-2 \left(\Theta ^2-3 \mu^{R}\right) f'+6 \dot{R} \Theta f''\right)}{\Theta  (f')^2 }\mathcal{R}\right]+\nn\\&&+\frac{k}{S^{2}}\left[\frac{4 w S^2 \Theta }{3 (w +1)}\Delta_{m}^{(k)}+\frac{2  S^2
   f''}{f'}\Re^{(k)}-\frac{2 S^2 \left(\Theta f''-3 \dot{R} f^{(3)}\right)}{3 f'}\mathcal{R}^{(k)}\right]\,,\label{eqCHarm}
    \end{eqnarray}
\begin{eqnarray}
 0&=& \frac{C^{(k)}}{S^2}+ \left(\frac{4  }{3}\Theta +\frac{2
\dot{R}
   f''}{f'}\right) Z^{(k)}-2\frac{
    \mu }{f'}{\Delta}_m^{(k)}+ \left[2 \dot{R}
   \Theta  \frac{f^{(3)}}{ f'}-\frac{f''}{ (f')^2} \left(f-2 \mu +2
   \dot{R} \Theta  f''\right)+2 \frac{ f''}{f'} \frac{k^{2}}{S^2}\right]\mathcal{R}^{(k)}+\frac{2 \Theta
   f''}{f'}\Re^{(k)}\;,\nn\\\label{constrCZDel}
\end{eqnarray}
which is a system of ordinary differential equations. This system takes a more manageable form if we reduce it to a pair of second order equations:
\begin{eqnarray}
   &&\nn\ddot{\Delta}_{m}^{(k)}+\left[\left(
   \frac{2}{3}-w\right) \Theta -\frac{\dot{R}
   f''}{f'}\right] \dot{\Delta}_{m}^{(k)}-\left[w  \frac{k^2}{S^{2}}-w  (3
   p^{R}+\mu^{R})-\frac{2 w  \dot{R} \Theta
   f''}{f'}-\frac{\left(3 w ^2-1\right) \mu }{f'}\right]\Delta_{m}^{(k)}=\\&&=
   \frac{1}{2}(w +1)\left[2 \frac{k^2}{S^2}f''-1+
   \left(f-2 \mu +2 \dot{R} \Theta  f''\right)\frac{f''}{f'^2}
   -2  \dot{R} \Theta
   \frac{f^{(3)}}{f'}\right] \mathcal{R}^{(k)} -\frac{(w +1) \Theta
   f'' }{f'}\dot{\mathcal{R}}^{(k)}\label{EqPerIIOrd1}\,,\\&&
   \nn f''\ddot{\mathcal{R}}^{(k)}+\left(\Theta f'' +2 \dot{R}
   f^{(3)}\right)
   \dot{\mathcal{R}}^{(k)}-\left[\frac{k^2}{S^2}f''+ 2 \frac{K}{S^2}f''
  +\frac{2}{9} \Theta^2 f''- (w +1) \frac{\mu}{2 f'}f''- \frac{1}{6}(\mu^{R}+ 3
p^{R})f''+\right.\\&&\nn\left.-\frac{f'}{3}+ \frac{f}{6 f'}f'' +
\dot{R} \Theta  \frac{f''^{2}}{6 f'} -
   \ddot{R} f^{(3)}- \Theta f^{(3)} \dot{R}- f^{(4)}\dot{R}^2
   \right]\mathcal{R}^{(k)}=-
   \left[ \frac{1}{3}(3 w -1) \mu+ \right.\\&&\left.+\frac{w}{1+w} \left(f^{(3)}
   \dot{R}^2+  (p^{R}+\mu^{R}) f'+ \frac{7}{3}\dot{R} \Theta f''
 +\ddot{R} f'' \right)\right]\Delta_{m}^{(k)}-\frac{(w -1) \dot{R} f''}{w +1} \dot{\Delta}_{m}^{(k)}\,.\label{EqPerIIOrd2}
\end{eqnarray}
In the $f(R)=R$ case these equations reduce to the standard equations for the evolution of the scalar perturbations in GR:
\begin{eqnarray}\label{eqIIordGR}
 &&\nn\ddot{\Delta}_{m}^{(k)}-\left(w-{\textstyle\frac{2}{3}}\right) \Theta\dot{\Delta}_{m}^{(k)}
 -  \left[w\frac{k^2}{S^{2}}-\left({\textstyle\frac{1}{2}}+w-{\textstyle\frac{3}{2}}w^2\right)\mu\right]\Delta_{m}^{(k)}=0\;,\\
&&\mathcal{R}^{(k)}=\left(3w-1\right)\mu\Delta_{m}^{(k)}\;.
\end{eqnarray}
If one compares the system (\ref{EqPerIIOrd1}-\ref{EqPerIIOrd2}) with the equations for the evolution of scalar perturbations for two interacting fluids in GR one notices that they have the same structure, i.e., one finds friction terms and source terms due to the interaction and  the gravitation of the two effective fluids. It is then natural to ask ourselves if this analogy can be useful to better understand the physics of these models. The answer is affirmative, but with some very important caveats. First of all a more correct way to draw this analogy would be to write the system of equations for  $\Delta_m$ and $\Delta_R=\frac{S^{2}\3nab^2\mu_R}{\mu_R}$ and analyze their structure rather than using the ones above. Also, as stated in  \cite{SantePert}, one has to be careful in remembering that we are dealing with {\it effective fluids} and, as such, they might violate some basic constraints that standard fluids  usually follow (such as the energy conditions) or present subtleties in the definition of their comoving frame.

However, in spite of  these differences one can still use the coefficients of the $(\Delta_m, \Delta_R)$  equations to  obtain information about the interaction between standard matter and the curvature fluid.  Unfortunately the length of this system makes it impossible for us to present it here (instead we will give their structure in the specific examples of Section VII). However, the  coefficients of these equations are found to behave as a ratio of polynomials in the wavenumber and have a non trivial behavior in $t$.  This kind of behavior is very different to what is found in a GR-two fluid system. In a photon-baryon system, for example, the dissipation terms grow as $k^2$ and  behave like $1/t$ in time. This implies that, unlike Thompson scattering of the baryon-photon system, the effect of the interaction between matter and non-linear gravitation can influence large and small scales alike, depending on the structure of the action. Therefore, from the distribution of the structures in the observed sky one can deduce constraints on the nature of the theory of gravity. In fact, we will find that there is a specific spectral signature of fourth order gravity which is associated with these features. Another important difference with GR is that  (\ref{EqPerIIOrd1}-\ref{EqPerIIOrd2})  are  scale dependent for any value of the barotropic factor. This means that whatever the equation of state of standard matter, the perturbation solutions will always depend on the scale for which they are calculated, even in the special case of dust which in GR is associated to a scale invariant spectrum.

\section{Perturbations and cosmological parameters}\label{Eq&Par}

An interesting way of understanding the properties of the perturbations equations given in the previous section is to write them in terms of the cosmological parameters:
\begin{equation}\label{VisserPar}
q=-9 \;\Theta^{-2}\;\frac{\ddot{S}}{S}\,,\qquad\Omega_K=-9\frac{K}{S^2} \;\Theta^{-2}\,,\qquad j=27\;\frac{S^{(3)}}{S} \;\Theta^{-3}\,,\qquad s=81\;\frac{S^{(4)}}{S}\; \Theta^{-4}\,,
\end{equation}
where $q$ is  the deceleration parameter, $\Omega_K$ the spatial curvature density parameter  and $j$ and $s$ are the higher order kinematical parameters jerk (or jolt) $j$ and  snap $s$  \cite{VisserPar}. These quantities were devised in order to characterize  the kinematics of a cosmological model in a way that is independent on  any assumption on the dynamics and as we will see, they will be very useful for our purposes.

Using $q, j, s$, and $\Omega_K$  defined in \rf{VisserPar} the Ricci scalar and its derivative can be rewritten as
\begin{eqnarray}
&&R=R(q,\Omega_K,\Theta^2)= \left(\frac{2 q}{3}-6 \Omega _K+\frac{2}{3}\right) \Theta^2,\\
&&\dot{R}= \left(\frac{2 j}{9}+\frac{2 q}{9}+4 \Omega _K-\frac{4}{9}\right) \Theta^3,\\
&&\ddot{R}= \left(\frac{2 q^2}{27}-\frac{16 q}{27}+\frac{2 s}{27}+\frac{4 \Omega _K q}{3}-4 \Omega _K+\frac{4}{9}\right) \Theta^4\,.
\end{eqnarray}
In terms of these quantities the coefficients of the (\ref{EqPerIIOrd1} - \ref{EqPerIIOrd2}) above can be written in the form $\Theta^{p_1}F(q,j,s,\Omega_{K},k_{i},\Theta^2)$ where $p_1$ is a suitable integer associated with the dimension of the coefficient and $k_{i}=\alpha_i^{-1}\Theta^{-p_2}$ are the wavenumber of the physical scales of the theory associated with the dimensional constants $\alpha_i$ of dimension $p_2$  in the action. In this way (\ref{EqPerIIOrd1} - \ref{EqPerIIOrd2}) become
\begin{eqnarray}
   &&\ddot{\Delta}_{m}^{(k)}+\mathcal{A}\; \Theta\;\dot{\Delta}_{m}^{(k)}+\mathcal{B}\;\Theta^{2}\;\Delta_{m}^{(k)}=
  \mathcal{C} \;\Theta^{2}\;\mathcal{R}^{(k)} +\mathcal{D}\;\Theta\;\dot{\mathcal{R}}^{(k)}\;, \label{EqPerIIOrdParGen1}\\&&
  \ddot{\mathcal{R}}^{(k)}+\mathcal{E}\,\Theta\,
   \dot{\mathcal{R}}^{(k)}+\mathcal{F}\;\Theta^{2}\;\mathcal{R}^{(k)}=-
   \mathcal{G} \;\Theta ^4 \Delta_{m}^{(k)}-\mathcal{H}\,\Theta^{3}\, \dot{\Delta}_{m}^{(k)} \label{EqPerIIOrdParGen2}\,,
\end{eqnarray}
with
\begin{eqnarray}
&&\mathcal{A}=\left(
   \frac{2}{3}-w\right)-\frac{2 \left(j-q+2 \Omega _K-2\right) f'' \Theta ^2}{9 f'}\,, \\
&&\mathcal{B}=\frac{k^2 w }{S^2\Theta^2} +\frac{2 (w -1) \left(j-q+2 \Omega _K-2\right) f'' \Theta ^2}{9 f'}+\frac{1}{3}
   q(1- w )+\frac{f (w +1)}{2  \Theta ^2 f'}\,,\\
&&\nn\mathcal{C}=-\frac{(w +1)f''}{f'}\frac{ k^2}{S^2 \Theta ^2 }+\frac{(w +1) \left(2 q^2+8 q+s-3 (q+2)w +j (-q+3 w +2)+(-4 q+6 w -2) \Omega _K+2\right)  f''}{3 \left(j-q+2 \Omega _K-2\right) f'}\\&&+\frac{(w +1) (q (3 w +2)-j)}{2\Theta ^2 \left(j-q+2  \Omega _K-2\right)}+\frac{9 f (w +1)^2}{4 \left(j-q+2 \Omega _K-2\right) \Theta ^4 f'}\,,\\
&&\mathcal{D}=-\frac{(w +1) \Theta  f''}{f'}\,,
\end{eqnarray}
\begin{eqnarray}
&&\mathcal{E}=\frac{1}{3}
   \left(2 q-6 w -\frac{2 (3 q+j (q+3)+s)}{j-q+2 \Omega _K-2}+5\right) -\frac{9 (w +1) f }{2 \left(j-q+2 \Omega _K-2\right) \Theta ^4 f''}
   -\frac{\left(3 w  q+q+2 \Omega _K-2\right) f'}{\left(j-q+2 \Omega _K-2\right)\Theta^2  f''}\,,\\
&&\nn\mathcal{F}= \frac{k^2}{S^2  \Theta^2}-\frac{1}{9}
   \left(q^2-3 w  q-q-2-2 \Omega _K\right)+\frac{4 \left(-j+q-2 \Omega _K+2\right){}^2 f^{(4)} \Theta ^4}{81 f''}+\frac{(3 q+j (q+3)+s) (2 q-3 w +1)}{9(-j+q-2 \Omega_K+2)}\\&&\nn -\frac{(3 q+j (q+3)+s)^2}{9\left(-j+q-2 \Omega _K+2\right){}^2}-\frac{3 f (w +1) \left(3 j+q (q+5)+s-2 q \Omega _K\right)}{4 \left(-j+q-2 \Omega _K+2\right){}^2 f'' \Theta ^2}+\frac{\left[2 j^2-(q (9 w +7)+2) j+4 (q+2) \Omega _k^2\right] f'}{6  \left(-j+q-2 \Omega _K+2\right){}^2 \Theta^{2} f''}\\&&-\frac{\left[q
   \left(q^2+q+s+3 (q (q+5)+s) w -18\right)-2 (s+4)+2 (-j+s+q (11-3 q w )+8) \Omega
   _K\right] f'}{6 \Theta^{2}\left(-j+q-2 \Omega _K+2\right){}^2 f''}\,,\\
&&\mathcal{G}=\frac{2 (w  (3 w -4)+1) \left(j-q+2 \Omega _K-2\right)}{27 (w +1)}+\frac{\left(3 q w ^2+4 \Omega _K w -4 w +q\right) f' }{9 (w +1) \Theta ^2 f''}+\frac{(1+3 w)f}{6 \Theta ^2 f''}  \,,\\ &&
\mathcal{H}= -\frac{2 (w -1) \left(j-q+2 \Omega _K-2\right) }{9 (w+1)}\,,
\end{eqnarray}
where  $f$ and its derivatives with respect to $R$ are considered functions of $R\left(q,\Omega_K, \Theta^{2}\right)$ and $k_i$.

There are  some  general remarks that we can make at this point. First of all, the fact that  the perturbation equations can be shown to depend on the higher order cosmological parameters is a symptom of the fact that fourth order gravity is much more sensitive to the features of the background than GR (whose equations depend only on $\Theta$). Also, since these parameters can (at least in principle) be measured, we have a natural way to constraint both the dynamics of the background and the formation of structure. Note that this would not be possible in a standard scalar tensor theory of gravity, due to the fact that the background equations in that case remain second order. This suggest that, although in the background a scalar field might be able, in some cases, to emulate the behavior of fourth order gravity, this become more difficult at first order in perturbation theory \cite{conserved, kunz}.

Another important point concerns the meaning of the long and short wavelength limit. As it is clear from \rf{eqIIordGR}, in GR these  limits  can be defined  by comparing  the values of  $\frac{k^2}{S^2 \Theta^2}$ and the matter term in the $\Delta_{m}$ coefficient. However, looking at \rf{eqIIordGR} it is clear that  the situation here is more delicate. For example, the short wavelengths regime cannot be defined  as simply $\frac{k^2}{S^2 \Theta^2}\gg 1$, but  $\frac{k^2}{S^2 \Theta^2}$  has to be bigger than {\it all} the other quantities appearing in the coefficients $\mathcal{B}$, $\mathcal{C}$ and $\mathcal{F}$. The same reasoning holds for  the long wavelengths: $\frac{k^2}{S^2 \Theta^2}$  has to be smaller than {\it all} the other quantities appearing in the coefficients $\mathcal{B}$, $\mathcal{C}$ and $\mathcal{F}$ \footnote{However, there is a difference between the two limits because in the long wavelength limit one can always set $\left|\frac{k^2}{S^2 \Theta^2}\right|\approx0$, while in the short wavelength the definition is completely dependent on the values of the coefficients and, as consequences, on the features of the background.}.  This effectively suggest the presence of  a least three different regimes in the evolution of the perturbations. Firstly,  the ``deep super-horizon" regime in which $k$ is effectively zero, an intermediate one (or two depending on the value of the barotropic factor $w$) which is determined by the details of the background and a ``deep subchorionic" regime in which one has effectively $k\rightarrow\infty$.

The situation seems to become more complicated when the fourth order gravity action posses dimensional constants (like in the case $f(R )=R +\alpha R^{n}$ that we will consider later). Since these constants are associated with the scales at which the different contributions to the action become dominant, one would expect  the introduction of additional scales into the theory, i.e., further possible evolution regimes for scalar perturbations.  It turns out, however,  that these additional constants lead only to changes in the power spectrum at intermediate scales, preserving scale invariance on large and small scales. In section VII we will see, using some examples, why this happens, and how these features can be used as a signature of fourth order gravity.

\section{Properties of the scalar perturbations and comparison with General Relativity}

In order to understand the details of the evolution of scalar perturbation in a specific $f(R)$ model we need to
analyze the behavior of the solutions of the (\ref{EqPerIIOrdParGen1}-\ref{EqPerIIOrdParGen2}) of that model.
This can be done by examining their time dependence,
but also by defining some characteristic quantities which help extract physical content
from these solutions. One of these quantities is the power spectrum, $P(k)$ of  $\Delta_{m}$
and $\mathcal R$ appearing in \rf{EqPerIIOrdParGen2},  i.e., the variance of the
amplitudes of their Fourier transform at a given value of $k$. In the case of $\Delta_{m}$
(on which we will focus our analysis) $P(k)$ is defined by the relation \cite{Coles}:
\begin{equation}
\langle\Delta_{m}({\mathbf k_1})\Delta_{m}({\mathbf k_2})\rangle=P(k_1) \delta({\mathbf k_1}+{\mathbf k_2})\;,
\end{equation}
where ${\mathbf k_i}$ are two wavevectors characterizing two Fourier components of the
solutions of (\ref{EqPerIIOrdParGen1}-\ref{EqPerIIOrdParGen2}) and $P({\mathbf k_1})=P(k_1)$ because of isotropy
in the distribution of the perturbations. This quantity tells us how the fluctuations of
matter depend on the wavenumber at a specific time and carries information about the amplitude
of the perturbations (but not on their spatial structure). In GR the power spectrum on
large scales is constant, while on small scales it is suppressed in comparison with the large
scales \cite{PaddyPert}. However, in the case of pure dust the matter fluctuations are scale invariant,
so the power spectrum can be considered constant. This quantity is a powerful tool for comparing the
predictions of the system (\ref{EqPerIIOrd1}-\ref{EqPerIIOrd2}) with observations and is able
to reveal a great deal of information on the physics of scalar perturbations.
In addition, the analysis of the time variation of $P(k)$ provides information on the ways in
which the perturbations evolve in time on different scales. The fact that different scales
evolve differently is a key feature of Geometric Dark Energy and if observed would allow one
to differentiate between these models and ones based on standard Dark Energy.
For our purpose we will normalize $P(k)$ such that it is unity on super-horizon scales.
This can then be scaled with current observations of the power spectrum on large scales (see \cite{WMAP}
for the latest constraints).

Another interesting way of exploring the properties of the equations
presented above is to compare the features of their solutions with the corresponding results
given in GR. This can be done by defining some suitable quantities whose value is associated
with specific properties of the perturbation evolution. The first quantity is the ratio of
the density perturbations in the modified theory with that of the corresponding result in the
Einstein-de Sitter model, i.e., $g=\Delta_{m}/\Delta_{EdS}$ . Since the evolution of
scalar perturbations in the Einstein-de Sitter cosmology is $\Delta_{EdS}=c_{+}S+c_{-}S^{-3/2}$,
at late times it is  approximately given by $g\approx\Delta_{m}/S$. The importance of the quantity
$g$ is due to the fact that $g^{GR}=1$ at all times. This means that calculating the
quantity $X=g-g^{GR}=g-1$ is a natural way of determining the deviation of the behavior
of scalar perturbations in fourth order gravity from the corresponding result in GR.

The second quantity is given by $\displaystyle{Y=\frac{d \ln(g)}{d\ln(S)}}$. This quantity is
also measure of the deviation from GR, but has the additional advantage of measuring the behavior
of scalar perturbations as a function of $a$ at late times. This is particularly useful when one
is forced to performed numerical integration of the equations above when exact solutions are
not available.

In what follows, we will use the quantities presented above to analyze the properties of two simple
classes of fourth order theories of gravity.

\section{Examples}
\subsection{The case $f(R)=\chi R^{n}$ ($R^{n}$-gravity)}
\begin{figure}[htbp]
\includegraphics[scale=1.2]{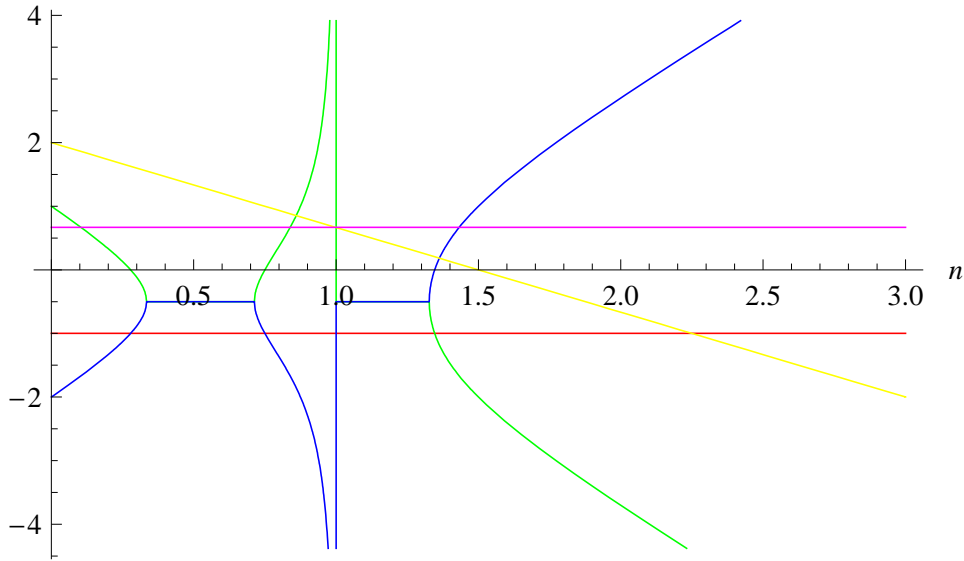}
\caption{Plot against $n$ of the real part of the long wavelength modes
for $R^n$-gravity in the dust case (blue, red green and yellow lines) together
with the GR modes (red and purple line). This graph is consistent with Figures
\ref{Fig2-3},  \ref{Fig4-5} and \ref{Fig6-7},  i.e.,  when one of the four perturbation
modes grows faster than GR $X$ is bigger than zero and when they are all smaller
than GR $X$ is negative.  Of course the cumulative effect of the combination of
more than one growing mode has to be taken in account in comparing the two graphs.
As one can see this produces a slight shift in the value of $n$ associated
with the change of sign in $X$ and the dominance of one of the $R^n$ growing mode
over the GR growing mode.}
 \label{Fig1}
\end{figure}

The case  $f(R)=\chi R^{n}$ also called sometimes $R^{n}$-gravity is characterized by the action
\begin{equation}\label{lagrRn}
L=\sqrt{-g}\left[\chi R^{n}+{\cal L}_{M}\right]\;,
\end{equation}
and constitutes the simplest possible example of fourth order gravity. Its homogeneous and isotropic cosmologies have been studied in detail using the dynamical system approach \cite{ellisbook,cdct:dynsys05,SanteGenDynSys} and the evolution of the large scale cosmological perturbations of a FLRW background has been investigated in  \cite{SantePert}  using the covariant gauge invariant approach. The results show some profound differences between this theory and GR. For example, in the transient Friedmann
background $a=t^{2n/3(1+w)}$, $\Delta_{m}$ was found to grow on long wavelength for almost all the
values of the parameter $n$, even when these values corresponded to backgrounds which undergo accelerated expansion (see Figure \ref{Fig1}
for a plot of the real part of the exponents of the modes of the $\Delta_{m}$ solution). In this section we will continue the investigation of the evolution of scalar perturbations for this model \footnote{Clearly we will consider only $n>0$ for this background. Negative values of $n$ would represent a contracting model.} focusing more specifically on the small scales. Substituting the form of $f(R)$  in the general equations  (\ref{EqPerIIOrd1}-\ref{EqPerIIOrd2}) we obtain  the system
(\ref{EqPerIIOrdRn1}-\ref{EqPerIIOrdRn2}) given in Appendix \ref{App3}.
Providing  the details of the background, the values of the parameter $n$, the barotropic factor $w$, the spatial curvature index $K$ and the wavenumber $k$ one is able to numerically integrate this system to obtain the behaviour of the matter fluctuations. We can then use the quantities described in the previous section to extract physical information about the evolution of density perturbations in this model.

\begin{figure}[htbp]
\subfigure{\includegraphics[scale=0.40]{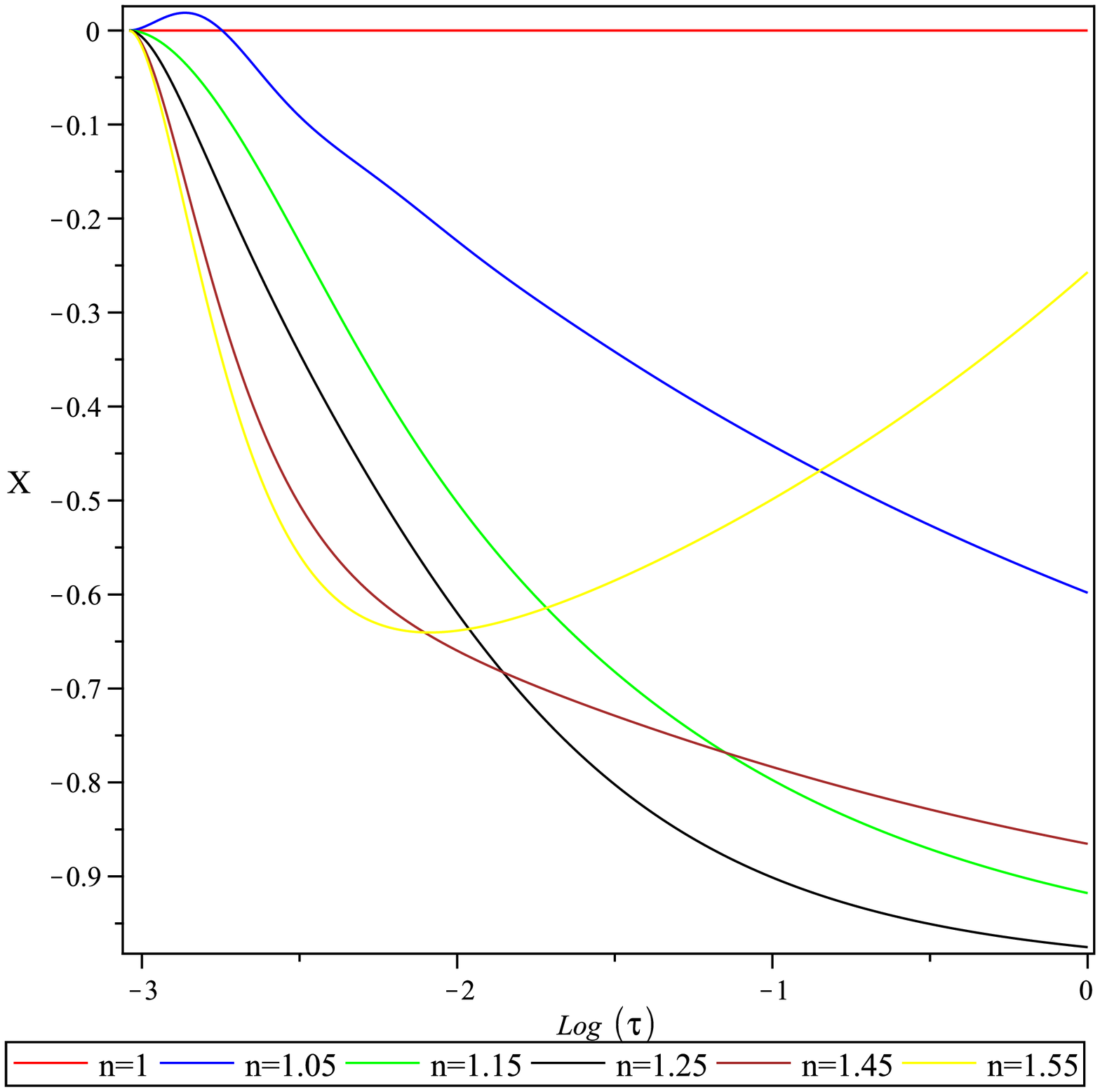}\label{Fig2-3a}}
\subfigure{\includegraphics[scale=0.41]{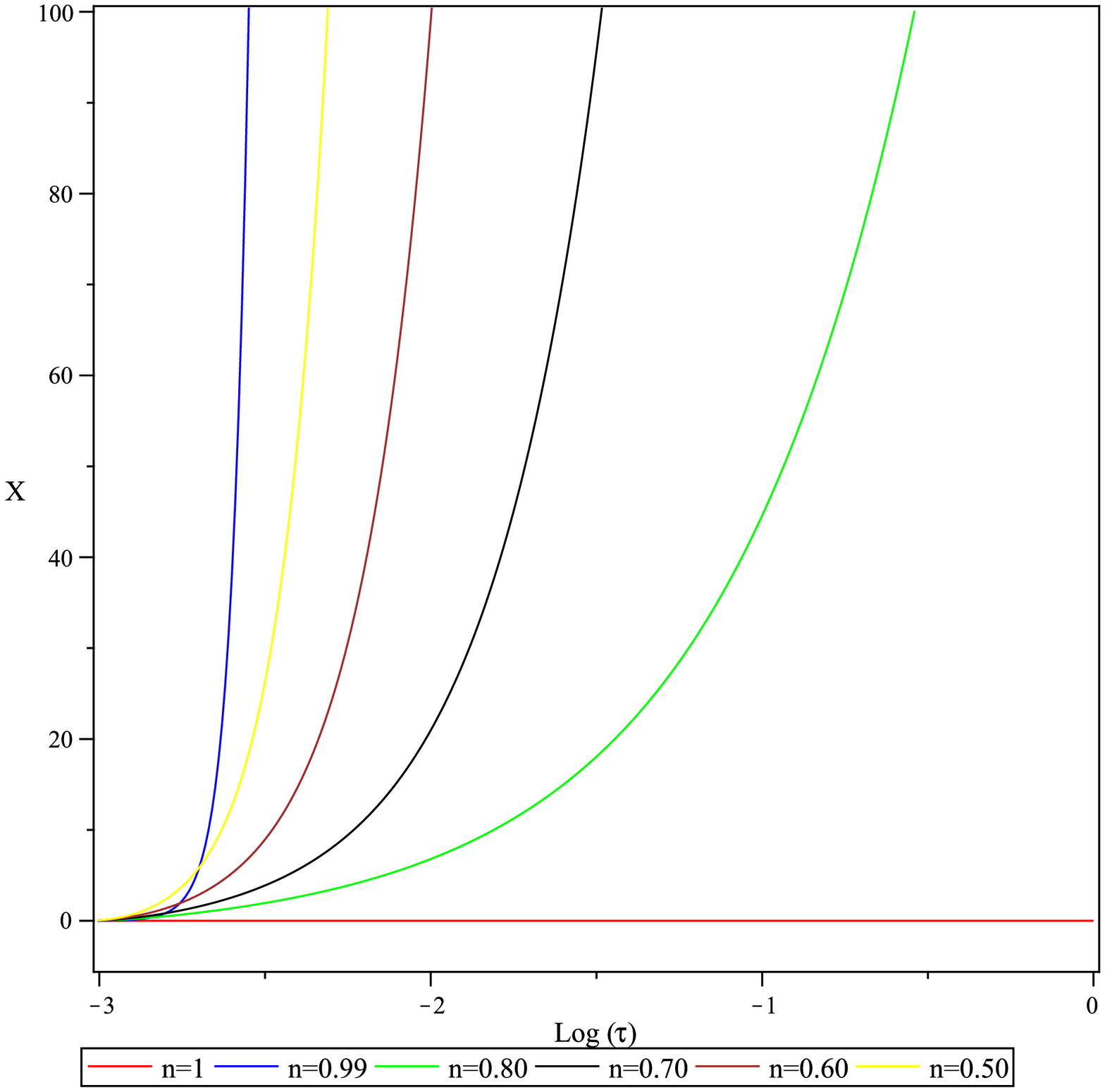}\label{Fig2-3b}}
\caption{Plot of $X$ as a function of $\tau=\log_{10}(S)$ for $R^n$-gravity in the long wavelength limit. The red line $X=0$ corresponds to GR.
Note the consistency between this picture and Figure \ref{Fig1}. \label{Fig2-3}}
\end{figure}

Figure \ref{Fig2-3} shows the behavior of the quantity $X$ as a function of the time
parameter $\tau=\log_{10}(S)$ on large scales ($k\approx 0$) for dust and different values of $n$.
If we start with values of $n$ close to $1^+$ the perturbations seems to first evolve faster
than the growing mode of the GR-Einstein-de Sitter case and then at a slower rate which
continues to decrease. For higher values of $n$, this first phase is absent and the rate
is always slower than in the GR-Einstein-de Sitter case. For $n \ge 3/2$ the rate of
growth of the perturbation is initially below the GR-Einstein-de Sitter mode, however, at
late times the mode starts to grow at a much faster rate. For $0<n<1$ the situation is
radically different: the rate of growth is always faster than GR.   As $n$ goes from zero
to $1^-$ the growth rate seems to decrease, but at $n\approx0.85$ this behavior
changes and by $n=0.90$ the perturbations grow much faster than for any other value of $n$. This indicates the possible existence of an instability of the theory in this region of parameter space.

\begin{figure}[htbp]
\subfigure[Plot of $\log_{10}(X)$  vs $n$ at $\tau=\log_{10}(S)=1$ ]{\includegraphics[scale=0.40]{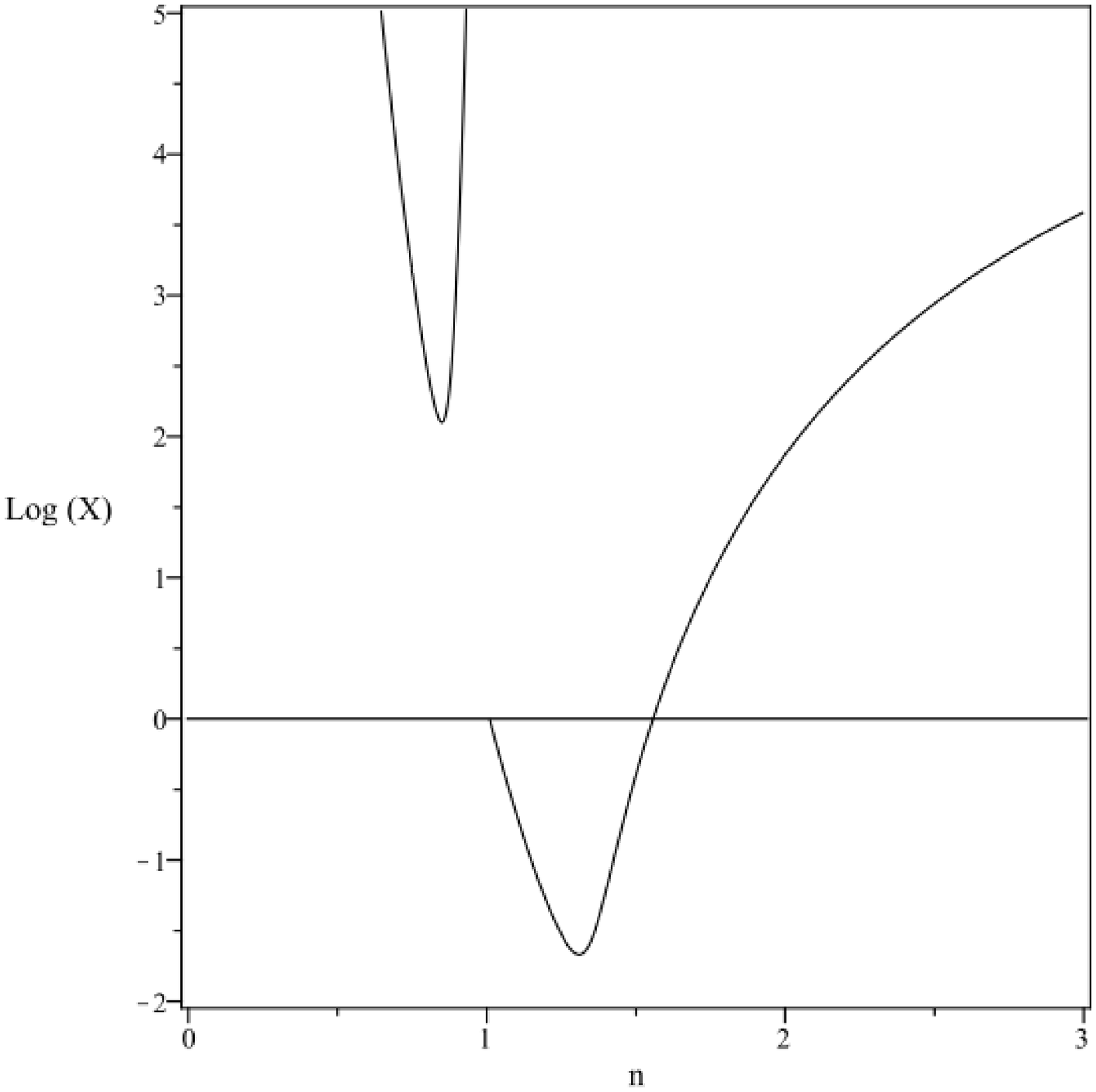}\label{Fig4-5a}}
\subfigure[Plot of $Y$  vs $n$ at $\tau=\log_{10}(S)=1$ ]{\includegraphics[scale=0.40]{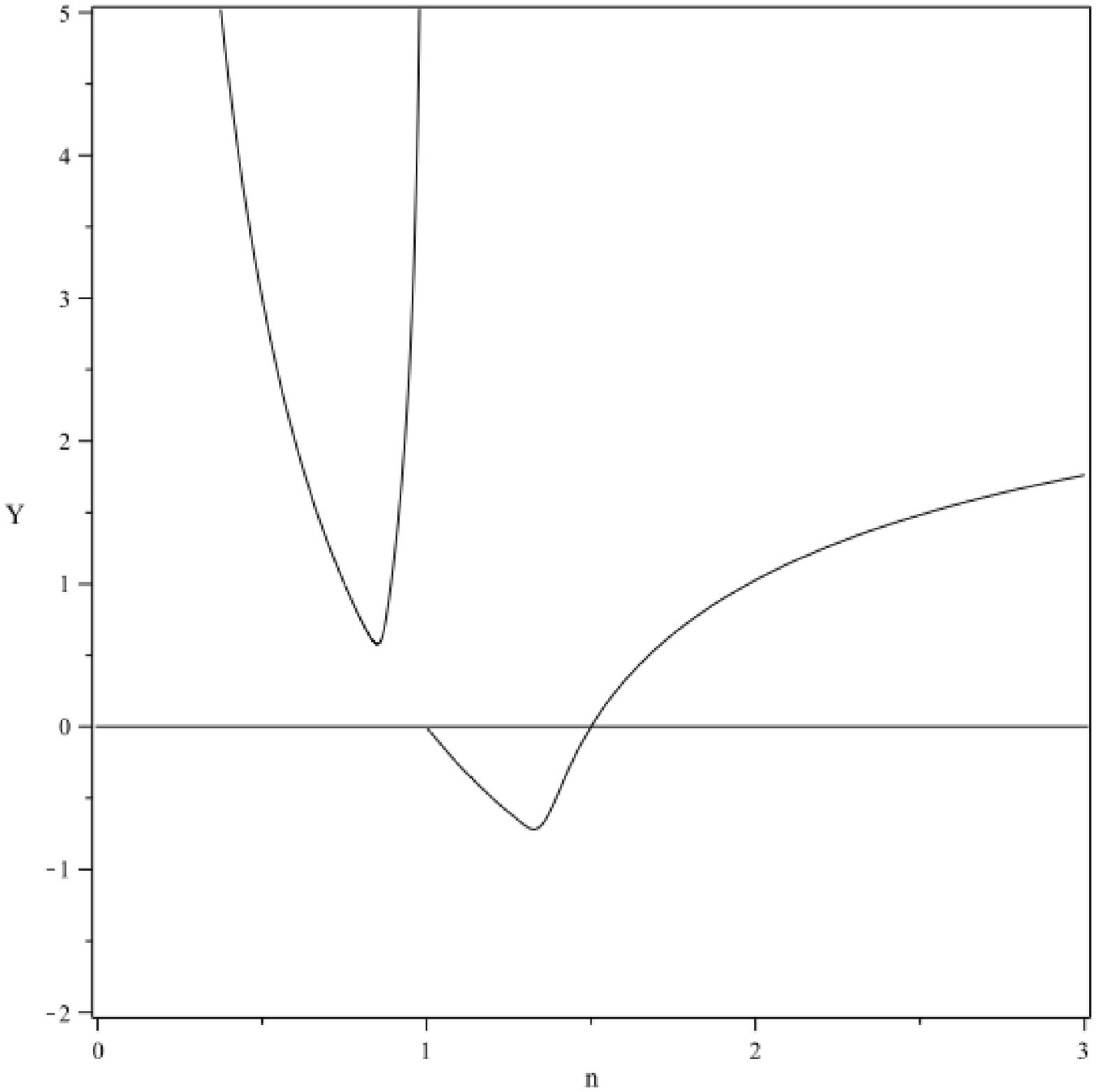}\label{Fig4-5b}}
\caption{Plots of $X$ and $Y$ as a function of $n$ evaluated at $\tau=\log_{10}(S)=1$
for $R^n$-gravity in the long wavelength limit. The horizontal lines $X,Y=0$ represent the value
of these quantities in GR. \label{Fig4-5}}
\end{figure}

These features are well summarized by  Figure \ref{Fig4-5a} which represents
$X$ at the present time  $\tau=1$ as function of $n$. When compared with Figure
\ref{Fig1} one can see that the two figures are consistent.

\begin{figure}[htbp]
\subfigure[Plot of $Y(\tau)$ for $n=1.4$  ]{\includegraphics[scale=0.40]{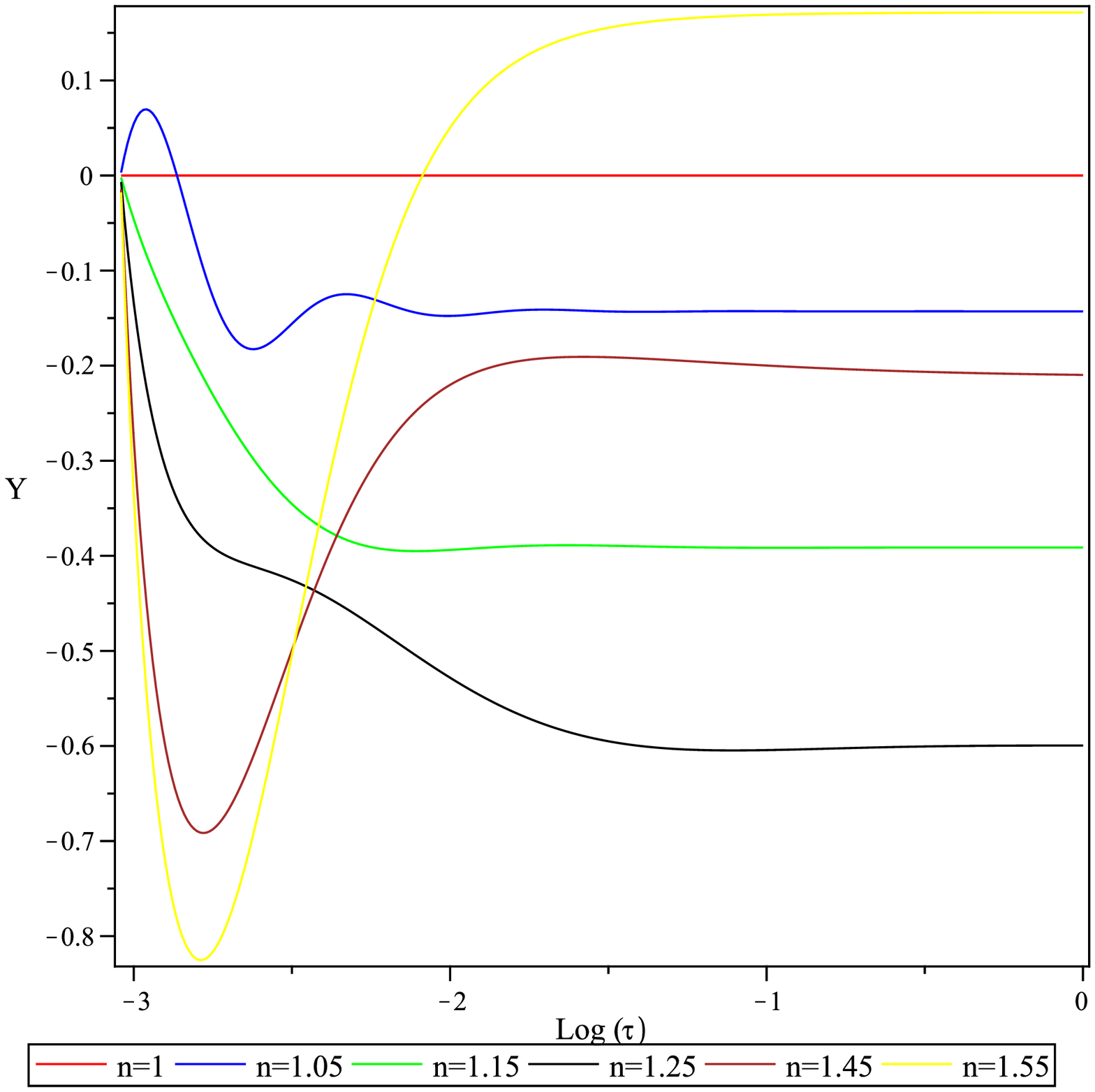}}
\subfigure[Plot of $Y(\tau)$ for $n=1.55$  ]{\includegraphics[scale=0.40]{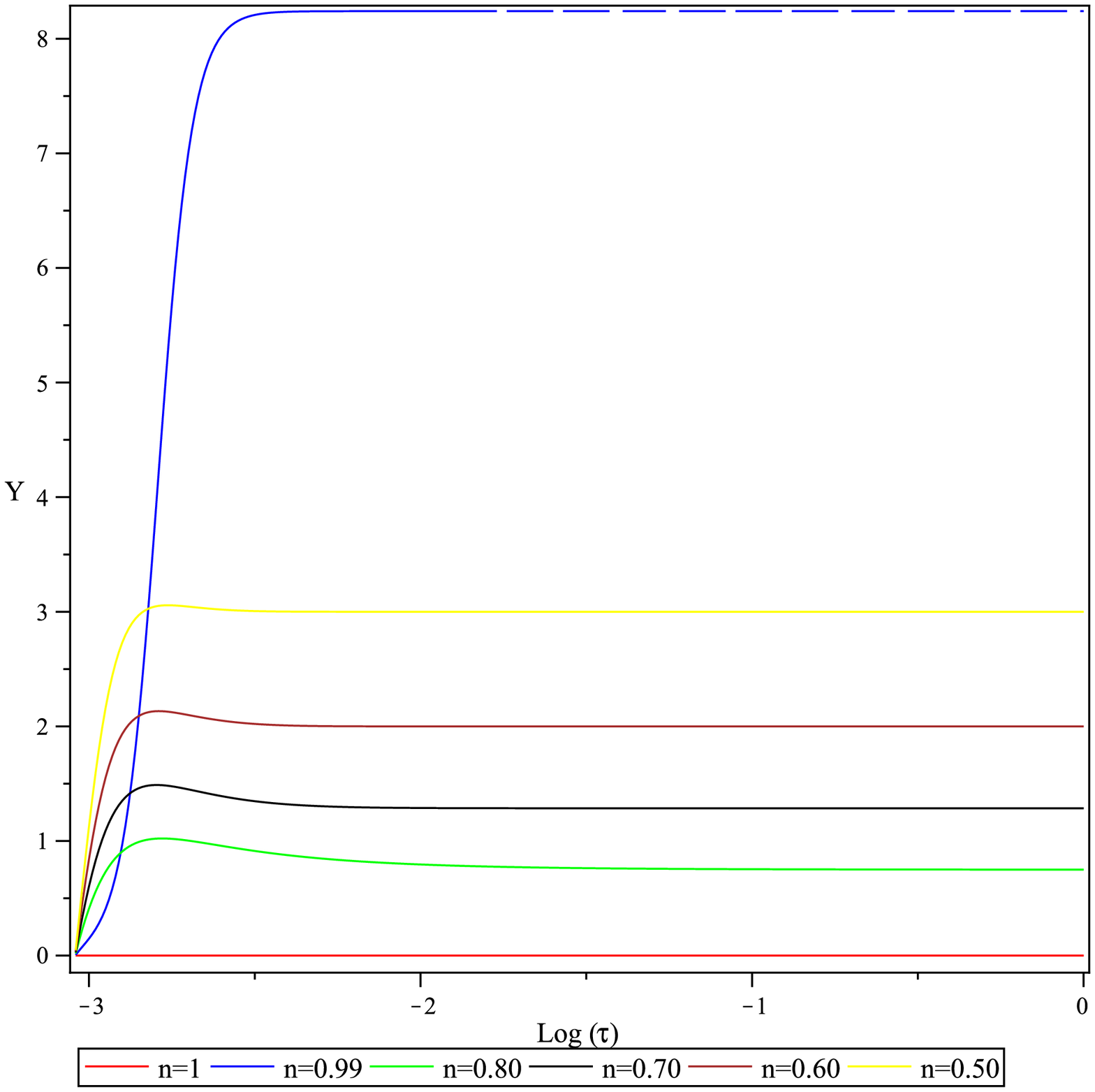}}
\caption{Plot of $Y(\tau)$ as a function of $\tau=\log_{10}(S)$ for $R^n$-gravity with $n>1$ (left)
and $0<n<1$ (right).  The red line $Y=0$ corresponds to GR. Note the consistency between this picture and Figure \ref{Fig1}. \label{Fig6-7}}
\end{figure}

In Figure \ref{Fig6-7} we see the behavior of $Y$ as a function of $\tau=\log_{10}(S)$  for large
scale perturbation. It is clear that
the growth rate of the perturbations approaches asymptotically a constant value, which
corresponds to the dominance of the fastest growing mode of the solutions of the system
(\ref{EqPerIIOrdRn1}-\ref{EqPerIIOrdRn2}). Again, we found consistency between
Figure \ref{Fig1} and both this plot and Figure \ref{Fig4-5b}.

This confirms that $X$ and $Y$ provide a very useful source of information
about the time evolution of scalar perturbations even when we do not have an
analytic solution as it often happens in fourth order gravity.

\begin{figure}[htbp]
\subfigure[ Plot of $X(\tau)$ for $n=1.4$ and different values of $k$ ]{\includegraphics[scale=0.37]{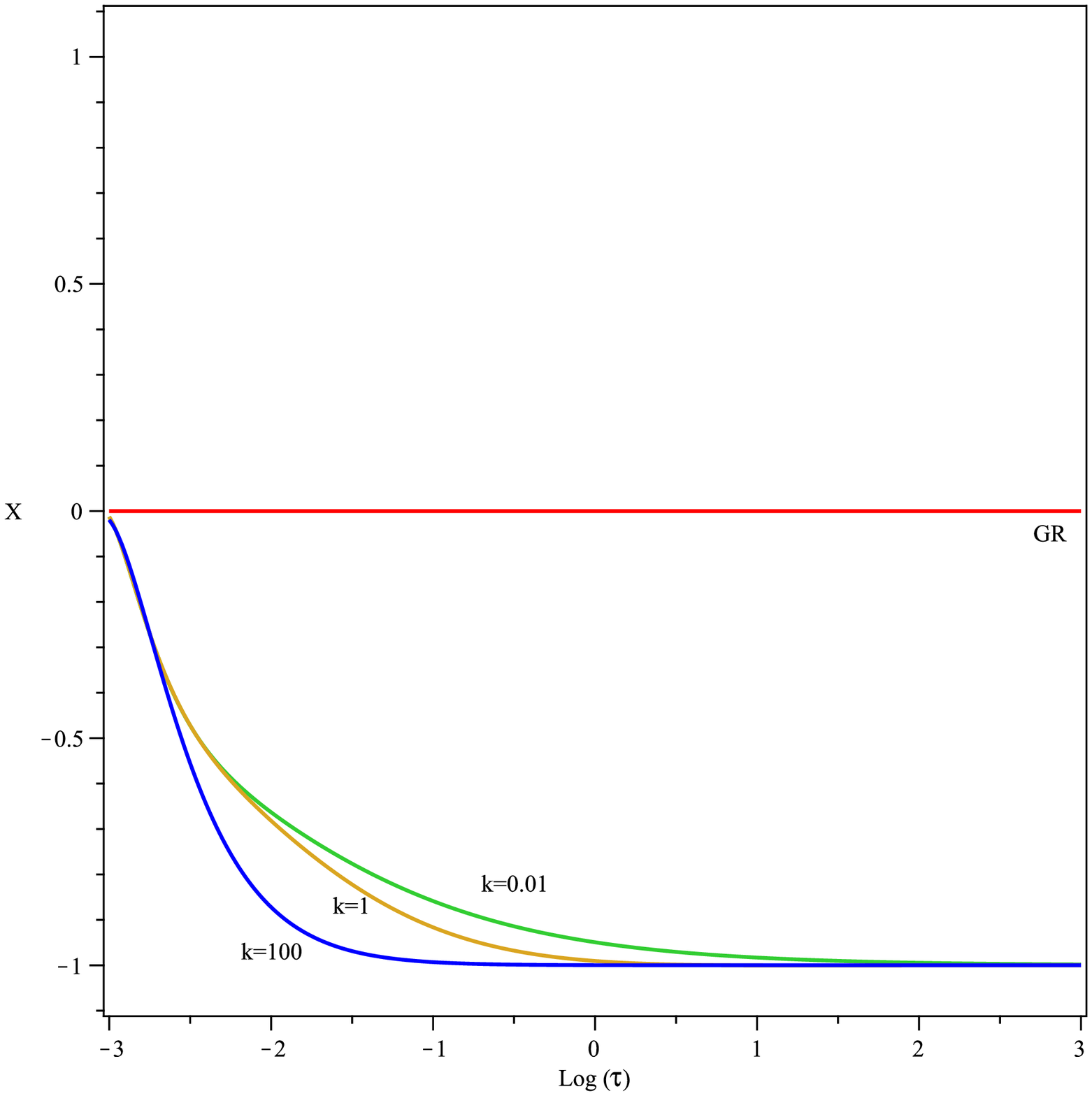}}
\subfigure[ Plot of $X(\tau)$ for $n=1.55$ and different values of $k$ ]{\includegraphics[scale=0.37]{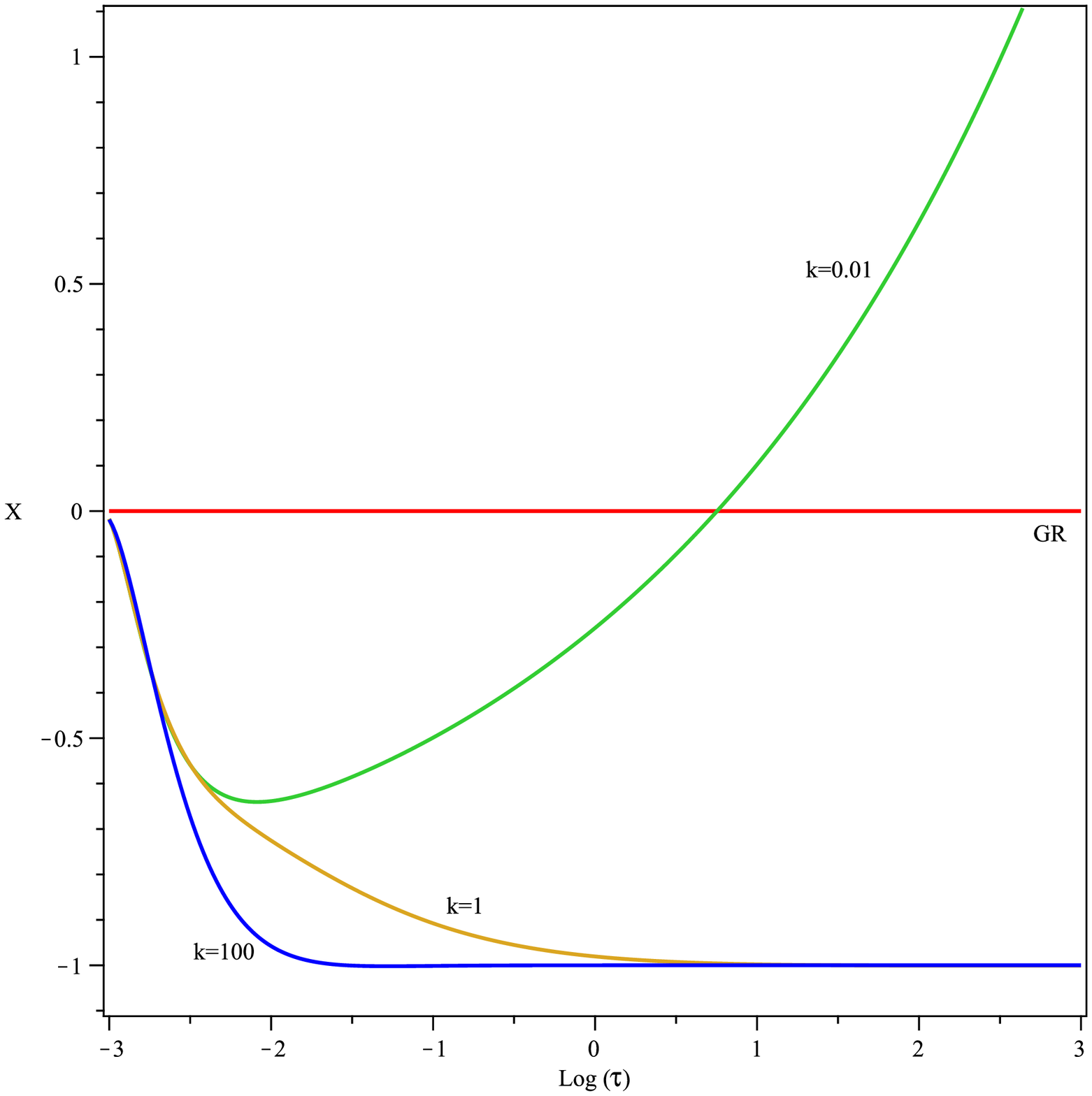}}
\subfigure[ Plot of $X(\tau)$ for $n=0.9$ and different values of $k$ ]{\includegraphics[scale=0.37]{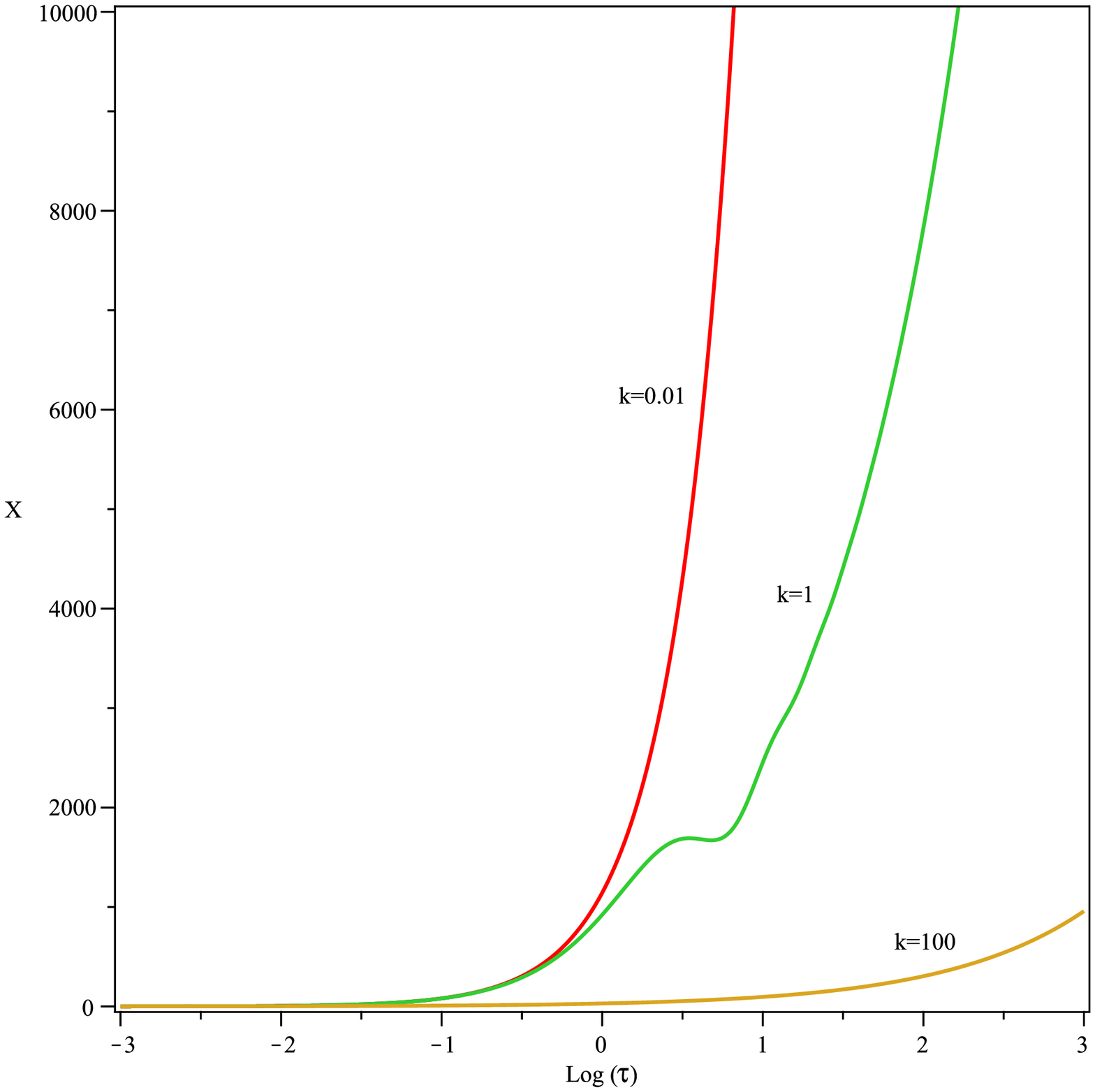}}
\caption{Plot of $X(\tau)$ as a function of $\tau=\log_{10}(S)$ for $R^n$-gravity and $k\neq 0$.
Note how the growth of the perturbations is suppressed when $k$ grows i.e. when smaller scales are considered.}\label{Fig8-10}
\end{figure}

Let us then use the quantity $X$ to analyze the growth rate of the matter fluctuation at different scales (see Figure \ref{Fig8-10}). One can see we see that the rate of growth decreases with respect to GR as
one tends towards smaller scales. In fact, from the analysis of the time evolution of the power spectrum, we will see  that this is associated to a dissipation of the perturbations at that range of scales.

These results are consistent with the conclusions already drawn in \cite{SantePert}, i.e. that the
dynamics of perturbation growth can be radically different from what happens in GR.  This, in turn, implies that the time needed for linear
structures to form and, as a consequence, the duration of the matter era, could be very different in this framework. Such features could prove to be
an interesting method of placing constraints on the theory of gravity by using data from existing and future large scale structure surveys.

It is also interesting to notice that the behavior of the perturbations for $n\approx 1$ is also radically
different from the GR case, i.e.,  {\it even small deviations from $n=1$ are able to produce
completely different dynamics}.  One can understand such behavior since even small deviations
from linearity in the action result in a change of order (from two to four) of the perturbation equations, so it is natural to expect significant differences in
their solutions.

As already mentioned the equations (\ref{EqPerIIOrd1}-\ref{EqPerIIOrd2}) and consequently (\ref{EqPerIIOrdRn1}-\ref{EqPerIIOrdRn2}) are similar
in structure to the equations one would obtain in the classic multi-fluid GR case and these
similarities can be used to infer the interaction properties of standard matter and the
curvature fluid once the proper variables have been chosen (i.e.\ $\Delta_{m}$ and $\Delta_{R}$ ).
The full expressions of these terms in the background given above is too long to be reported in full.
However it is instructive to examine their structure.  In the $\Delta_{m}$ equation, the
dissipation terms (i.e.\ the $\Delta'_{m}$ and the $\Delta'_{R}$ terms) are given by
\begin{equation}\label{DissDelm}
{\mathds C}_{\Delta'_{m}}=\frac{A_1(n,w)}{t}\left[w+\frac{A_2(n,w)}{1+A_3(n,w) k^{2} t^{2-\frac{4n}{3(1+w)}}+A_3(n,w)\left(1+A_5(n,w)k^{2} t^{2-\frac{4n}{3(1+w)}} \right)^{-1}}\right]\,,
\end{equation}
and
\begin{equation}\label{DissDelR}
{\mathds C}_{\Delta'_{R}}=\frac{E_1(n,w)}{t}\left[ 1+E_2(n,w) k^{2} t^{2-\frac{4n}{3(1+w)}}(1+E_3(n,w)k^{2} t^{2-\frac{4n}{3(1+w)}})\right]^{-1}\,,
\end{equation}
respectively,  while the source terms are\begin{equation}\label{SourceDelm}
{\mathds C}_{\Delta_{m}}=\frac{B_1(n,w)}{t^{2}}\left[\frac{1+B_2(n,w) k^{2} t^{2-\frac{4n}{3(1+w)}}(1+B_3(n,w)k^{2} t^{2-\frac{4n}{3(1+w)}})}{1+B_4(n,w) k^{2} t^{2-\frac{4n}{3(1+w)}}(1+B_5(n,w)k^{2} t^{2-\frac{4n}{3(1+w)}})}\right]\,,
\end{equation}
and
\begin{equation}\label{SourceDelR}
{\mathds C}_{\Delta_{R}}=\frac{C_1(n,w)}{t^{2}}\left[\frac{1+C_2(n,w) k^{2} t^{2-\frac{4n}{3(1+w)}}(1+C_3(n,w)k^{2} t^{2-\frac{4n}{3(1+w)}})}{1+C_4(n,w) k^{2} t^{2-\frac{4n}{3(1+w)}}(1+C_5(n,w)k^{2} t^{2-\frac{4n}{3(1+w)}})}\right]\,.
\end{equation}
where $A_i$, $B_i$, $C_i$, $E_i$ are functions of $n$ and $w$ only.
The structure of the coefficients of $\Delta'_{R}$ is, as expected, similar.

For large scales ($k\rightarrow 0$) we  have
\begin{equation}\label{LWLRn}
{\mathds C}_{\Delta'_{m}}\rightarrow\frac{A_1(n,w)\left[w+A_2(n,w)\right]}{t}\,, \quad {\mathds C}_{\Delta'_{R}}\rightarrow\frac{E_1(n,w)}{t}\,,\quad {\mathds C}_{\Delta_{m}}\rightarrow\frac{B_1(n,w)}{t^{2}}
\,,\quad {\mathds C}_{\Delta_{R}}\rightarrow\frac{C_1(n,w)}{t^{2}}\,,
\end{equation}
which corresponds to the equations that one would obtain in GR with two collisional
fluids. For small
scales  ($k\rightarrow \infty$) we have
\begin{equation}\label{SWLRn}
{\mathds C}_{\Delta'_{m}}\rightarrow w \frac{A_1(n,w)}{t}\,, \quad {\mathds C}_{\Delta'_{R}}\rightarrow 0 \,,\quad {\mathds C}_{\Delta_{m}}\rightarrow\frac{B_1(n,w)}{t^{2}}\left[\frac{B_3(n,w)}{B_5(n,w)}\right]
\,,\quad{\mathds C}_{\Delta_{R}}\rightarrow\frac{C_1(n,w)}{t^{2}}\left[\frac{C_3(n,w)}{C_5(n,w)}\right]\;,
\end{equation}
which again corresponds to the equations that one would obtains in the case of GR with two fluids, with the
difference that this time, since ${\mathds C}_{\Delta'_{R}}=0$,  they are non-collisional \footnote{Note the fact that, in the case of dust ($w=0$) the term ${\mathds C}_{\Delta'_{m}}$ becomes zero. Although it might appear strange at first sight this is not characteristic of
fourth order gravity. In fact the same happens in GR with a mixture of matter and radiation \cite{DBE,Challinor}.}.

The form of these coefficients is very different to the ones obtained, for example, in
the case of a GR baryon-photon system. As we mentioned in this last case the
dissipation terms grow as $k^2$. This means that, as expected, the effect of the
interaction between the two fluids becomes more important when one considers
smaller scales. In the above case the situation is different because the interaction
scale of standard matter and curvature fluid is a non-trivial function of $k$ which
is peaked at a certain value of $k$. This means that the
effect of the interaction is maximized around this specific scale.

\begin{figure}[htbp]\label{PkRn}
\subfigure[Plot of  the power spectrum at $\tau=1$ for $R^n$-gravity and $n>1$. Note that the spectrum is composed of three parts corresponding to three different evolution regimes for the perturbations. ]{\includegraphics[scale=0.75]{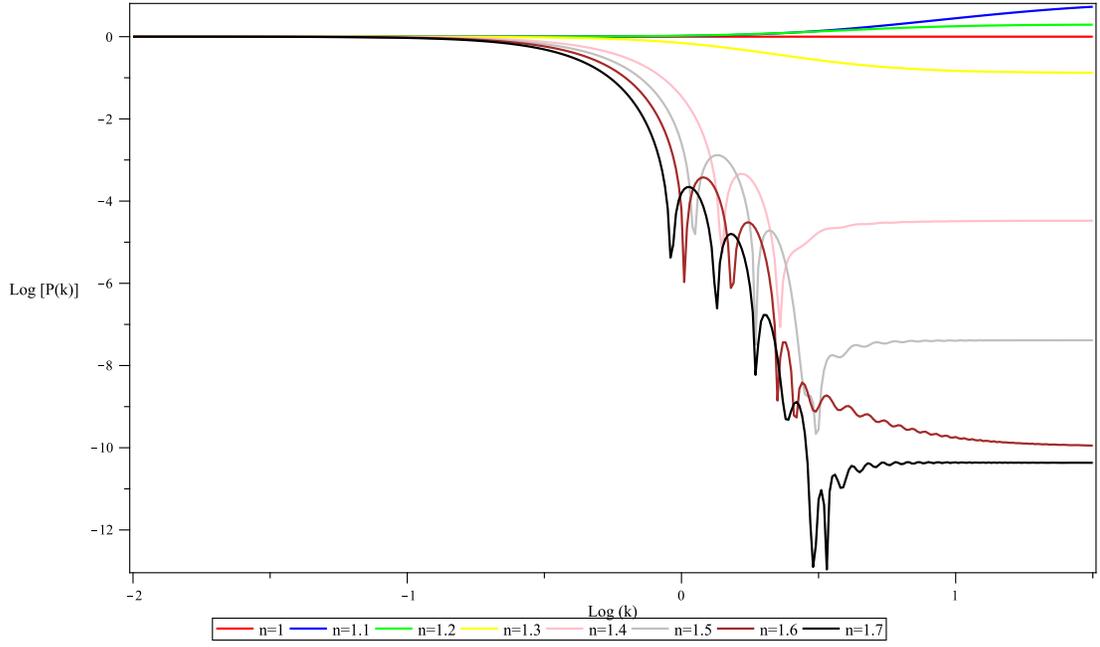}\label{Fig11}}
\subfigure[Plot of the Power spectrum as a function of $k$ for $R^n$-gravity at $\tau=1$ and $0<n<1$. The spectra for $n=0.6$ and $n=0.5$
only approach the scale invariant plateau at extremely high $k$ when compared to the other curves. Note the behavior of the spectrum for $n\approx 0.8$, differently from all the other cases, there is basically no loss of power in the spectrum at large $k$. ]{\includegraphics[scale=0.50
]{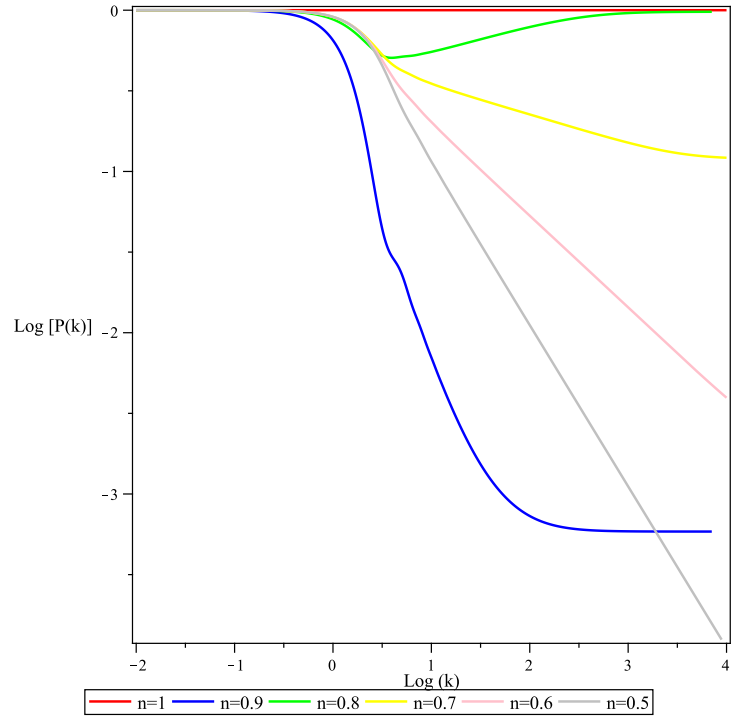}\label{Fig12}}
\caption{Power spectra for $R^n$-gravity}
\end{figure}

\begin{figure}[htbp]
\begin{center}
\includegraphics[scale=0.50]{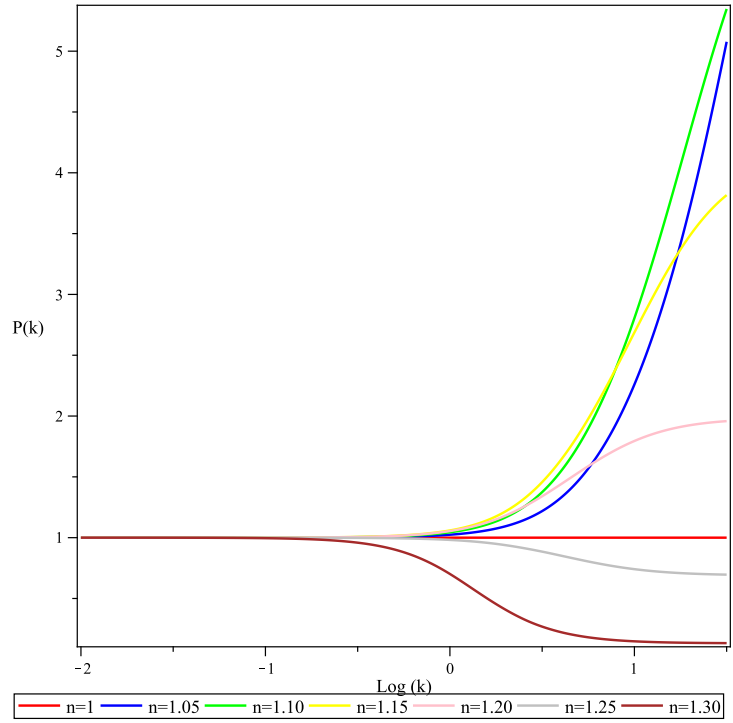}
\caption{Detail of the plot of the power spectrum for $R^n$-gravity at $\tau=1$. As expected, for these values of $n$ we find the presence of the three regimes mentioned in the text. Note also that for $n\approx 1^{+}$ the small scales are characterized by an excess of power. Such features is compatible with the results found in  \cite{Li:2008ai,HuSawicki,Bertschinger:2008zb}. \label{Fig13}}
\end{center}
\end{figure}

Let us now focus on the impact of these features on the matter power spectrum
(Figures \ref{Fig11}, \ref{Fig12}, \ref{Fig13}).  As mentioned in Section  \ref{Eq&Par}  the k-structure of Equations (\ref{EqPerIIOrd1}-\ref{EqPerIIOrd2})  suggest that in fourth order gravity there exist at least three different growth  regimes  of the perturbations. This is confirmed by our results. In particular, in the case of
dust we have  three regimes for any values of the remaining parameters:  (i) on very large scales the spectrum
goes like GR i.e. it is scale invariant;  (ii) as $k$ becomes bigger the scale invariance is broken and oscillations in the spectrum appear;  (iii)   for even larger $k$ the spectrum becomes again scale invariant. However, on these scales the spectrum can contain either an excess or deficit of power depending on the value of  $n$. In particular for $n\approx 1^+$ small scales have more power than large scales,
but, as one moves towards larger values of $n$, the small scale modes are suppressed.
For $0<n<1$, instead the drop in power seems to decrease as one moves from $n=0$ towards
$1^-$ and we see a sudden increase for $n\approx 1^-$. It is worth noting
the case $n\approx 0.8$ for which there is basically no difference in power between
large and small scales and there are no significant oscillations in the spectrum.

\begin{figure}[htbp]
\begin{center}
\includegraphics[scale=0.80]{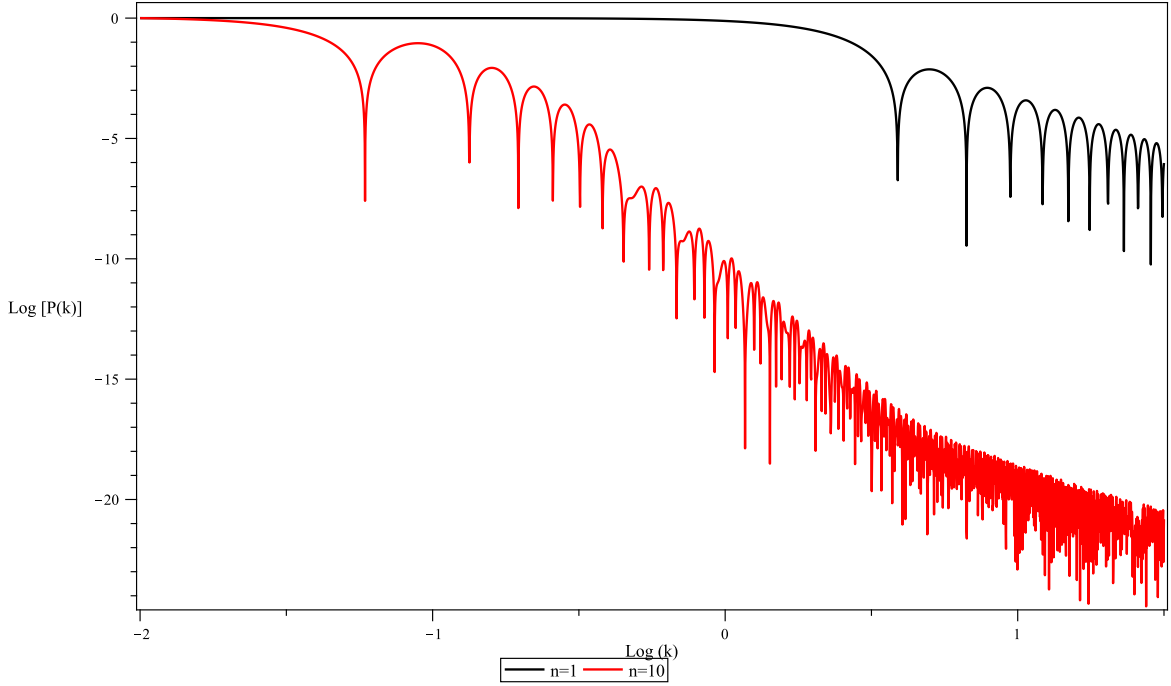}
\caption{The power spectrum for $R^n$-gravity at $\tau=1$ in the
case of radiation for GR and $R^n$-gravity with $n=10$. As expected in this last plot  we find four different
regimes instead of the three of the dust case: a first regime for $k\rightarrow0$ which is scale invariant; a second and a third regime which correspond to the two different slopes between $k\approx 10^{-2}$ and  $k\approx 10^{1/2}$ and a fourth regime which has the same slope of the GR plot. \label{Fig14}}
\end{center}
\end{figure}

Further  indication of the link between the $k$ structure of (\ref{EqPerIIOrd1}-\ref{EqPerIIOrd2}) and the different regimes of the matter power spectrum can be seen if one analyzes  the power spectrum in a radiation dominated era. In this case the perturbations equations contain an additional term which is not present in the dust case.  This means that one would expect {\it four} different regimes, rather than three. In Figure \ref{Fig14} we have plotted the matter power spectrum of $R^n$-gravity in the case of radiation and $n=10$ (this value of $n$ is chosen only for convenience and it does not have any physical motivation), and as expected one can recognize four different regimes. Note that we obtain the same $k$ scaling as in GR on small scales.

\begin{figure}[htbp]
\begin{center}
\includegraphics[scale=0.5]
{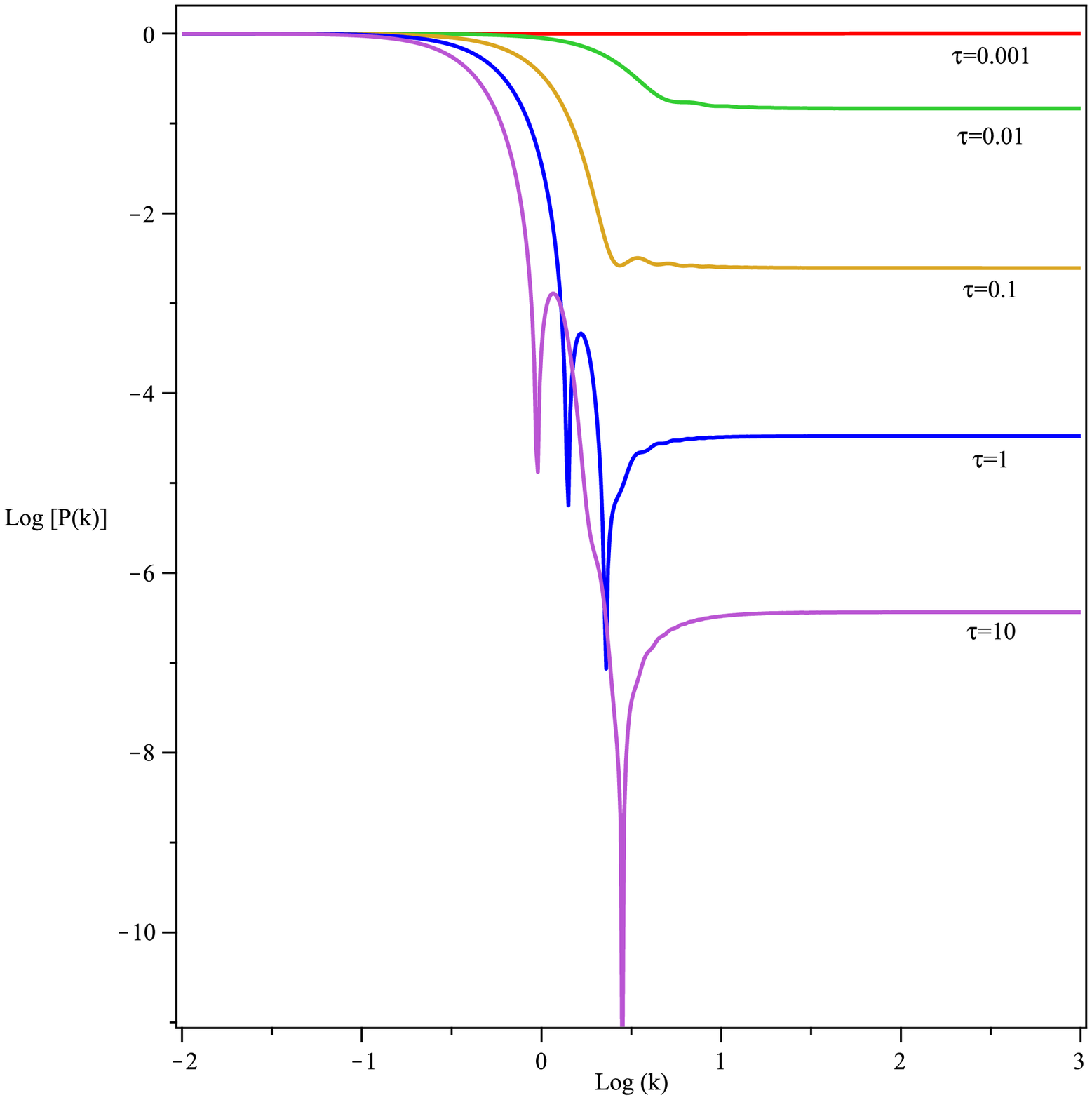}
\caption{Evolution the Power spectrum for $R^n$-gravity for $n=1.4$. The spectrum has been normalized in such a way  that the curves coincide at large scales. Note how, as time passes, small scale perturbations are dissipated and oscillation appear. \label{Fig15}}
\end{center}
\end{figure}

Finally further information on the dynamics of the matter perturbations can be obtained examining the time evolution of the power spectrum. In Figure  \ref{Fig15} we give the power spectrum for $n=1.4$ at different times. One can see that, as the universe expands, the small scale part of the spectrum is more and more suppressed and oscillations start to form. This is in agreement with what one finds from the analysis of $X$  and suggests that in this model small scale perturbations tend to be dissipated in time. On the other hand the large scales do not seem to be evolving, which might appear in contrast with what mentioned above. However  this is a byproduct of the normalization:  for clarity we have normalized the spectrum in such a way that every curve has the same power in long wavelength limit.

The features of the spectrum that we have derived can be then interpreted in terms of the interaction between the curvature fluid and standard matter. On very large and very small scales, the coefficients (\ref{DissDelm}-\ref{SourceDelR})  become independent from $k$ so that the evolution of the perturbations does not change with the scale and the power spectrum is scale invariant. On intermediate scales the interaction between the two fluids  is maximized and the curvature fluid acts as a relativistic component whose pressure is responsible for the oscillations and the dissipation of the small scale perturbations in the same way in which the photons operate in a baryon-photon system \footnote{This suggests the following interesting interpretation for the perturbation variables $\mathcal{R}$ and $\Re$. These quantities can be thought to represent  the modes associated with the  contribution of the additional scalar degree of freedom typical of $f(R)$-gravity. In this sense the spectrum can be explained physically as a consequence of the interaction between  these scalar modes and standard matter.}. The result is a considerable loss of power for a relatively small variation of the parameter $n$. For example, in the case $n=1.4$ the difference in power between he two scale invariant parts of the spectrum for $n=1.1$ is of  one order of magnitude while for $n=1.6$  is about ten orders of magnitude.

Probably the most important consequence of the form of the spectrum presented above is the fact that the effect of these type of fourth order corrections is evident only for a special range of scales, while the rest of the spectrum has the same $k$ dependence of GR (but different amplitude). This implies that we have a spectrum that both satisfies the requirement for scale invariance and has distinct features that one could in principle detect, by combining future Cosmic  Microwave Background (CMB) and large scale surveys (LSS) \cite{Planck, SDSS}.

\subsection{The case $f(R)=R+\alpha R^{n}$}\label{SecRRn}

We will now consider a theory with the following action,
\begin{equation}\label{lagr RRn}
L=\sqrt{-g}\left[R+\alpha R^{n}+{\cal L}_{M}\right]\;.
\end{equation}
This theory has gained much popularity as a fourth order gravity model within the
context of both inflation and dark energy \cite{Ottewill, star80, Suen, Carroll}.

Unlike $R^{n}$-gravity, \rf{lagr RRn} includes explicitly the
Hilbert-Einstein term, so that one can consider it as the result of an additive correction to GR.
This also means that this model introduces an explicit physical scale which is determined
by the relative magnitude of the two terms in the action, making it the simplest fourth order
gravity theory with an additional scale for the gravitational interaction.

Some work on estimating the parameter $\alpha$ for the specific case of $n=2$
was completed during the eighties. This work was mainly based on arguments pertaining
to black hole physics and quantum gravity (see for example \cite{Suen}). In the following
we will take $\alpha$, which in our units is the ratio between the coupling constant of
the fourth order corrections, to be positive definite. Of course, we expect this model
to behave much in the same way as the model discussed in the previous section for $\alpha\rightarrow\infty$,
and to recover GR for $\alpha=0$. This also means that any new feature in this model will emerge for intermediate values of the coupling.

The $f(R)=R+\alpha R^{n}$ model has been analyzed at the level of the background using
many different approaches (see for example \cite{Ottewill, star80, Suen, Carroll, Teyssandier:1989dw}),
but probably some of the most interesting results for cosmology have been found using the dynamical system
approach \cite{Barrow Hervik,SanteGenDynSys,Amendola:2006kh}. The dynamical systems analysis proved
that this class of models has, like $R^{n}$-gravity, an unstable fixed point associated
with the Friedmann-like solution  $a=t^{2n/3(1+w)}$.
Substituting the form of the action \rf{lagr RRn}, the system of perturbation equations (\ref{EqPerIIOrd1}-\ref{EqPerIIOrd2}) takes the form (\ref{EqPerIIOrdRRn1}-\ref{EqPerIIOrdRRn2}) given in Appendix \ref{App4} and one can analyzed the evolution of the scalar perturbations in this background  \footnote{Unlike the previous example, the structure of the phase space for this model is not well known. This means that, although in this theory one has a fixed point that resembles the one of $R^n$-gravity, it is not obvious that it plays the same role. In addition, this background is not, in general, a physical solution of the cosmological equations \cite{SanteGenDynSys}. This means that there are cosmic histories in which the general integral of these equations approximates the behavior we consider, but it will never be exactly the same. We choose to treat this as a further approximation in our investigation. }.

In spite of the fact that both this theory and $R^n$-gravity have a Friedmann-like solution, dealing with the perturbation equations for this background in $f(R)=R+\alpha R^n$ is considerably more
complicated than in $R^{n}$-gravity. As a consequence, one is unable to find exact
solutions for the perturbation equations, even when one adopts the long wavelength
limit. To progress one must numerically integrate the system of perturbation
equations and extract useful information relating to the perturbation
dynamics using the quantities $X$, $Y$ and the power spectrum.

\begin{figure}[htbp]
\subfigure[ Plot of  $X(\tau)$ for $f(R)=R+\alpha R^{n}$ for large scales ($k=0$), $n>1$, dust and $\alpha=10$. Note the similarity with the plots of Figure \ref{Fig2-3}.]{\includegraphics[scale=0.35
]{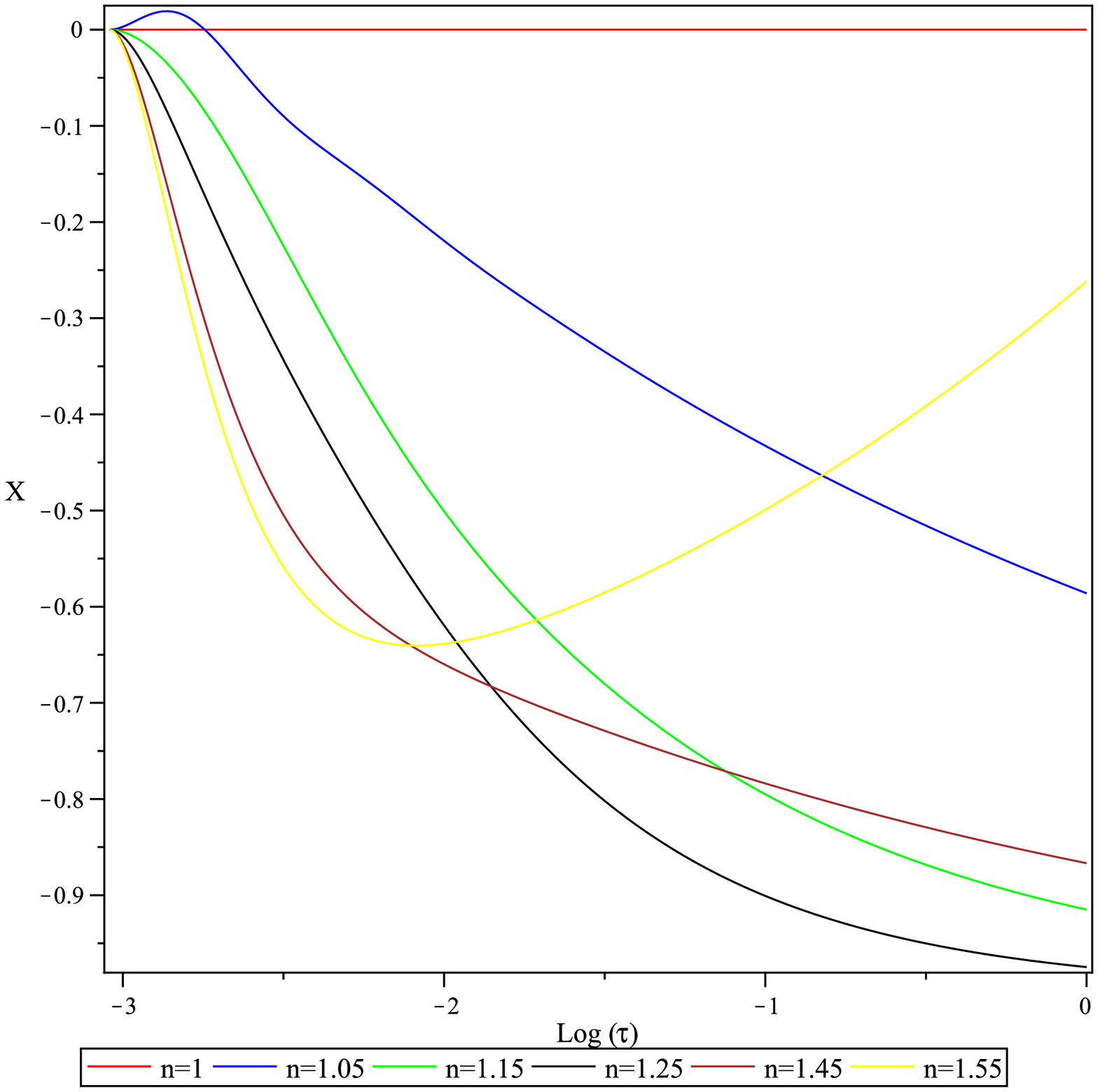}\label{Fig16}}
\subfigure[ Plot of  $X(\tau)$ for $f(R)=R+\alpha R^{n}$ for large scales ($k=0$), $n>1$, dust and $\alpha=1$. Differently from  the plot in Figure \ref{Fig16} in this case the curve $n=1.55$ reaches a maximum and then decreases.]{\includegraphics[scale=0.35]{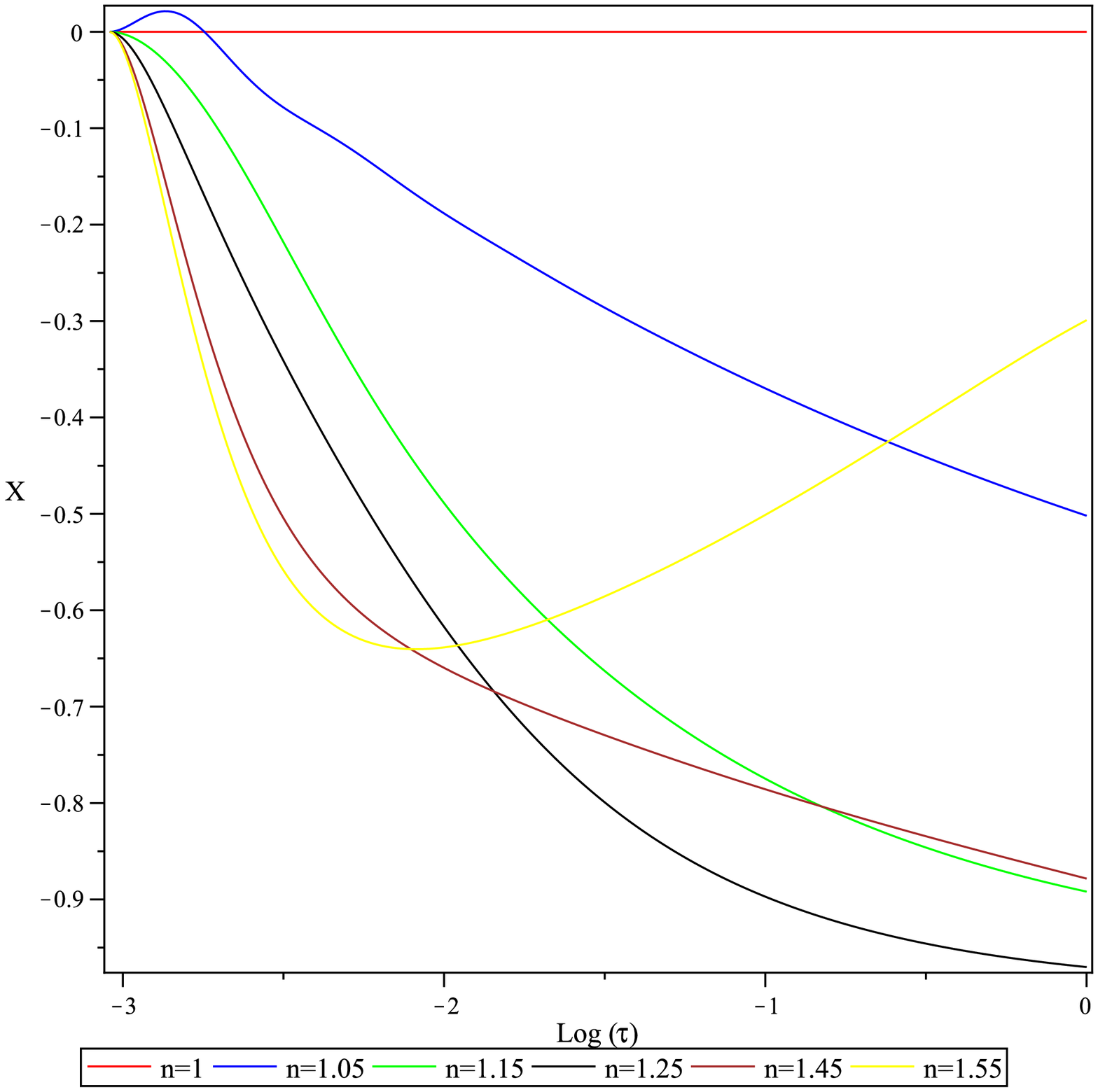}\label{Fig17}}
\subfigure[ Detail of the plot \ref{Fig16} for $1<n<1.05$, The oscillations in the curves are the feature that differentiate $f(R)=R+\alpha R^{n}$  from $R^n$-gravity.]{\includegraphics[scale=0.35]{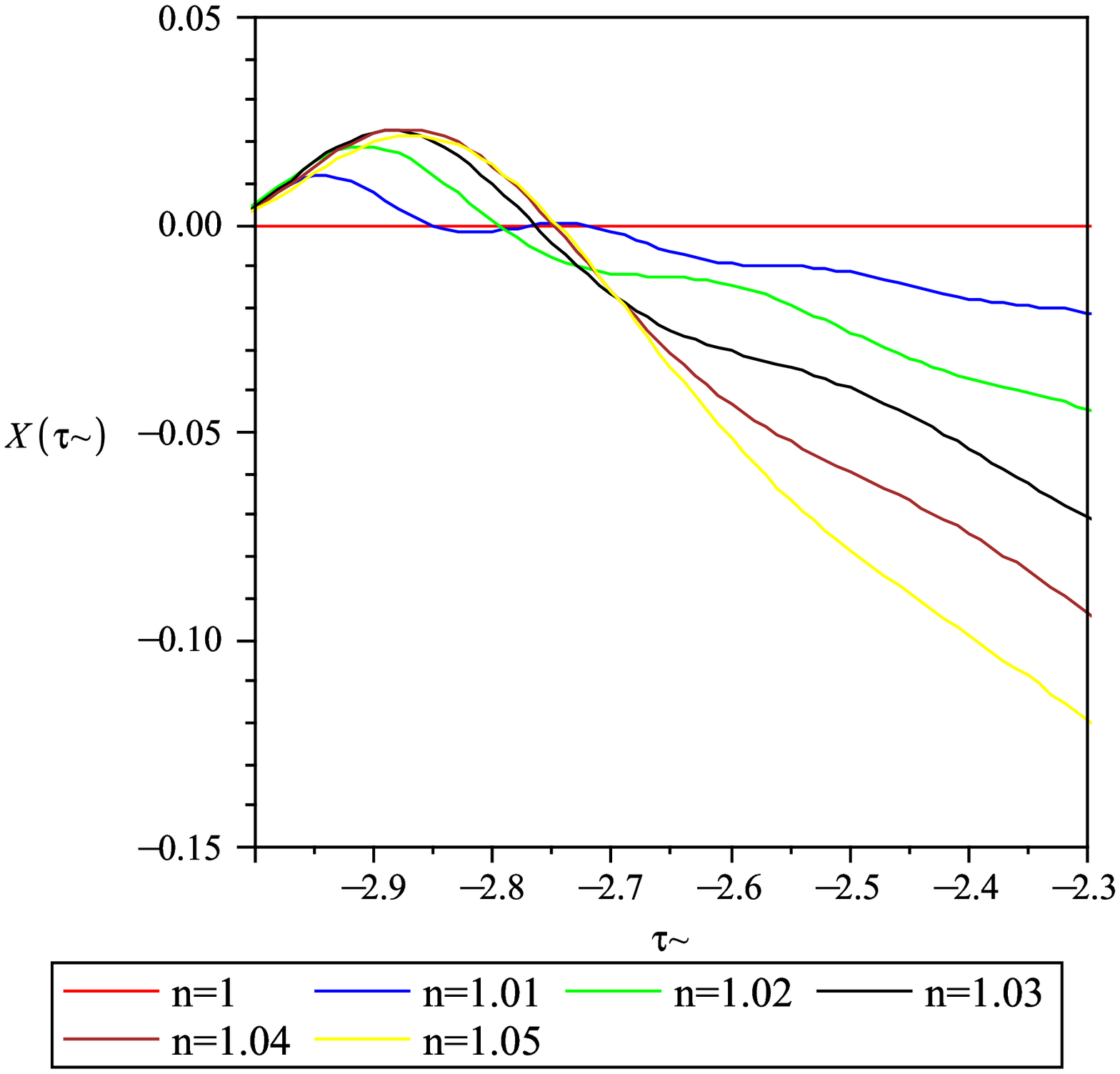}\label{Fig18}}
\subfigure[ Detail of the plot \ref{Fig17} for $1<n<1.05$, The oscillations in the curves are the feature that differentiate $f(R)=R+\alpha R^{n}$  from $R^n$-gravity.]{\includegraphics[scale=0.35]{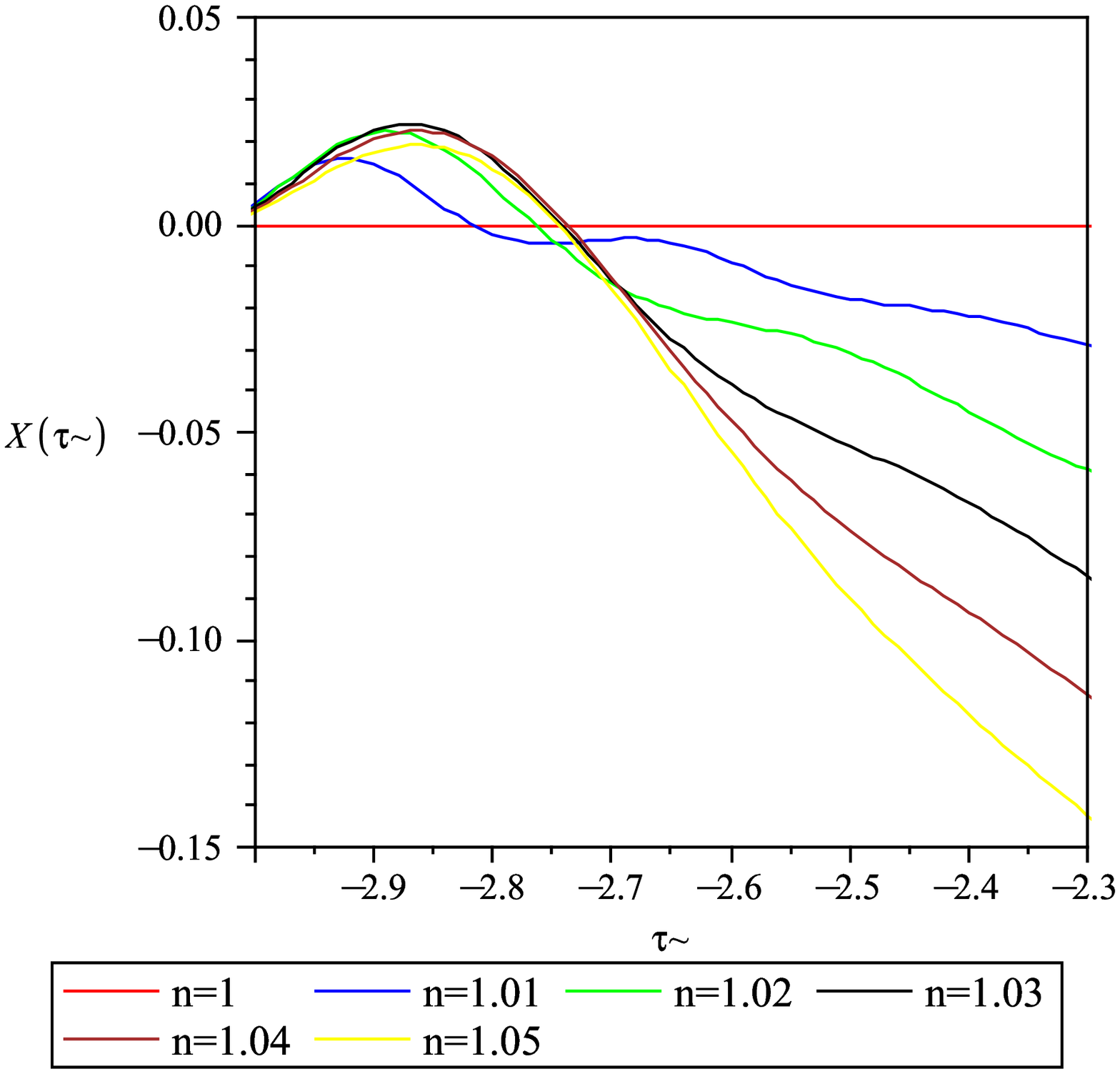}\label{Fig19}}
\subfigure[ Plot of  $X(\tau)$ for $f(R)=R+\alpha R^{n}$ for large scales ($k=0$), $n>1$, dust and $\alpha=0.1$.]{\includegraphics[scale=0.35]{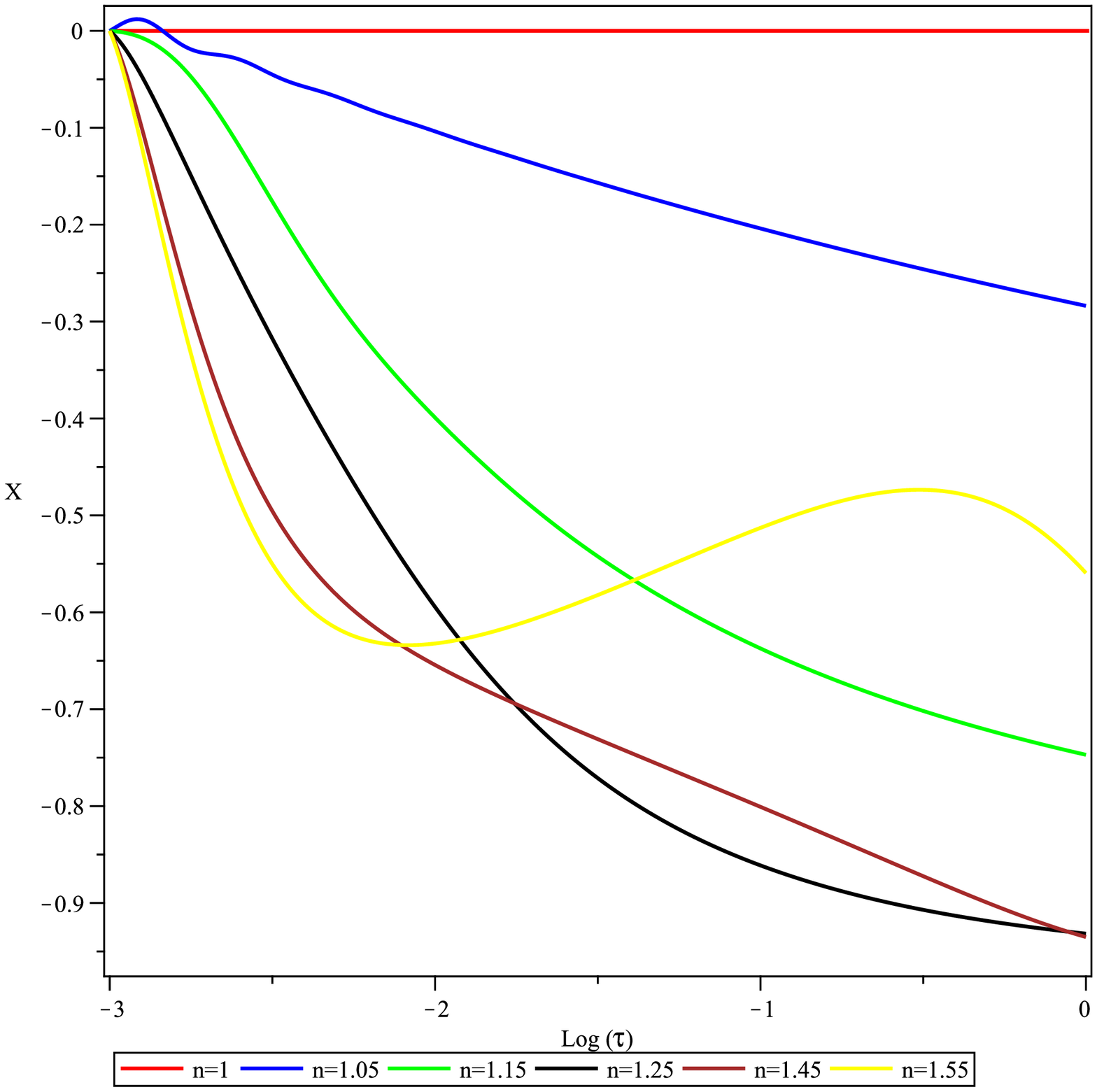}\label{Fig20}}
\subfigure[ Plot of  $X(\tau)$ for $f(R)=R+\alpha R^{n}$ for large scales ($k=0$), $n>1$, dust and $\alpha=0.01$.]{\includegraphics[scale=0.35]{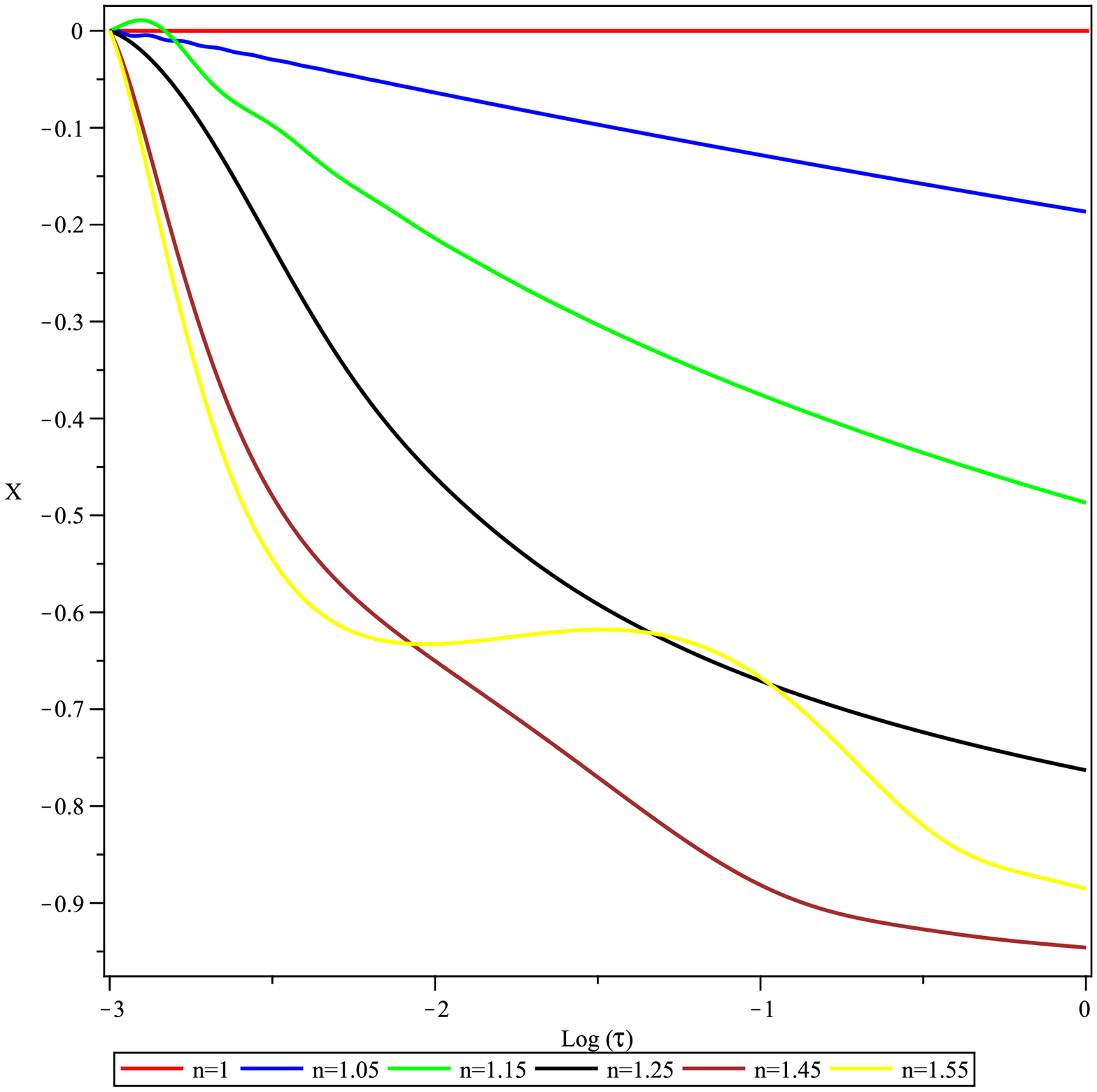}\label{Fig21}}
\caption{Plots of  $X(\tau)$ for large scales in  $R+\alpha R^n$-gravity}\label{XRRnLWL}
\end{figure}

If we look at Figure \ref{XRRnLWL} we notice that for $\alpha\ge1$ the dynamics
of the large scale perturbations in this model are  similar to those found
in $R^n$-gravity but contain some small differences. For example, from  Figure \ref{Fig16} one sees that  when $\alpha=10$ and $n\approx 1^+$  the growth rate has some oscillations at small $\tau$ followed by the decay typical of $R^n$-gravity. However, at smaller values of $\alpha$ the situation changes. In particular, the curves that did not show  oscillations at small $\tau$ are characterized by  the onset of late time oscillations. Because of that the value of $n$ at which the growth rate starts to increase at late time is different. These features reveal the important dynamical differences between this model, $R^n$-gravity and GR. Such differences will be even more evident when we will examine the time evolution of the power spectrum.

The analysis of the behavior of the quantity $Y$  on large scales also reveal
similarities between this model and $R^n$-gravity. In particular, one sees that, again,  for large values of $\alpha$ (i.e. $\alpha=10)$ there is very little difference between the previous example and this model, but when $\alpha$ becomes smaller the solution takes more time to saturate and oscillates with a wider amplitude.  In addition, the asymptotic values of $Y$ that correspond to the dominant perturbation mode  changes with $\alpha$ and $n$.

Let us now consider the evolution of the perturbations on smaller scales. The behavior of $X$ at different scales (Figures \ref{Fig26-28}) reveals changes in the growth rate. However  for different values of $\alpha$  and $n$ these changes are not always associated with suppression of the growth rate like in the case of $R^n$-gravity. Figure \ref{Fig26-28d} gives an example of these differences.

\begin{figure}[htbp]
\subfigure[ Plot of $Y(\tau)$ for $f(R)=R+\alpha R^{n}$ for large scales ($k=0$), $n>1$, dust and $\alpha=10$.]{\includegraphics[scale=0.45]{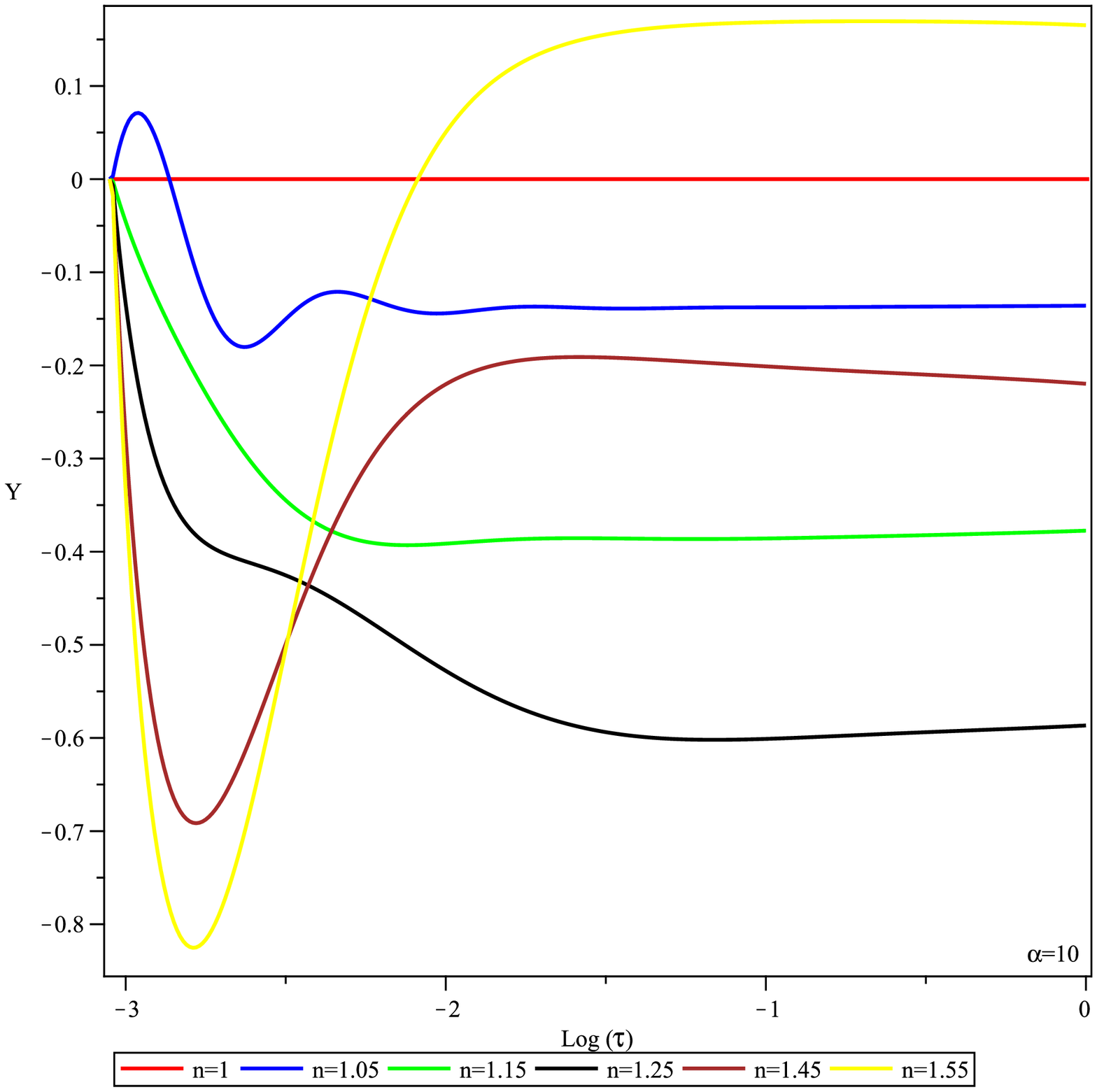}}\label{Fig22}
\subfigure[ Plot of $Y(\tau)$ for $f(R)=R+\alpha R^{n}$ for large scales ($k=0$), $n>1$, dust and $\alpha=1$.]{\includegraphics[scale=0.45]{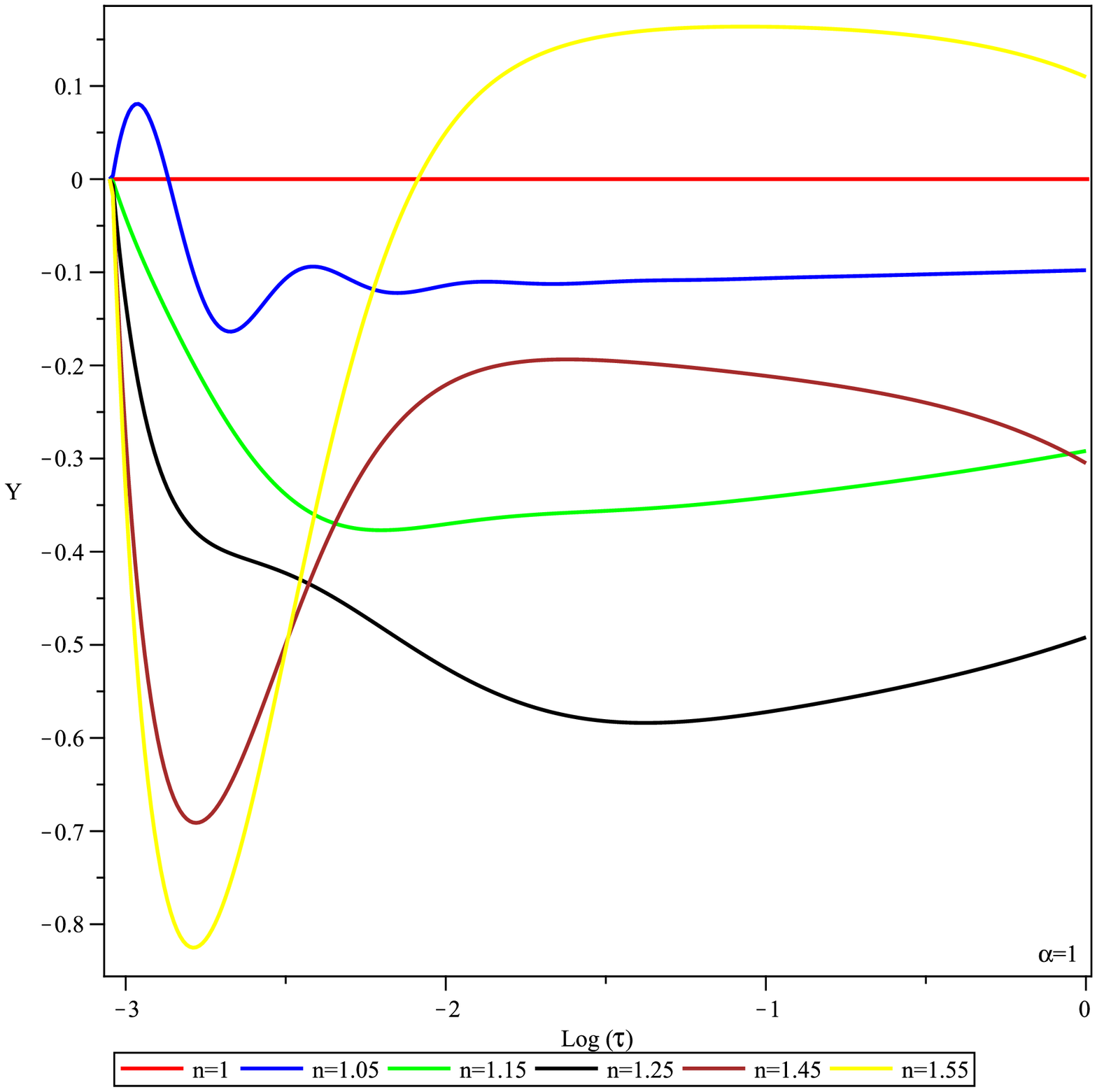}}\label{Fig23}
\subfigure[ Plot of $Y(\tau)$ for $f(R)=R+\alpha R^{n}$ for large scales ($k=0$), $n>1$, dust and $\alpha=0.1$]{\includegraphics[scale=0.45]{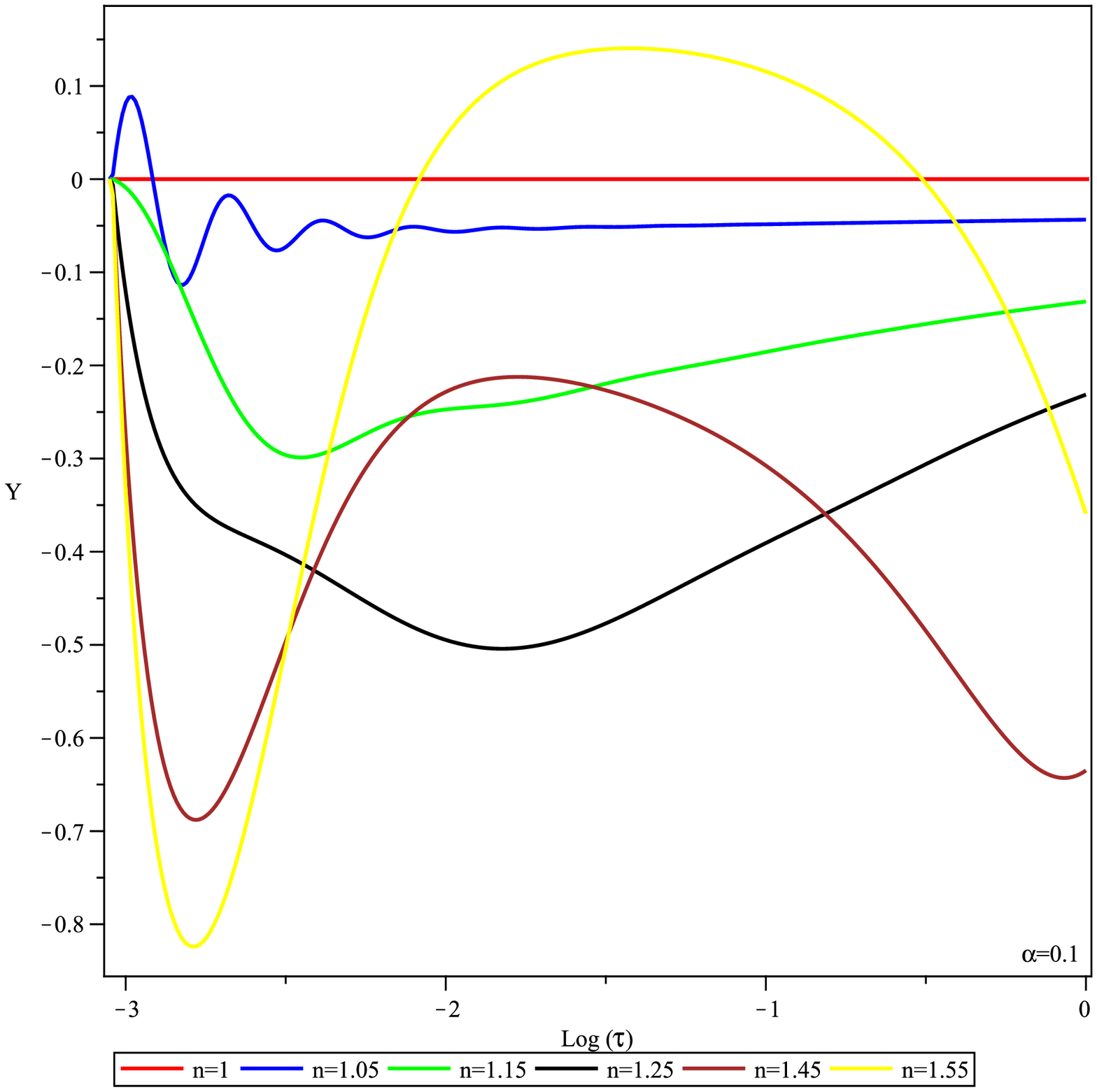}}\label{Fig24}
\subfigure[ Plot of $Y(\tau)$ for $f(R)=R+\alpha R^{n}$ for large scales ($k=0$), $n>1$, dust and $\alpha=0.01$.]{\includegraphics[scale=0.45]{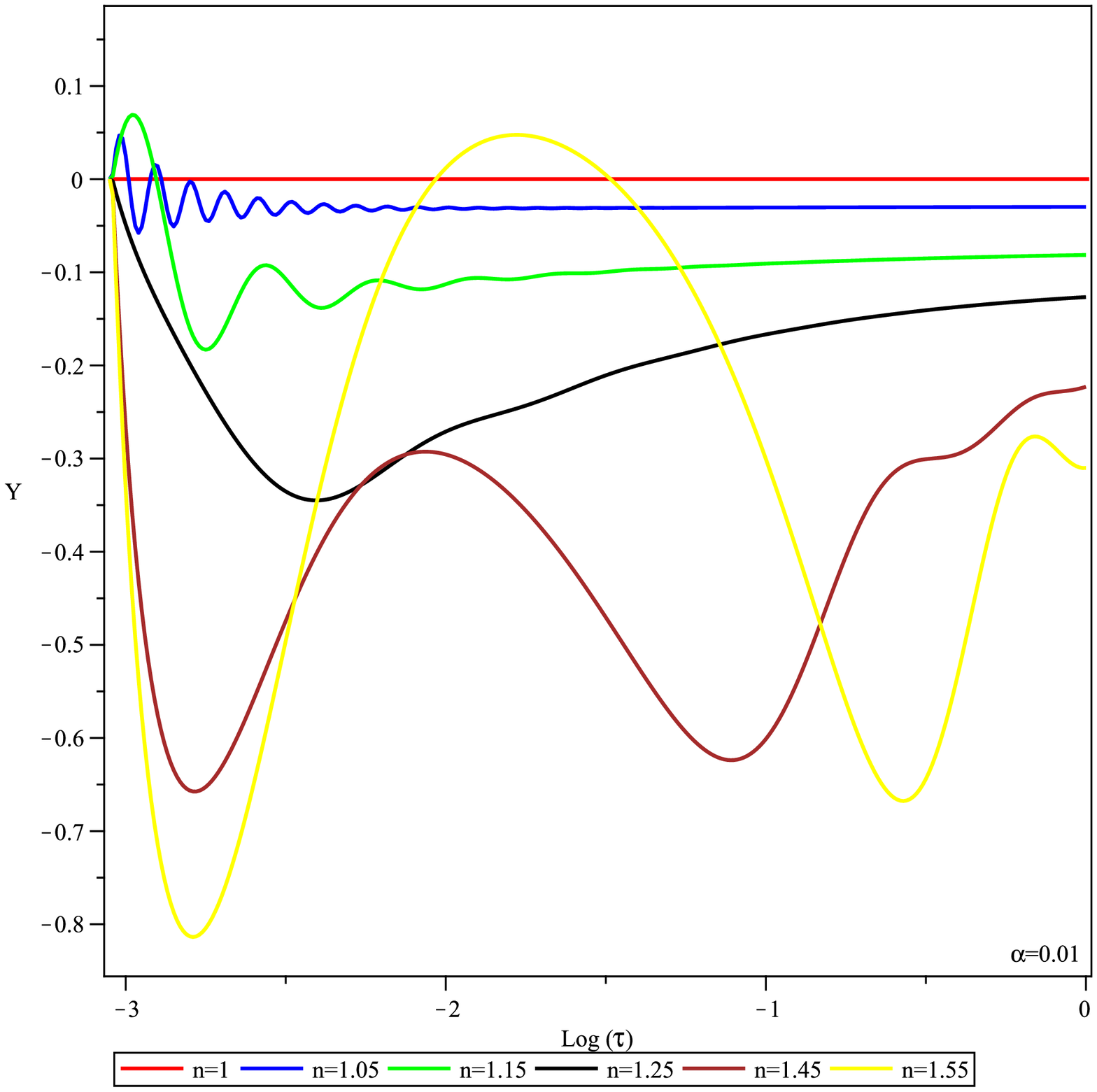}}\label{Fig25}
\caption{Plots of $Y(\tau)$ for large scales in $R+\alpha R^n$-gravity}
\end{figure}

\begin{figure}[htbp]
\subfigure[ Plot of $X(\tau)$ for $n=1.4$, $\alpha=100$ and various values of $k$ ]{\includegraphics[scale=0.37]{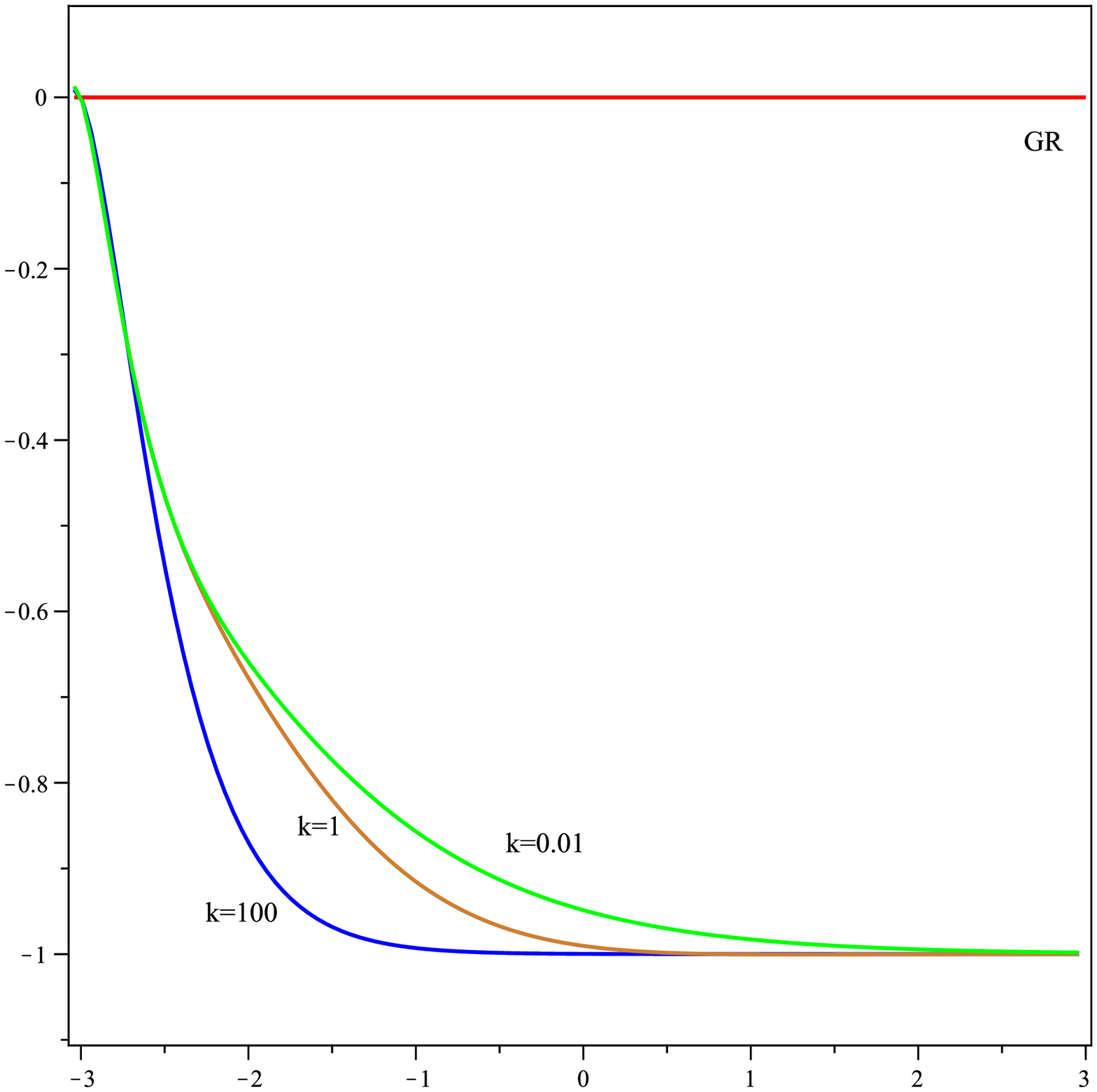}\label{Fig26-28a}}
\subfigure[ Plot of $X(\tau)$ for $n=1.4$, $\alpha=1$ and various values of $k$ ]{\includegraphics[scale=0.37]{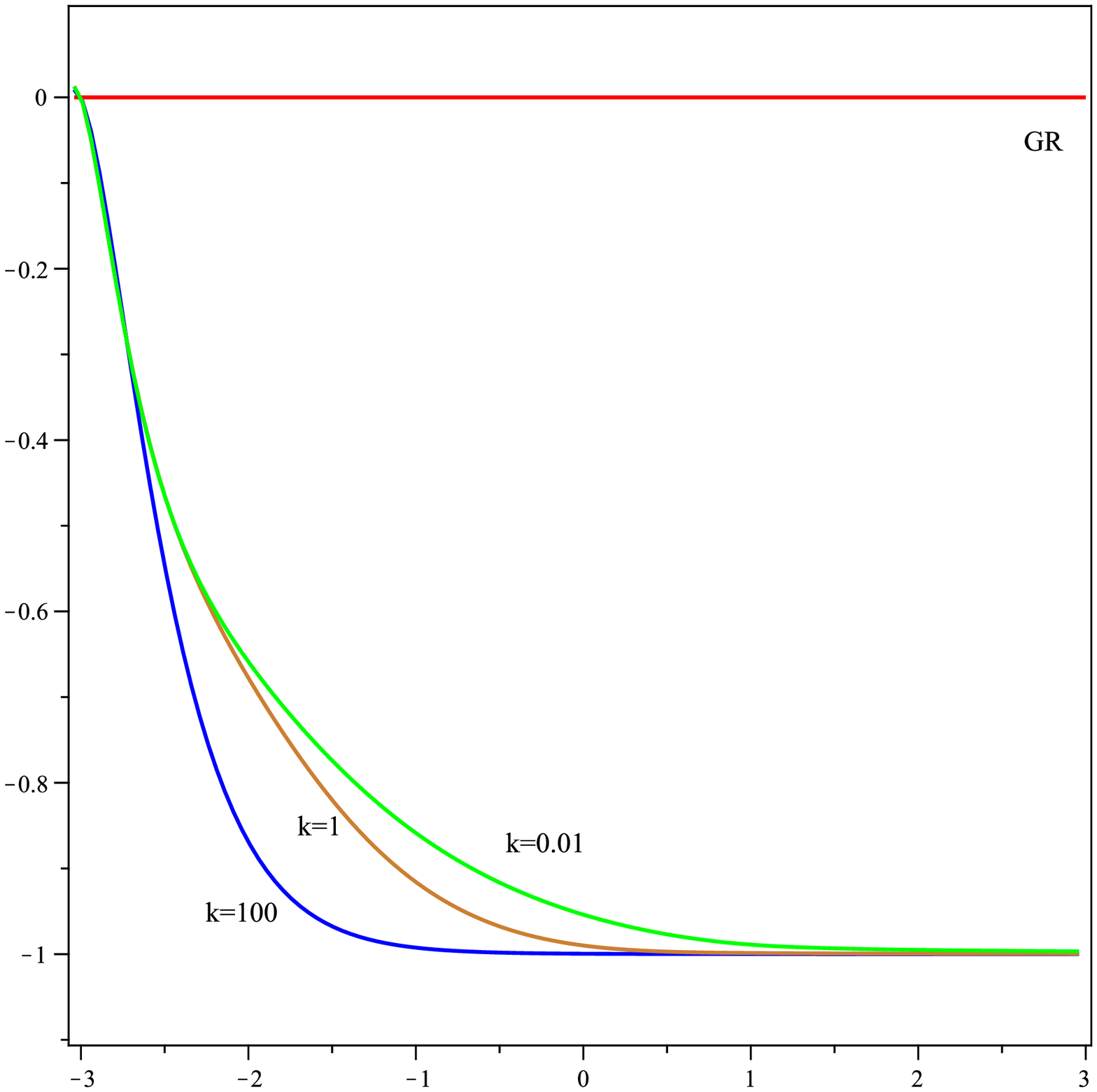}\label{Fig26-28b}}
\subfigure[ Plot of $X(\tau)$ for $n=1.4$, $\alpha=0.01$ and various values of $k$ ]{\includegraphics[scale=0.37]{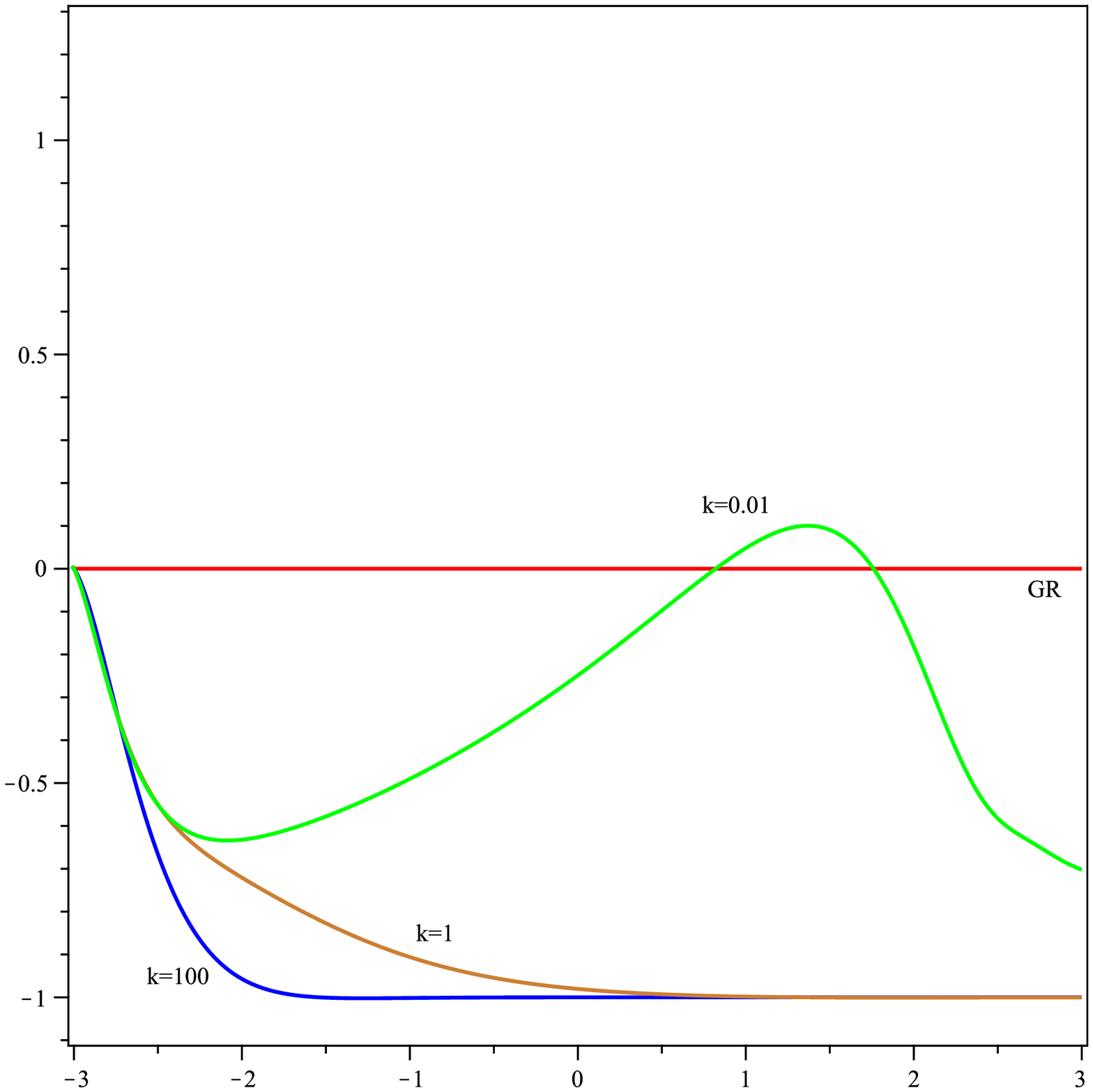}\label{Fig26-28c}}
\subfigure[ Plot of $X(\tau)$ for $n=1.55$, $\alpha=0.01$ and various values of $k$ ]{\includegraphics[scale=0.37]{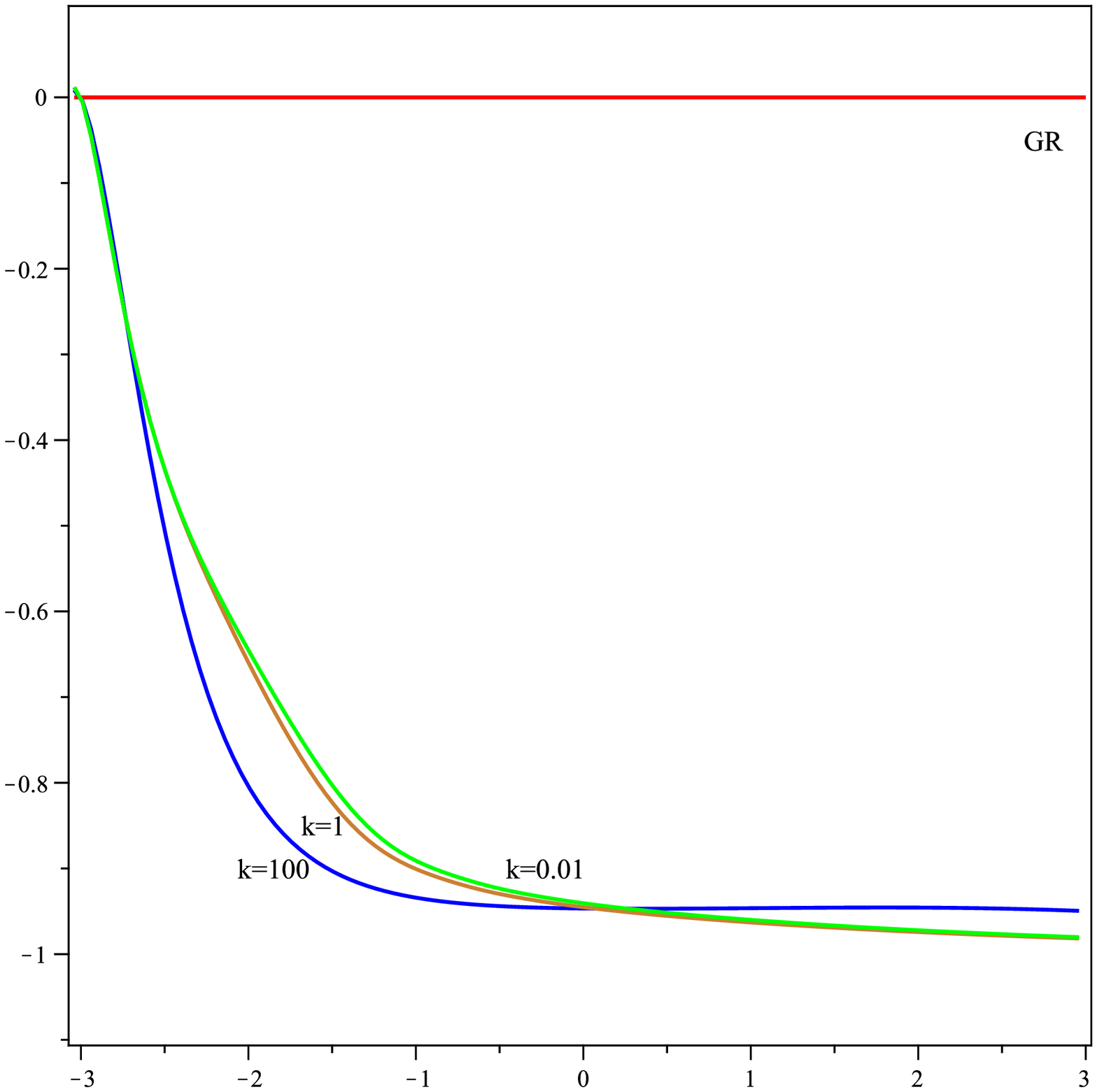}\label{Fig26-28d}}
\caption{Plot of $X(\tau)$ as a function of $\tau=\log_{10}(S)$ for $R+\alpha R^n$-gravity and $k\neq 0$.
Note how the growth of the perturbations is suppressed when $k$ grows and when $\alpha$ decreases.\label{Fig26-28}}
\end{figure}

When one calculates the structure of the coefficients for the $\Delta_{m}$ equation in
the system $(\Delta_{m},\Delta_{R})$ one obtains
\begin{equation}\label{DissDelm2}
{\mathds C}_{\Delta'_{m}}=\frac{A_1(n, w ,\mathds{A})}{t}\left[w+\frac{A_2(n, w ,\mathds{A})}{1+A_3(n, w ,\mathds{A}) k^{2} t^{2-\frac{4n}{3(1+w)}}+A_3(n, w ,\mathds{A})\left(1+A_5(n, w ,\mathds{A})k^{2} t^{2-\frac{4n}{3(1+w)}} \right)^{-1}}\right]\,,
\end{equation}
and
\begin{equation}\label{DissDelR2}
{\mathds C}_{\Delta'_{R}}=\frac{E_1(n, w ,\mathds{A})}{t}\left[ 1+E_2(n, w ,\mathds{A}) k^{2} t^{2-\frac{4n}{3(1+w)}}(1+E_3(n, w ,\mathds{A})k^{2} t^{2-\frac{4n}{3(1+w)}})\right]^{-1}\,.
\end{equation}
respectively,  while the source terms are\begin{equation}\label{SourceDelm2}
{\mathds C}_{\Delta_{m}}=\frac{B_1(n, w ,\mathds{A})}{t^{2}}\left[\frac{1+B_2(n, w ,\mathds{A}) k^{2} t^{2-\frac{4n}{3(1+w)}}(1+B_3(n, w ,\mathds{A})k^{2} t^{2-\frac{4n}{3(1+w)}})}{1+B_4(n, w ,\mathds{A}) k^{2} t^{2-\frac{4n}{3(1+w)}}(1+B_5(n, w ,\mathds{A})k^{2} t^{2-\frac{4n}{3(1+w)}})}\right]\,,
\end{equation}
and
\begin{equation}\label{SourceDelR2}
{\mathds C}_{\Delta_{R}}=\frac{C_1(n, w ,\mathds{A})}{t^{2}}\left[\frac{1+C_2(n, w ,\mathds{A}) k^{2} t^{2-\frac{4n}{3(1+w)}}(1+C_3(n, w ,\mathds{A})k^{2} t^{2-\frac{4n}{3(1+w)}})}{1+C_4(n, w ,\mathds{A}) k^{2} t^{2-\frac{4n}{3(1+w)}}(1+C_5(n, w ,\mathds{A})k^{2} t^{2-\frac{4n}{3(1+w)}})}\right]\,,
\end{equation}
 where $A_i$, $B_i$, $C_i$, $E_i$ are functions of $n$ and $w$ and $\mathds{A}=\alpha\; t^{\,2n-2}$. Although the time behavior of these coefficients
is always different from GR (unless of course $\alpha=0$), their $k$ behavior at a fixed time is
similar to the one found in the case $f(R)=R^n$. This means that this model also
behaves like a  two fluids system for large and small $k$ and that the deviation
from scale invariance occurs only in a specific $k$ interval, which this time is
determined by both the parameters $\alpha$ and $n$.

\begin{figure}[htbp]
\subfigure[ Plot of $P(k)$ for $n>1$, $\alpha=10$ and various values of $k$]{\includegraphics[scale=0.45]{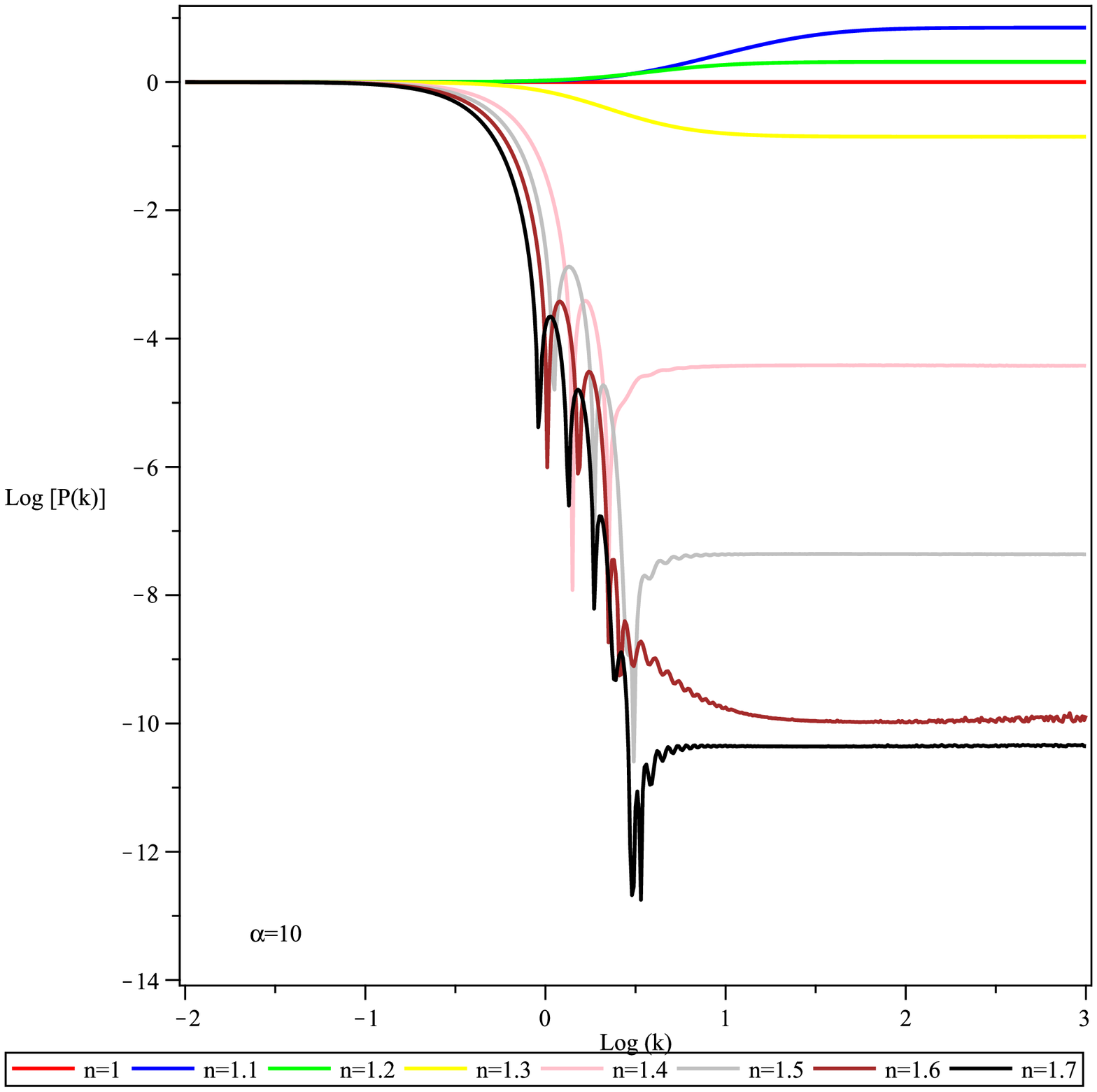}\label{PkRRnX1}}
\subfigure[ Plot of $P(k)$ for $n>1$, $\alpha=1$ and various values of $k$]{\includegraphics[scale=0.45]{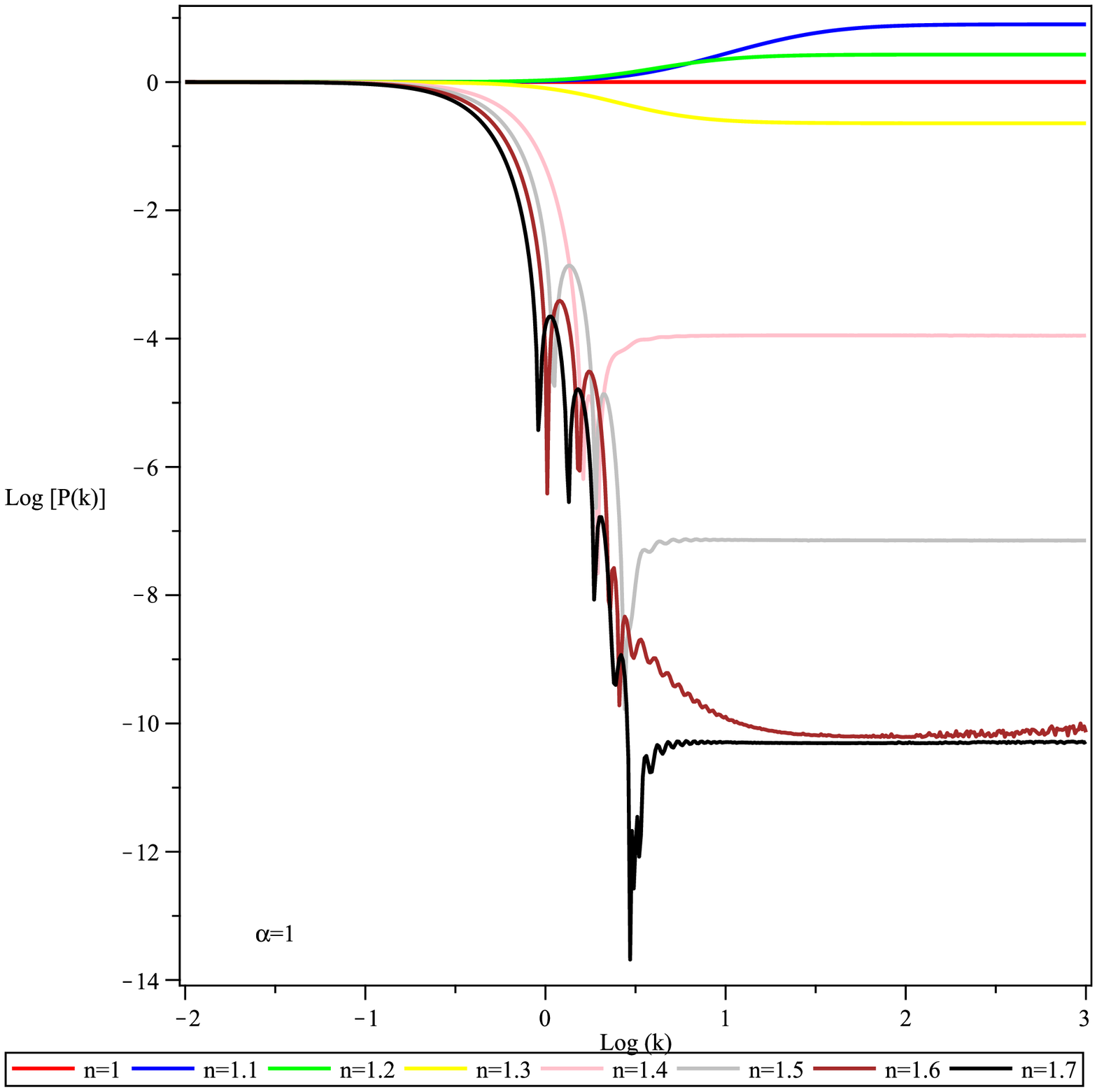}\label{PkRRnX2}}
\subfigure[ Plot of $P(k)$ for $n>1$, $\alpha=0.1$ and various values of $k$]{\includegraphics[scale=0.45]{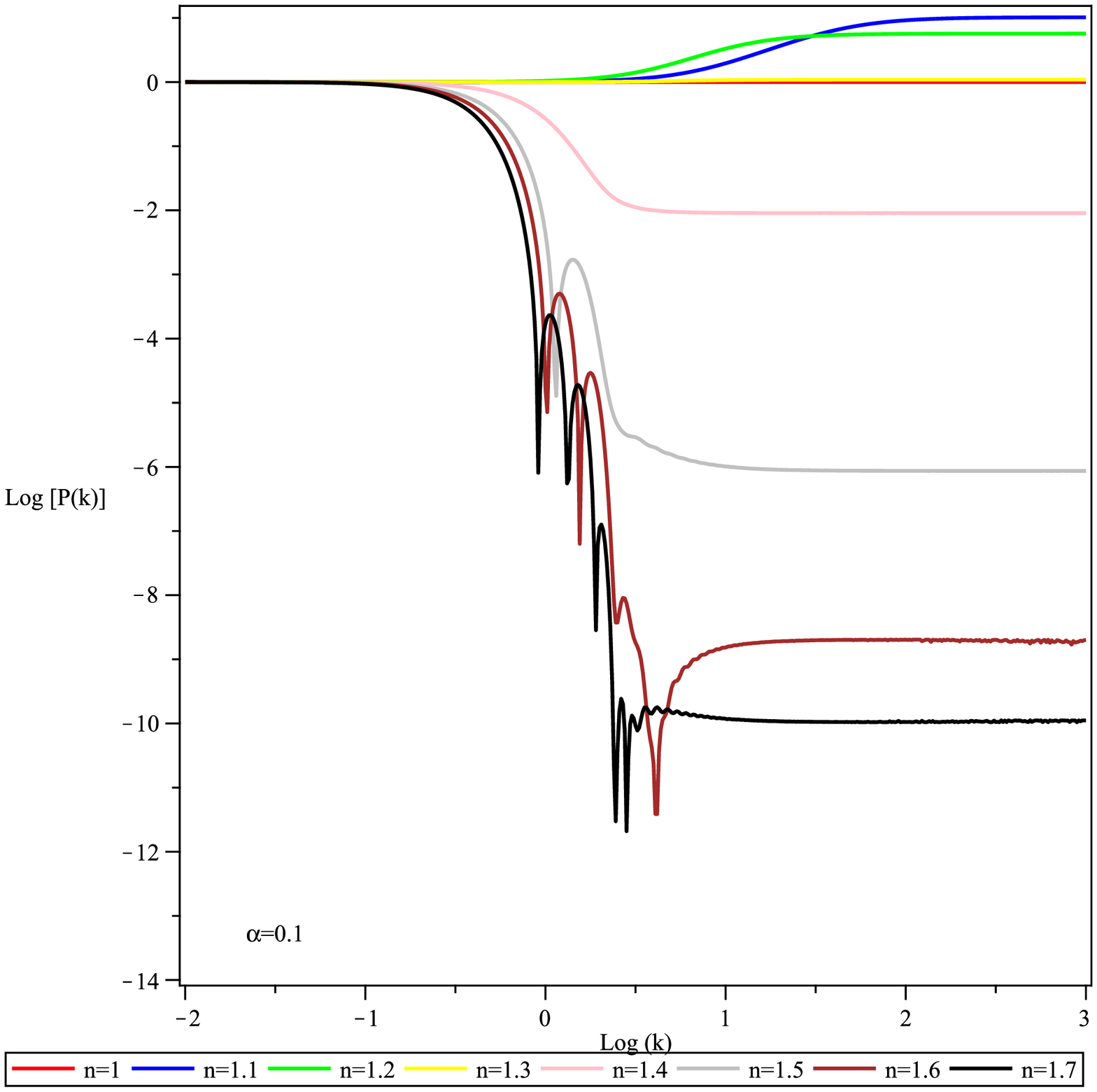}\label{PkRRnX3}}
\subfigure[ Plot of $P(k)$ for $n>1$, $\alpha=0.01$ and various values of $k$]{\includegraphics[scale=0.45]{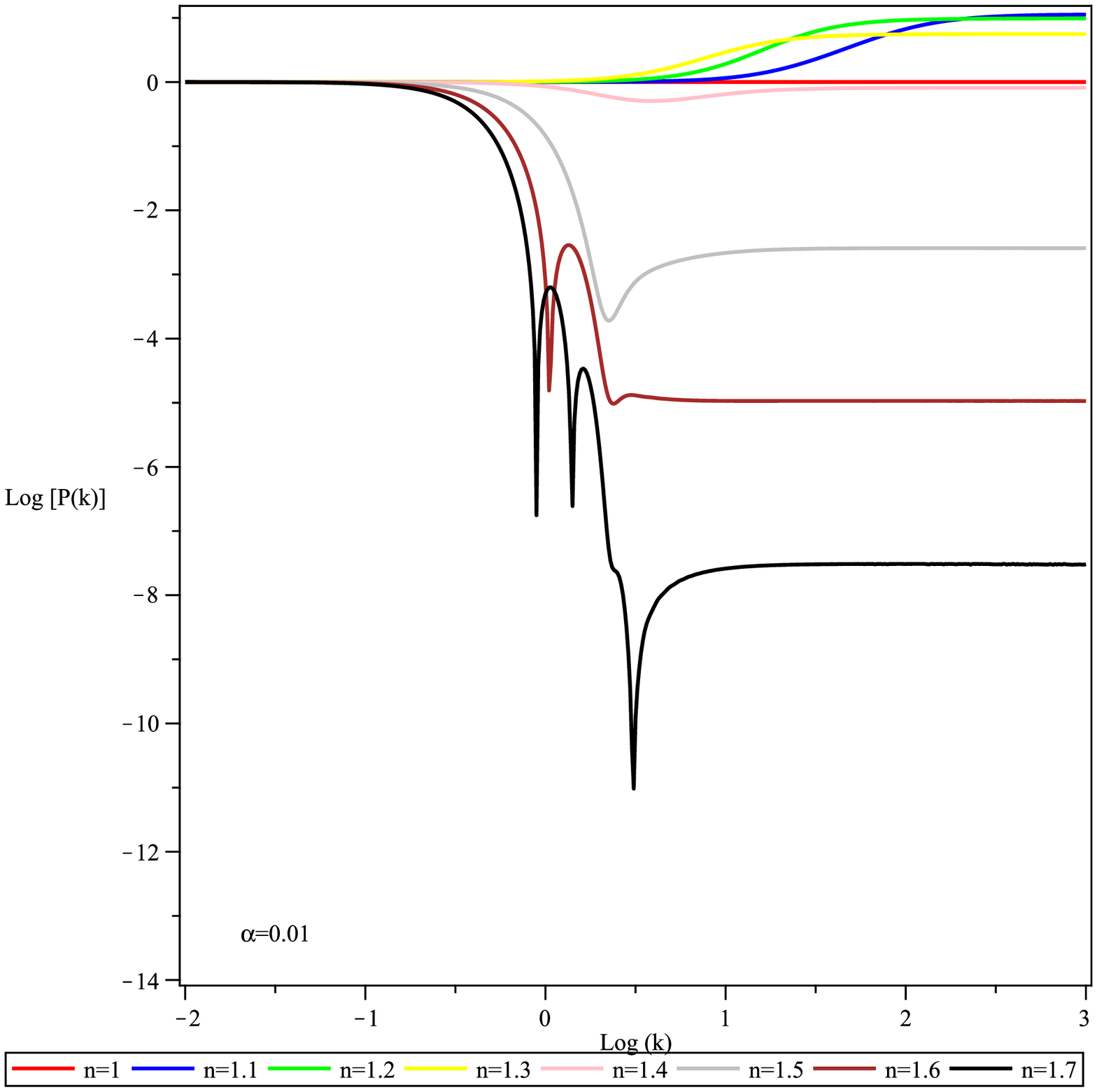}\label{PkRRnX7}}
\caption{Plot of the Power spectrum as a function of $k$ for $R+\alpha R^n$-gravity at $\tau=1$ for $n>1$.}\label{PkRRn}
\end{figure}

These features characterize the spectra shown in Figures \ref{PkRRn}. As one can clearly see, these plots resemble the ones we have  derived in the previous example. There are, of course, differences in the position of the oscillations and the amount of the power drop, but one finds again three different regimes in the case of dust and two of them ($k\rightarrow 0$ and $k\rightarrow \infty$ ) correspond to scale invariance. Particularly interesting is the fact that in principle the values of $\alpha$ and $n$ can be fine tuned in such a  way  to obtain a spectrum in which the small scales have the same power as the large ones. In a situation like this most of the spectrum would be scale invariant and all the deviations would be concentrated around a specific scale.

\begin{figure}[htbp]
\subfigure[Plot of $P(k)$ for $n=1.4$, $\alpha=10$ and evaluated at various values of $\tau$]{\includegraphics[scale=0.40]{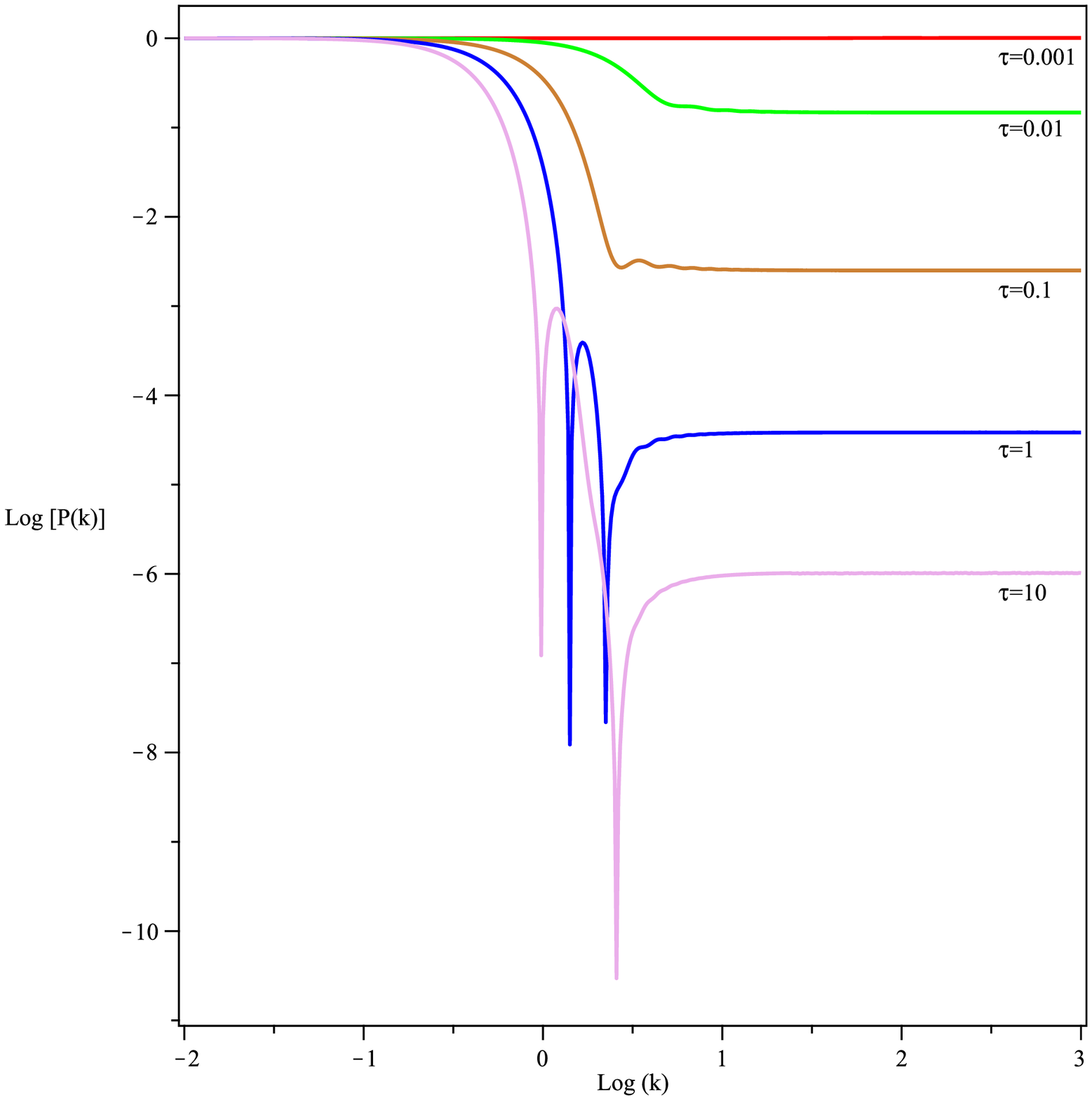}\label{Fig33}}
\subfigure[Plot of $P(k)$ for $n=1.4$, $\alpha=0.01$ and evaluated at various values of $\tau$]{\includegraphics[scale=0.40]{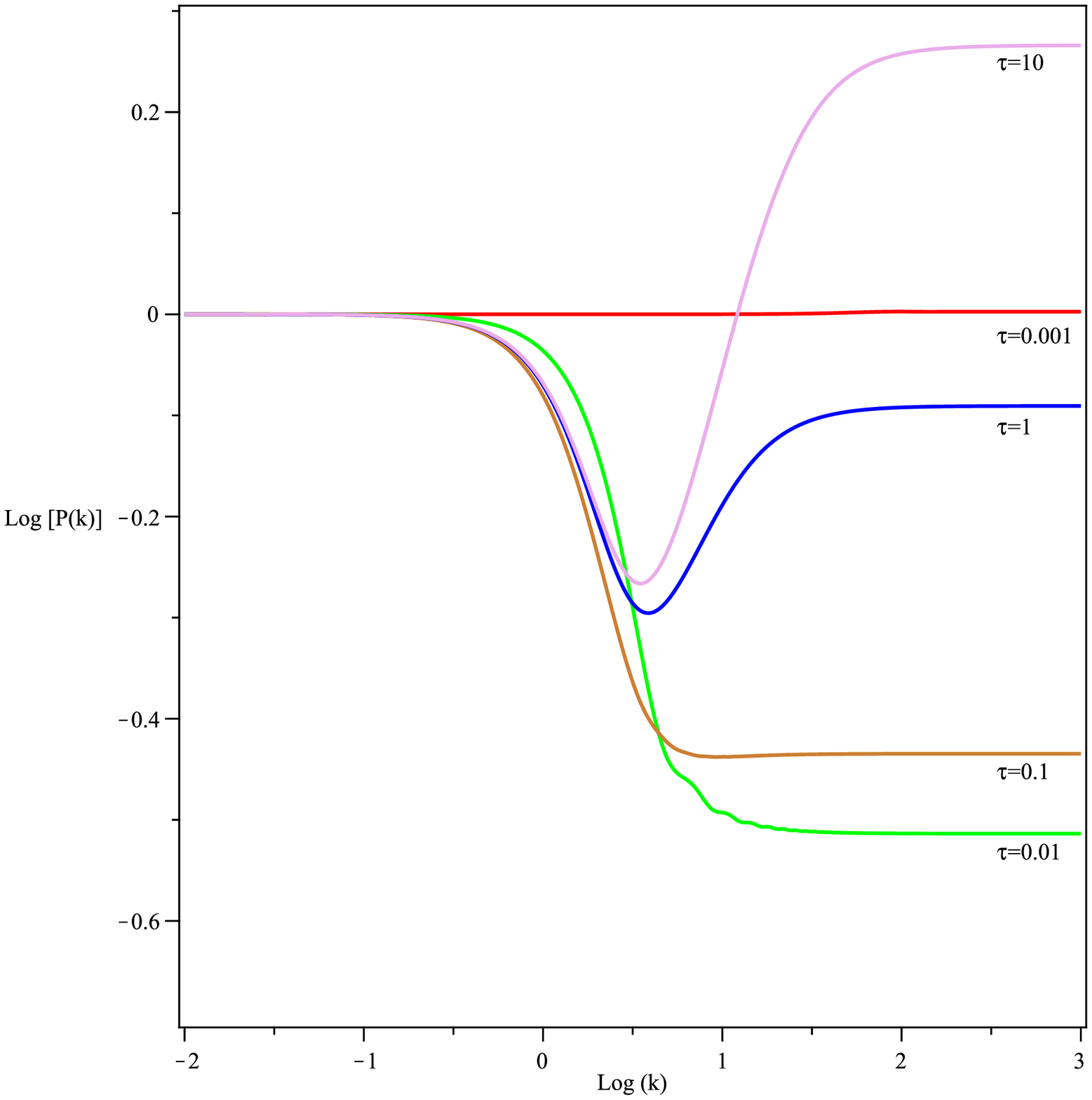}\label{Fig34}}
\caption{The time evolution of the Power spectrum in $R+\alpha R^n$-gravity for $n=1.4$ and differing values of $\alpha$. Note the
drastically differing vertical scales in the plots. Additionally, note the rise in power at small scales at late times in the case of
$n=1.4$ and $\alpha=0.01$.}
\end{figure}

The time evolution of these spectra also reveals some interesting insight into the dynamics of the matter fluctuations. As usual for large values of $\alpha$ the evolution is very similar to the one obtained for $R^n$-gravity  as it is shown in Figure \ref{Fig33}. However, when the value of the coupling changes the behavior of the perturbations can change dramatically. An example is given in figure \ref{Fig34} in which the evolution for the power spectrum of the model $(n=1.4,\alpha=00.1)$ in which the small scale perturbations are first dissipated and successively start to grow again. This means that in principle one could choose $n$ and $\alpha$, such that for example the small scale perturbation grow at different rates at different times. This property could be useful in  the resolution of open problems in GR structure formation, like the cosmological dark matter or the excess of red galaxies.

In conclusion,  in spite of all the differences in the dynamics of  perturbations, the power spectrum in this class of model seems to preserve most of the main structure of the one in $R^n$-gravity. This implies  that all the considerations made in the previous section concerning the physical mechanisms behind the form of the spectrum  can be made also in this case. This result was not expected and suggest that  we might have encountered a characteristic signature of $f(R)$-gravity which would be crucial to investigate the validity of these models.

\section{Discussion and Conclusion}
In this paper we have discussed the properties of scalar perturbations in fourth order gravity and described some useful methods for extracting physical information from the complex equations that govern their evolution. These tools are devised in  such a way to gain as clear an understanding as possible of the behavior of scalar perturbations on all scales and to facilitate a direct comparison of these results with the corresponding ones in GR. Two simple models:  $f(R)=R^{n}$ and $f(R)=R+\alpha R^{n}$ were analyzed in detail because of their simplicity and because  we have a relatively good understanding of their background via the dynamical systems approach.

The results obtained show profound differences in the dynamics of the perturbations compared to what occurs in GR: in these models we find the growth rate of the perturbations is in general different from the GR one and always dependent on the scale. This implies, for example, that depending on the value of the parameters the perturbation can grow or dissipate (or both) at different rates, with obvious consequences for the global cosmic history.

The fourth order system of differential equations that governs the behavior of the scalar perturbations of these models, in particular when written in terms of   $\Delta_m$ and $\Delta_R$, has a structure that resembles closely one of a two fluids GR model. Although the $\Delta_m$ and $\Delta_R$  system of equations is enormously more complicated than (\ref{EqPerIIOrd1}-\ref{EqPerIIOrd2}), one can still use it to get an idea of the nature of the interaction between the non-Einstein part of the gravitational interaction and standard matter.

In section VII we found that the wavenumber structure of these coefficients is such that  for very large or very small $k$ they become scale invariant.  This implies, in turn,  that the matter power spectrum is  scale invariant for $k\rightarrow 0,\infty$ and can present some characteristic features on scales that depend on the different parameters of the model. Another way of interpreting the form of the spectrum without necessarily using the curvature fluid idea is to interpret
$\mathcal{R}$ and  $\Re$ as being associated with the propagation of the scalar degree of freedom of the theory, or a scalar gravitational mode, whose interaction with matter is able to emulate the effect of a relativistic component, which, like photons in GR, induces power loss in the oscillations that appear in the spectrum.

The picture that emerges from our results is that in our examples, fourth order gravity influences deeply the structure formation process, but the modifications are only detectable around a specific value of $k$. Everywhere else in $k$ space the results are very close (although dynamically different) to GR. This is particularly interesting since it means that for a suitable choice of values for the parameters, the oscillations can be positioned beyond the observational boundary of currently available data. Hence our results seems to imply that not only could these models be compatible with the observed matter power spectrum, but that we also have a systematic way of constraining their parameters using data coming from  a range of scales, for example by combining CMB and LSS data \cite{WMAP, SDSS}.

It is also worth commenting briefly about the compatibility of our analysis with the existing literature. For example in  \cite{Li:2008ai} a class of models which is very similar to the one we analyzed in Section \ref{SecRRn} is considered. Using a specific background the authors showed that the matter power spectrum  is characterized by an excess of power at small scales when the theory is very close to $\Lambda$CDM. A similar result is found in \cite{HuSawicki}  in which classes of models are considered which  contain additive corrections to the Hilbert -Einstein action (i.e. they have the form  $R+g(R)$ ). These general corrections are parameterized by  a quantity $B$, which measures the deviation from GR. Instead in \cite{Bertschinger:2008zb}, generic modifications are considered (both scale-dependent and scale-independent), using a more flexible parameterization. In the sub-case of scale dependent modifications (like in the $f(R)$ case) examples are given in which one finds  once again excess of power on small scales in the power spectrum. Also one should note that in the same sub-case examples were given when one finds a deficit in power.

Interestingly, both the examples considered in this paper exhibit  excess power on small scales  for $n\rightarrow 1^{+}$ in the case of $R^{n}$-gravity and for $n\rightarrow 1^{+}$  and $\alpha\ll 1$ in the case $R+\alpha R^n$. This indicates that, for small corrections  to GR, excess power at large scale  seems to be a generic feature in $f(R)$-gravity. However there are situations in which one might want to analyze theories which are not necessarily close to GR. For example in \cite{Salv06,Salv07,Salucci} a fit of  $R^n$-gravity  with the data coming from the rotation curves of galaxies and supernovae type Ia leads to values of $n$ in the range $[1.7, 3.5]$. For these values of $n$, both our examples indicate a loss of power  at small scales, which means that we are provided with an opportunity to rule out this model by testing it against available data. Although the presence of an excess or deficit of power on small scales seems closely related  to the value of the specific parameters of the model itself, there are  some indications that other features of the spectrum  found in our examples  are indeed general. For example, the $k$ structure of the general equations (\ref{EqPerIIOrdParGen1}-\ref{EqPerIIOrdParGen2}) suggests that the evolution of perturbations in a generic $f(R)$ theory  presents at least three different regimes. Also looking at the derivation of the  $(\Delta_m, \Delta_R)$ equations (which can be performed in general, provided a sufficiently large amount of paper and time) one realizes that the $k$ dependence of the dissipation and source coefficients we have found in our examples is expected to be common to any $f(R)$ Lagrangian because it originates from  the $\3nab^2\3nab^2$ terms in the perturbation equations.

We end by commenting that these results together with the dynamical systems analysis of the background cosmological history presented in other papers \cite{OurDynSys,SanteGenDynSys} provides a unified and consistent approach to the combined study of FLRW observational constraints and a complete analysis of linear structure growth in the context of $f(R)$ gravity.

\vfill {\noindent{\bf Acknowledgements:}\\
The authors wish to thank Dr J. Larena for useful discussion and suggestions. SC wish to thanks J Donkers for useful discussion and support during the development of this paper. KNA and SC are supported by Claude Leon Foundation fellowships. This work was supported by the National Research Foundation (South
Africa) and the {\it Ministrero degli Affari Esteri - DIG per la
Promozione e Cooperazione Culturale} (Italy) under the joint
Italy/South Africa science and technology agreement.
\newpage
\appendix
\section{General propagation and constraint equations of the 1+3 covariant formalism.}\label{CovID}
\noindent Expansion propagation (generalized Raychaudhuri equation):
\begin{eqnarray}\label{1+3eqRayHO}
&&\dot{\Theta}+{\textstyle\frac{1}{3}}\Theta^2+\sigma_{{{a}}{{b}}}
\sigma^{{{a}}{{b}}} -2\omega_{{a}}\omega^{{a}} -\tilde{\nabla}^a
\dot{u}_{{a}}+ \dot{u}_{{a}}
\dot{u}^{{a}}+{\textstyle\frac{1}{2}}(\tilde{\mu}^{m} +
3\tilde{p}^{m}) =-{\textstyle\frac{1}{2}}({\mu}^{R} +
3{p}^{R})\;.
\end{eqnarray}
Vorticity propagation:
\begin{equation}\label{1+3VorHO}
\dot{\omega}_{\langle {{a}}\rangle }
+{\textstyle\frac{2}{3}}\Theta\omega_{{a}} +{\textstyle\frac{1}{2}}\curl
\dot{u}_{{a}} -\sigma_{{{a}}{{b}}}\omega^{{b}}=0 \;.
\end{equation}
Shear propagation:
\begin{equation}\label{1+3ShearHO}
\dot{\sigma}_{\langle {{a}}{{b}} \rangle }
+{\textstyle\frac{2}{3}}\Theta\sigma_{{{a}}{{b}}}
+E_{{{a}}{{b}}}-\D_{\langle {{a}}}\dot{u}_{{{b}}\rangle }
+\sigma_{{c}\langle {{a}}}\sigma_{{{b}}\rangle }{}^{c}+
\omega_{\langle {{a}}}\omega_{{{b}}\rangle} - \dot{u}_{\langle
{{a}}}\dot{u}_{{{b}}\rangle}
\,=\,{\textstyle\frac{1}{2}}\pi^{R}_{{{a}}{{b}}}\;.
\end{equation}
Gravito-electric propagation:
\begin{eqnarray}\label{1+3GrElHO}
 && \dot{E}_{\langle {{a}}{{b}} \rangle }
+\Theta E_{{{a}}{{b}}} -\curl H_{{{a}}{{b}}}
+{\textstyle\frac{1}{2}}(\tilde{\mu}^{m}+\tilde{p}^{m})\sigma_{{{a}}{{b}}}
-2\dot{u}^{c}\ep_{{c}{d}({{a}}}H_{{{b}})}{}^{d} -3\sigma_{{c}\langle
{{a}}}E_{{{b}}\rangle }{}^{c} +\omega^{c}
\ep_{{c}{d}({{a}}}E_{{{b}})}{}^{d}
\nonumber\\&&~~{}=-{\textstyle\frac{1}{2}}(\mu^{R}+p^{R})\sigma_{{{a}}{{b}}}
-{\textstyle\frac{1}{2}}\dot{\pi}^{R}_{\langle {{a}}{{b}}\rangle  }
-{\textstyle\frac{1}{2}}\D_{\langle {{a}}}q^{R}_{{{b}}\rangle }
-{\textstyle\frac{1}{6}}
\Theta\pi^{R}_{{{a}}{{b}}}-{\textstyle\frac{1}{2}}\sigma^{c}{}_{\langle
{{a}}}\pi^{R}_{{{b}}\rangle {c}} -{\textstyle\frac{1}{2}}
\omega^{c}\ep_{{c}({{a}}}^{d}\pi^{R}_{b )d}\;.
\end{eqnarray}
Gravito-magnetic propagation:
\begin{eqnarray}\label{1+3GrMagHO}
 &&\dot{H}_{\langle
{{a}}{{b}} \rangle } +\Theta H_{{{a}}{{b}}} +\curl E_{{{a}}{{b}}}-
3\sigma_{{c}\langle {{a}}}H_{{{b}}\rangle }{}^{c} +\omega^{c}
\ep_{{c}{d}({{a}}}H_{{{b}})}{}^{d}
+2\dot{u}^{c}\ep_{{c}{d}({{a}}}E_{{{b}})}{}^{d}
\nonumber\\&&~~{}={\textstyle\frac{1}{2}}\curl\pi^{R}_{{{a}}{{b}}}-{\textstyle\frac{3}{2}}\omega_{\langle
{{a}}}q^{R}_{{{b}}\rangle
}+{\textstyle\frac{1}{2}}\sigma^{c}{}_{({{a}}}
\ep_{{{b}}){c}}^{\;\;\;\;d}q^{R}_{d}\;.
\end{eqnarray}
Vorticity constraint:
\begin{equation}\label{1+3VorConstrHO}
\D^{{a}}\omega_{{a}} -\dot{u}^{{a}}\omega_{{a}} =0\;.
\end{equation}
Shear constraint:
\begin{equation}\label{1+3ShearConstrHO}
\D^{{b}}\sigma_{{{a}}{{b}}}-\curl\omega_{{a}}
-{\textstyle\frac{2}{3}}\D_{{a}}\Theta +2[\omega,\dot{u}]_{{a}} =
-q^{R}_{a}\;.
\end{equation}
Gravito-magnetic constraint:
\begin{equation}\label{1+3GrMagConstrHO}
 \curl\sigma_{{{a}}{{b}}}+\D_{\langle {{a}}}\omega_{{{b}}\rangle  }
 -H_{{{a}}{{b}}}+2\dot{u}_{\langle {{a}}}
\omega_{{{b}}\rangle  }=0 \;.
\end{equation}
Gravito-electric divergence:
\begin{eqnarray}\label{1+3GrElConstrHO}
&& \D^{{b}} E_{{{a}}{{b}}}
-{\textstyle\frac{1}{3}}\D_{{a}}\tilde{\mu}^{m} -[\sigma,H]_{{a}}
+3H_{{{a}}{{b}}}\omega^{{b}}={\textstyle\frac{1}{2}}\sigma_{{{a}}}^{{b}}q^{R}_{{b}}-
{\textstyle\frac{3}{2}}
[\omega,q^{R}]_{{a}}-{\textstyle\frac{1}{2}}\D^{{b}}\pi^{R}_{{{a}}{{b}}}
 +{\textstyle\frac{1}{3}}\D_{{a}}\mu^{R}
-{\textstyle\frac{1}{3}}\Theta q^{R}_{{a}}\;.
\end{eqnarray}
Gravito-magnetic divergence:
\begin{eqnarray}\label{1+3GrMagDivHO}
 &&\D^{{b}} H_{{{a}}{{b}}}
-(\tilde{\mu}^{m}+\tilde{p}^{m})\omega_{{a}} +[\sigma,E]_{{a}}
 -3E_{{{a}}{{b}}}\omega^{{b}}=-{\textstyle\frac{1}{2}}\curl q^{R}_{{a}}
+(\mu^{R}+p^{R})\omega_{{a}} -{\textstyle\frac{1}{2}}
[\sigma,\pi^{R}]_{{a}} -{\textstyle\frac{1}{2}}\pi^{R}_{{{a}}{{b}}}
\omega^{{b}}\;.
\end{eqnarray}
Standard Matter Conservation
\begin{eqnarray}
&&\dot{\mu}^m\,=\, - \,\Theta\,(\mu^m+{p^m})\;,\label{eq:cons1}\\
&&\3nab^{a}{p^m} =  - (\mu^m+{p^m})\,\udot^{a}\,.
\end{eqnarray}
Curvature fluid Conservation
\begin{eqnarray}
&&\l{eq:cons2} \dot{\mu^R} + \3nab^{a}q^R_{a} = - \,\Th\,(\mu^R+p^R)
- 2\,(\udot^{a}q^R_{a}) -
(\sig^{a}\!^{b}\pi^R_{b}\!_{a})+\mu^{m}\frac{f''\,\dot{R}}{f'^{2}}\;,\\
&& \l{eq:cons3} \dot{q}^R_{\lgl a\rgl} + \3nab_{a}p^R +
\3nab^{b}\pi^R_{ab} = - \,{\textstyle\frac{4}{3}}\,\Th\,q^R_{a} -
\sig_{a}\!^{b}\,q^R_{b} - (\mu^R+p^R)\,\udot_{a} -
\udot^{b}\,\pi^R_{ab} -
\eta_{a}^{bc}\,\om_{b}\,q^R_{c}+\mu^{m}\frac{f''\,\D_{a}{R}}{f'^{2}}
\  \end{eqnarray}

As usual angle brackets applied to a vector  denote
the projection of this vector on the tangent 3-spaces
\begin{equation}
V_{\langle{{a}}\rangle}=h_{{a}}{}^{{b}} V_{{b}}\;.
\end{equation}
Instead when applied to a tensor they denote the projected,
symmetric and trace free part of this object
\begin{equation}
W_{\langle{{a}}{{b}}\rangle}=\left[h_{({{a}}}{}^{c}
h_{{{b}})}{}^{d}-
{\textstyle\frac{1}{3}}h^{{c}{d}}h_{{{a}}{{b}}}\right]W_{{c}{d}}\,.
\end{equation}
The spatial curl of a vector and a tensor is
\begin{equation}
(\c\,X)^{a} = \epsilon^{abc}\,\3nab_{b}X_{c}\qquad \qquad (\c\,X)^{ab} = \epsilon^{cd\lgl a}\,\3nab_{c}X^{b\rgl}\!_{d}
\end{equation}
respectively, where $\epsilon_{abc}=u^d\eta_{abcd}$ is the spatial volume.
Finally $\omega_{{a}}=\frac{1}{2}\ep_{a}{}^{{b}{c}}\omega_{bc}$ and the
covariant tensor commutator is
\[
[W,Z]_{{a}} =\ep_{{{a}}{c}{d}}W^{c}{}_{e} Z^{{d}{e}}\,.
\]

The 1+3 equations above are completely equivalent to the Einstein equation and govern the
dynamics of the matter and gravitational fields in fourth order gravity. As we will see the  new source terms in their R.H.S. will  modify the evolution of the perturbations in a  non-trivial way. The standard GR equations are obtained by setting $f(R)=R$ which corresponds to setting all these sources to zero.
\section{The Covariant Gauge Invariant gradient equations}\label{App2}
In the following we give, for completeness, the equations for the evolution of the gradient variables
\begin{equation}\label{s3}
{\cal D}^{m}_{{a}}=\frac{S}{\mu^{m}}\D_{{a}}\mu^{m}\,,\qquad
Z_{{a}}=S\D_{{a}}\Theta\,,\qquad C_{{a}}=S\D_{{a}}\tilde{R}\;,
\qquad{\cal R}_{{a}}=S\D_{{a}} R\,,\qquad\Re_a=S\D_{{a}} \dot{R}\;.
\end{equation}

They read
\begin{eqnarray}
\dot{{\cal D}}^m_{{a}} &=&w\Theta{\cal D}^m_{{a}}-(1+w)Z_{{a}}\,,\label{s5}\\
\nn\dot{Z}_{{a}} &=& \left(\frac{\dot{R}f''}{f'}-\frac{2 \Theta }{3}\right)Z_a+
   \left[\frac{3 (w -1) (3 w +2)}{6 (w +1)} \frac{\mu}{ f'} + \frac{2 w \Theta ^2
   +3 w (\mu^{R}+3  p^{R}) }{6 (w +1) }+\frac{2 w}{w +1}\frac{k}{S^2}\right]
   {\cal D}^m_{a}+\frac{\Theta f''}{2f'}\Re_{a}\\&&+
   \left[\frac{1}{2}+2\frac{f''}{f'}\frac{k}{S^2}-\frac{1}{2} \frac{f}{f'}\frac{ f''}{f'}+ \frac{f''}{f'} \frac{\mu}{
   f'} + \dot{R} \Theta  \left(\frac{f''}{f'}\right)^{2}+ \dot{R} \Theta \frac{ f^{(3)}}{ f'}\right]\mathcal{R}_a
   -\frac{w}{w +1} \3nab^{2}{\cal
   D}^m_{a}-\frac{ f''}{ f'}\3nab^{2}\mathcal{R}_{a}\,,\\
\dot{{\cal R}}_a&=&\Re_{a}-\frac{w }{w +1}\dot{R}\;{\cal D}^m_{{a}}\,,\label{eqZa}\\
\nn\dot{\Re}_a&=&- \left(\Theta + 2\dot{R} \frac{
   f^{(3)}}{f''}\right)\Re_{a}- \dot{R} Z_{a} -
   \left[\frac{ (3 w -1)}{3} \frac{\mu}{f''} + 3\frac{w}{w +1}
   (p^{R}+\mu^{R}) \frac{f'}{f''}+ \frac{w}{3(w +1)} \dot{R} \left(\Theta
   -3 \dot{R} \frac{f^{(3)}}{f''}\right)\right]{\cal D}^m_{{a}}\\&&\nn+\left[\frac{3}{2} (1+w) \frac{k}{S^2}-\left(
   \frac{1}{3}\frac{f'}{f''}+\frac{f^{(4)}}{f'} \dot{R}^2+\Theta  \frac{f^{(3)}}{f'}
   \dot{R}-\frac{2}{9} \Theta ^2 +\frac{1}{3}(\mu^{R}+3 p^{R}) + \ddot{R} \frac{f^{(3)}}{f''}\right.\right.\\&&\left.\left.
   - \frac{1}{6}\frac{f}{f'}+\frac{1}{2} (w +1) \frac{\mu}{f'} -\frac{1}{3} \dot{R} \Theta
 \frac{f''}{f'}\right)\right]\mathcal{R}_{a}+\3nab^{2}\mathcal{R}_{a}\,,
\end{eqnarray}
together with the constraint
\begin{equation}\label{Gauss}
  \frac{C_a}{S^2}+ \left(\frac{4  }{3}\Theta +\frac{2 \dot{R}
   f''}{f'}\right) Z_a-2\frac{
    \mu }{f'}{\cal D}^m_{{a}}+ \left[2 \dot{R}
   \Theta  \frac{f^{(3)}}{ f'}-\frac{f''}{ f'^{2}} \left(f-2 \mu +2
   \dot{R} \Theta  f''-4\frac{k}{S^{2}}\right)\right]\mathcal{R}_a+\frac{2 \Theta
   f''}{f'}\Re_a-\frac{2 f''}{f'}\3nab^{2}\mathcal{R}_a=0\,.
\end{equation}
The propagation equation for the variable C is
\begin{eqnarray}
&& \dot{C}_a=\nonumber \frac{6 k^2 }{S^2 \Theta } \left(5\frac{ f'' }{
   f'}\mathcal{R}_a-3\dd^{m}_a \right)
   +k\left\{\frac{3}{S^2
   \Theta  }C_a +
   \left(\frac{ 6
   \mu^{R}}{ \Theta  }-\frac{2 (3\omega)  \Theta }{3(\omega +1)}\right)\dd^m_a-\frac{6  f''}{ \Theta f'}\3nab^{2} \mathcal{R}_a\right.\\&&\left.\nonumber+\left[-\frac{6 R' f''^2}{f'^2}+\frac{\left(2
   \left(\Theta ^2-3 \mu^R\right)
   f'-3 f\right) f''}{\Theta f'^2}+\frac{6 R' f^{(3)}}{f'}\right] \mathcal{R}_a\right\} \\&&+\3nab^{2}\left[\frac{4 \omega
    S^2 \Theta }{3 (\omega +1)}\dd^m_a+\frac{2  S^2
   f''}{f'}\Re_a-\frac{2 S^2 \left(\Theta
    f''-3 \dot{R} f^{(3)}\right)}{3 f'}\mathcal{R}_a\right]\,,
\end{eqnarray}
These equations were already given in \cite{SantePert}, but contain some typos  that we have been corrected in the equations above.
\section{Perturbation equations in the case $f(R)=\chi R^n$}\label{App3}
The general perturbation equations in the case $f(R)=\chi R^n$ read
\begin{eqnarray}
   &&\ddot{\Delta}_{m}^{(k)}+\mathcal{A}\; \Theta\;\dot{\Delta}_{m}^{(k)}-\mathcal{B}\;\Theta^{2}\;\Delta_{m}^{(k)}=
  \mathcal{C} \mathcal{R}^{(k)} +\mathcal{D}\;\Theta^{-1}\;\dot{\mathcal{R}}^{(k)}\,,\label{EqPerIIOrdRn1}\\&&
   \ddot{\mathcal{R}}^{(k)}+\mathcal{E}\,\Theta\,
   \dot{\mathcal{R}}^{(k)}-\mathcal{F}\;\Theta^{2}\;\mathcal{R}^{(k)}=-
   \mathcal{G} \;\Theta ^4 \Delta_{m}^{(k)}-\mathcal{H}\,\Theta^{3}\, \dot{\Delta}_m^{(k)}\label{EqPerIIOrdRn2}\,,
\end{eqnarray}
where
 \begin{eqnarray}
&&\mathcal{A}=w +\frac{(n-1) \left(j+q+18 \Omega _K-2\right)}{3 \left(q-9 \Omega
   _K+1\right)}-\frac{2}{3}\,,\\
&&\nn\mathcal{B}=\frac{w  k^2}{S^2 \Theta^2}+\frac{2 \left(3 (n-2)
   w ^2+(n-1) w -n+2\right) \left(q-9 \Omega _K+1\right)}{3 n} \\&&\nonumber-\frac{(n-1) \left(3 w
   ^2+w -1\right) \left(j+q+18 \Omega _K-2\right)}{q-9 \Omega _K+1}-\frac{((n-3) n+2) \left(3 w ^2+w -1\right)
   \left(j+q+18 \Omega _K-2\right){}^2}{3 \left(q-9 \Omega _K+1\right){}^2} \\&&
   -\frac{(n-1) \left(3 w
   ^2+w -1\right) \left((q-8) q+s+18 (q-3) \Omega _K+6\right)}{3 \left(q-9 \Omega
   _K+1\right)}\,,\\
&&\nn\mathcal{C}= \frac{3 (n-1) (w +1)}{2 \left(q-9 \Omega _K+1\right)}\left[\frac{k^2}{S^2 \Theta ^2}+\frac{(2 (n-3) n+3) \left(q-9 \Omega _K+1\right)}{3 (n-1) n}-\frac{(n (3 n-7)+4)
   \left(j+q+18 \Omega _K-2\right)}{3 (n-1) \left(q-9 \Omega _K+1\right)}\right.\\ && \left.-\frac{(n-2) (n-1) \left(j+q+18 \Omega _K-2\right){}^2}{3 \left(q-9 \Omega
   _K+1\right){}^2}-\frac{(n-1) \left((q-8)
   q+s+18 (q-3) \Omega _K+6\right)}{3 \left(q-9 \Omega _K+1\right)}\right]\,,\\
&&\mathcal{D}=\frac{3 (n-1) (w +1)}{2 \left(q-9 \Omega _K+1\right) }\,,
\end{eqnarray}
\begin{eqnarray}
&&\mathcal{E}=1+\frac{2 (n-2) \left(j+q+18 \Omega _K-2\right)}{3 \left(q-9 \Omega _K+1\right)}\,,\\
&&\nn\mathcal{F}=\frac{k^2}{S^2 \Theta^2}+\frac{(n-2) (2 n+3 (n-1) w ) \left(q-9
   \Omega _K+1\right)}{9 (n-1) n}-\frac{(-3 w +n (3 w +2)-4) \left(j+q+18 \Omega
   _K-2\right)}{6 \left(q-9 \Omega _K+1\right)}\\ &&\frac{(2-n) (n (3 w +2)-3 (w +2)) \left(j+q+18 \Omega
   _K-2\right){}^2}{18 \left(q-9 \Omega _K+1\right){}^2} +\frac{(3 w+4-n (3 w +2)) \left((q-8)
   q+s+18 (q-3) \Omega _K+6\right)}{18 \left(q-9 \Omega _K+1\right)}\,,\\
&&\nn \mathcal{G}=\frac{2}{27}\left[\frac{2 (n-2) (3 w -1) \left(q-9 \Omega _K+1\right){}^2}{(n-1)
   n}-\frac{3 (w  (3 w +4)-1) \left(j+q+18 \Omega _K-2\right)}{w +1}\right.\\ && \left.-\frac{(n-2) (w  (3
   w +4)-1) \left(j+q+18 \Omega _K-2\right){}^2}{(w +1) \left(q-9 \Omega
   _K+1\right)}-\frac{(w
   (3 w +4)-1) \left((q-8) q+s+18 (q-3) \Omega _K+6\right)}{w +1}\right]\,,\\
&&\mathcal{H}=\frac{2 (w -1) \left(j+q+18 \Omega _K-2\right)}{9 (w +1)}\,.
\end{eqnarray}
\section{Perturbation equations in the case $f(R)=R+\alpha R^n$ }\label{App4}

The general perturbation equations in the case $f(R)=R+\alpha R^n$ read
\begin{eqnarray}
   &&\ddot{\Delta}_{m}^{(k)}+\mathcal{A}\; \Theta\;\dot{\Delta}_{m}^{(k)}-\mathcal{B}\;\Theta^{2}\;\Delta_{m}^{(k)}=
  \mathcal{C} \mathcal{R}^{(k)} +\mathcal{D}\;\Theta^{-1}\;\dot{\mathcal{R}}^{(k)} \,, \label{EqPerIIOrdRRn1}\\&&
    \ddot{\mathcal{R}}^{(k)}+\mathcal{E}\,\Theta\,
   \dot{\mathcal{R}}^{(k)}-\mathcal{F}\;\Theta^{2}\;\mathcal{R}^{(k)}=-
   \mathcal{G} \;\Theta ^4 \Delta_{m}^{(k)}-\mathcal{H}\,\Theta^{3}\, \dot{\Delta}_{m}^{(k)} \label{EqPerIIOrdRRn2}\,,
\end{eqnarray}
where
\begin{eqnarray}
&&\mathcal{A}=w-\frac{2}{3} +\frac{4 \left(j+q+18 \Omega _K-2\right)}{3 \left(3 k_{\alpha }+4 \left(q-9 \Omega
   _K+1\right)\right)}\,,\\
&&\nn\mathcal{B}=\frac{w  k^2}{S^2  \Theta^2}+\frac{2 \left(\left(3-9
   w ^2\right) k_{\alpha }+2 w  \left(q-9 \Omega _K+1\right)\right) \left(q-9 \Omega
   _K+1\right)}{3\left(3 k_{\alpha }+4 \left(q-9 \Omega _K+1\right)\right)}-\frac{4
   \left(3 w ^2+w -1\right) \left(j+q+18 \Omega _K-2\right)}{ \left(3
   k_{\alpha }+4 \left(q-9 \Omega _K+1\right)\right)}\\&&-\frac{4 \left(3 w ^2+w -1\right)
   \left((q-8) q+s+18 (q-3) \Omega _K+6\right)}{3 \left(3 k_{\alpha }+4 \left(q-9
   \Omega _K+1\right)\right)}\,,\\
&&\nn\mathcal{C}=\frac{6 (w +1) }{3
   k_{\alpha }+4 \left(q-9 \Omega _K+1\right)} \left[\frac{k^2}{S^2 \Theta^2}+\frac{1}{24}
   \left(\frac{27 k_{\alpha }^2}{4 q+3 k_{\alpha }-36 \Omega _K+4}-4 \left(q-9 \Omega
   _K+1\right)\right)-15\right.\\ &&\left.-\frac{8 \left(j+q+18 \Omega _K-2\right)}{3 \left(3
   k_{\alpha }+4 \left(q-9 \Omega _K+1\right)\right)}-\frac{4 \left((q-8) q+s+18 (q-3) \Omega
   _K+6\right)}{3  \left(3 k_{\alpha }+4 \left(q-9 \Omega _K+1\right)\right)}\right]\,,\\
&&\mathcal{D}=-\frac{6 (w +1)}{\left(3 k_{\alpha }+4 \left(q-9 \Omega _K+1\right)\right)  }\,,
\end{eqnarray}
\begin{eqnarray}
&&\mathcal{E}=1\,,\\
&&\nn\mathcal{F}=\frac{k^2}{S^2 k_{\alpha } \Theta^2}+\frac{1}{12} \left(\frac{9 k_{\alpha
   } w }{4 q+3 k_{\alpha }-36 \Omega _K+4}-3 w -2\right)-\frac{2 w  \left(j+q+18\Omega _K-2\right)}{3 k_{\alpha }+4 \left(q-9 \Omega
   _K+1\right)}\\ &&-\frac{2 w  \left((q-8) q+s+18 (q-3) \Omega _K+6\right)}{3
   \left(3 k_{\alpha }+4 \left(q-9 \Omega _K+1\right)\right)}\,,\\
&&\nn \mathcal{G}=\frac{1}{6} (1-3 w ) k_{\alpha } \left(\frac{2 q}{3}-6 \Omega
   _K+\frac{2}{3}\right)+\frac{(1-w  (3 w +4)) \left(\frac{2 j}{9}+\frac{2 q}{9}+4 \Omega
   _K-\frac{4}{9}\right)}{w +1}\\&&+\left(-3 w +\frac{2}{w +1}-1\right) \left(\frac{2
   q^2}{27}+\frac{4 \Omega _K q}{3}-\frac{16 q}{27}+\frac{2 s}{27}-4 \Omega _K+\frac{4}{9}\right)\,,\\
&&\mathcal{H}=\left(\frac{2}{w +1}-1\right) \left(\frac{2 j}{9}+\frac{2 q}{9}+4 \Omega _K-\frac{4}{9}\right)\,.
\end{eqnarray}


\begin{thebibliography}{999}

\bibitem{stringhe}D.~G.~Boulware and S.~Deser,
  Phys.\ Rev.\ Lett.\  {\bf 55}, 2656 (1985); J.~Z.~Simon,
  Phys.\ Rev.\  D {\bf 41} (1990) 3720; K.~Forger, B.~A.~Ovrut, S.~J.~Theisen and D.~Waldram,
  Phys.\ Lett.\  B {\bf 388}, 512 (1996)
  [arXiv:hep-th/9605145]; G.~Cognola, E.~Elizalde, S.~Nojiri, S.~Odintsov and S.~Zerbini,
  Phys.\ Rev.\  D {\bf 75}, 086002 (2007)
  [arXiv:hep-th/0611198].

\bibitem{birrell} N.D. Birrell and P.C.W. Davies, {\it Quantum Fields
in Curved Space}, Cambridge Univ. Press, Cambridge (1982).

\bibitem{SalvSolo}
  S.~Capozziello,
  Int.\ J.\ Mod.\ Phys.\  D {\bf 11}, 483 (2002)
  [arXiv:gr-qc/0201033].

\bibitem{revnostra}
 S.~Capozziello, S.~Carloni and A.~Troisi,
 ``Recent Research Developments in Astronomy \&  Astrophysics"-RSP/AA/21 (2003)
  [arXiv:astro-ph/0303041]; S.~Capozziello, V.~F.~Cardone, S.~Carloni and A.~Troisi,  Int.\ J.\ Mod.\ Phys.\ D {\bf 12} (2003) 1969 [arXiv:astro-ph/0307018]; S.~Capozziello, V.~F.~Cardone and A.~Troisi,  JCAP {\bf 0608} (2006) 001 [arXiv:astro-ph/0602349].

\bibitem{Odintsov}
  S.~Nojiri and S.~D.~Odintsov,
  Phys.\ Rev.\  D {\bf 68} (2003) 123512
  [arXiv:hep-th/0307288]
   \bibitem{Carroll}
    S.~M.~Carroll, V.~Duvvuri, M.~Trodden and M.~S.~Turner,
  Phys.\ Rev.\  D {\bf 70}, 043528 (2004).
  [arXiv:astro-ph/0306438].

  \bibitem{star2007} A.~A.~Starobinsky,
  JETP Lett.\  {\bf 86}, 157 (2007)
  [arXiv:0706.2041 [astro-ph].
  \bibitem{Cognola1}  G.~Cognola, E.~Elizalde, S.~Nojiri, S.~D.~Odintsov, L.~Sebastiani and S.~Zerbini,
  Phys.\ Rev.\  D {\bf 77}, 046009 (2008)
  [arXiv:0712.4017 [hep-th]].

  \bibitem{kerner1} J.P. Duruisseau and R. Kerner, Gen. Rel. Grav. 15, 797-807 (1983).

  \bibitem{star80}  A. A. Starobinsky, Phys. Lett. {\bf B 91}, 99 (1980);K.~S.~Stelle,
  Gen.\ Rel.\ Grav.\  {\bf 9} (1978) 353.

\bibitem{OurDynSys}
   M.~Abdelwahab, S.~Carloni and P.~K.~S.~Dunsby,
  arXiv:0706.1375 [gr-qc];  J.~A.~Leach, S.~Carloni and P.~K.~S.~Dunsby,
  ``Shear dynamics in Bianchi I cosmologies with $R^n$-gravity'',
  Class.\ Quant.\ Grav.\  {\bf 23}, 4915 (2006)
  [arXiv:gr-qc/0603012];
  S.~Carloni, P.~K.~S.~Dunsby and D.~M.~Solomons,
  ``Bounce conditions in f(R) cosmologies'',
  Class.\ Quant.\ Grav.\  {\bf 23}, 1913 (2006)
  [arXiv:gr-qc/0510130];S.~Carloni, P.~K.~S.~Dunsby, S.~Capozziello and A.~Troisi,
  ``Cosmological dynamics of $R^n$ gravity'',
  Class.\ Quant.\ Grav.\  {\bf 22}, 4839 (2005)
  [arXiv:gr-qc/0410046].

 \bibitem{BarrowDyn} T.~Clifton and J.~D.~Barrow,
  Phys.\ Rev.\  D {\bf 72}, 103005 (2005)
  [arXiv:gr-qc/0509059];

 \bibitem{Barrow Hervik}  J.~D.~Barrow and S.~Hervik,
  Phys.\ Rev.\  D {\bf 74} (2006) 124017
  [arXiv:gr-qc/0610013].

\bibitem{OdintRec} S.~Nojiri, S.~D.~Odintsov and M.~Sami,
  Phys.\ Rev.\  D {\bf 74}, 046004 (2006)
  [arXiv:hep-th/0605039].

\bibitem{CognolaDyn} G.~Cognola and S.~Zerbini,
  arXiv:0802.3967 [hep-th]; G.~Cognola, M.~Gastaldi and S.~Zerbini,
  Int.\ J.\ Theor.\ Phys.\  {\bf 47}, 898 (2008)
  [arXiv:gr-qc/0701138].

\bibitem{Li:2008ai}
B.~Li, J.~D.~Barrow, D.~F.~Mota and H.~Zhao,
arXiv:0805.4400 [gr-qc];

 \bibitem{HuSawicki}Y.~S.~Song, W.~Hu and I.~Sawicki,
Phys.\ Rev.\  D {\bf 75}, 044004 (2007)
[arXiv:astro-ph/0610532];

\bibitem{Bertschinger:2008zb}
  E.~Bertschinger and P.~Zukin,
  Phys.\ Rev.\  D {\bf 78}, 024015 (2008)
  [arXiv:0801.2431 [astro-ph]].

\bibitem{otherperts}
W.~Hu and I.~Sawicki,
Phys.\ Rev.\  D {\bf 76}, 104043 (2007)
[arXiv:0708.1190 [astro-ph]];
H.~Oyaizu, M.~Lima and W.~Hu,
 arXiv:0807.2462 [astro-ph];

\bibitem{SantePert}
  S.~Carloni, P.~K.~S.~Dunsby and A.~Troisi,
  ``The evolution of density perturbations in $f(R)$ gravity,'' Phys. Rev. D 77, 024024 (2008)
  arXiv:0707.0106 [gr-qc]

\bibitem{K1}  K.~N.~Ananda, S.~Carloni and P.~K.~S.~Dunsby,
  arXiv:0708.2258 [gr-qc] Phys. Rev. D77 (2008) 024033

 \bibitem{EllisCovariant} G.~F.~R.~Ellis \& H van Elst,
``Cosmological Models", Carg\`{e}se Lectures 1998, in Theoretical
and Observational Cosmology, Ed. M Lachièze-Rey, (Dordrecht: Kluwer,
1999), 1. [arXiv:gr-qc/9812046].


\bibitem{EB} G.~F.~R.~Ellis \& M.~Bruni
Phys Rev D {\bf 40} 1804 (1989).

\bibitem{BDE}
M.~Bruni,  P.~K.~S.~Dunsby \& G.~F.~R.~Ellis,
Ap. J. {\bf 395} 34 (1992).

\bibitem{EBH} G.~F.~R.~Ellis, M.~Bruni and J.~Hwang,
  Phys.\ Rev.\  D {\bf 42} (1990) 1035 (1990).

\bibitem{DBE}  P.~K.~S.~Dunsby, M.~Bruni and G.~F.~R.~Ellis,
  Astrophys.\ J.\  {\bf 395}, 54 (1992)

\bibitem{BED} M.~Bruni, G.~F.~R.~Ellis and P.~K.~S.~Dunsby,
 Class.\ Quant.\ Grav.\  {\bf 9}, 921 (1992).

\bibitem{DBBE}   P.~K.~S.~Dunsby, B.~A.~C.~Bassett and G.~F.~R.~Ellis,
  Class.\ Quant.\ Grav.\  {\bf 14}, 1215 (1997)
  [arXiv:gr-qc/9811092].

\bibitem{conserved}
  P.~K.~S.~Dunsby and M.~Bruni,
  Int.\ J.\ Mod.\ Phys.\  D {\bf 3}, 443 (1994)
  [arXiv:gr-qc/9405008].

\bibitem{kunz}
M. Kunz, [arXiv:astro-ph/0612452v]

\bibitem{WMAP} E.~Komatsu {\it et al.}  [WMAP Collaboration],
  arXiv:0803.0547 [astro-ph].

\bibitem{Planck}
    [Planck Collaboration],
  arXiv:astro-ph/0604069.

\bibitem{SDSS} see the webpage http://www.sdss.org/

\bibitem{Salv06}  S.~Capozziello, V.~F.~Cardone and A.~Troisi,
  JCAP {\bf 0608}, 001 (2006)
  [arXiv:astro-ph/0602349].

\bibitem{Salv07} S.~Capozziello, V.~F.~Cardone and A.~Troisi,
  Mon.\ Not.\ Roy.\ Astron.\ Soc.\  {\bf 375}, 1423 (2007)
  [arXiv:astro-ph/0603522].

 \bibitem{Salucci}
  C.~F.~Martins and P.~Salucci,
  Mon.\ Not.\ Roy.\ Astron.\ Soc.\  {\bf 381}, 1103 (2007)
  [arXiv:astro-ph/0703243].

\bibitem{Challinor}
A.~Challinor and A.~Lasenby,
Astrophys.\ J.\  {\bf 513}, 1 (1999)
 [arXiv:astro-ph/9804301].

  \bibitem{VisserPar}
  M.~Visser,
  Class.\ Quant.\ Grav.\  {\bf 21} (2004) 2603
  [arXiv:gr-qc/0309109].

 \bibitem{Coles}
  P.~Coles and F.~Lucchin,
{\it  Chichester, UK: Wiley (1995) 449 p}

 \bibitem{PaddyPert} T.~Padmanabhan,
  AIP Conf.\ Proc.\  {\bf 843} (2006) 111
  [arXiv:astro-ph/0602117]; T.~Padmanabhan ``Structure Formation in the Universe" Cambridge university press (Cambridge)

 \bibitem{cdct:dynsys05} S.~Carloni, P.~Dunsby, S.~Capozziello \& A.~Troisi  \cqg 22, 4839 (2005).

 \bibitem{ellisbook} {\em Dynamical System in Cosmology} edited by
Wainwright J and  Ellis G F R (Cambridge: Cambridge  Univ. Press
1997) and references therein.


 \bibitem{Ottewill} J.~D.~Barrow and A.~C.~Ottewill,
  J.\ Phys.\ A  {\bf 16} (1983) 2757.

  \bibitem{Suen} M.~B.~Mijic, M.~S.~Morris and W.~M.~Suen,
  Phys.\ Rev.\  D {\bf 34} (1986) 2934.

 \bibitem{Teyssandier:1989dw}
  P.~Teyssandier,
  Class.\ Quant.\ Grav.\  {\bf 6}, 219 (1989).

 \bibitem{SanteGenDynSys}  S.~Carloni, A.~Troisi and P.~K.~S.~Dunsby,
  arXiv:0706.0452 [gr-qc]

\bibitem{Amendola:2006kh}
  L.~Amendola, D.~Polarski and S.~Tsujikawa,
  Phys.\ Rev.\ Lett.\  {\bf 98}, 131302 (2007)
  [arXiv:astro-ph/0603703].

    \end{thebibliography}
\end{document}